\documentclass[11pt]{report}

\usepackage[utf8]{inputenc}
\usepackage[T1]{fontenc}
\usepackage{times}
\usepackage[scaled]{berasans}
\usepackage{kpfonts}
\usepackage{mathpazo}
\usepackage{array}
\usepackage{listings}

\usepackage[english]{babel}

\usepackage[full]{textcomp}
\usepackage{eurosym}
\usepackage{gensymb}

\usepackage{graphicx}
\usepackage{tikz}
\usepackage{eso-pic, rotating}
\usepackage[labelfont={color=arfima,bf}]{caption}
\usepackage{subcaption}
\usepackage{tabularx}
\usepackage{longtable}
\usepackage{booktabs}
\usepackage{multirow}
\usepackage{threeparttable}
\usepackage{hhline}

\usepackage[backend=biber,style=numeric]{biblatex}
\usepackage{geometry}
\usepackage{setspace}
\usepackage{ragged2e}
\usepackage{xcolor, colortbl}
\usepackage{sectsty}
\usepackage{fancyhdr}
\usepackage{hyphenat}
\usepackage[hang]{footmisc}
\usepackage{lipsum}

\usepackage{amsmath}
\usepackage{amsfonts}
\usepackage{amssymb}
\usepackage{upgreek}
\usepackage{mathtools}

\usepackage{tcolorbox}
\usepackage{framed}
\usepackage{adjustbox}
\usepackage[linkcolor=blue,colorlinks=true, citecolor=blue, urlcolor=blue]{hyperref}
\usepackage[nodayofweek,level]{datetime}
\usepackage{afterpage}
\usepackage{float}
\usepackage{calc}
\usepackage{lscape}
\usepackage{lastpage}
\usepackage{csquotes}
\usepackage{enumerate}
\usepackage{rotating}

\nocite{*}

\definecolor{color1}{HTML}{000060}
\definecolor{color2}{HTML}{333333}
\definecolor{rahmen}{RGB}{0,73,114}
\definecolor{grund}{RGB}{238,241,251}          
\definecolor{schrift}{RGB}{0,73,114}
\definecolor{arfima}{RGB}{55,96,146}
\definecolor{pantone}{HTML}{001489}
\chapterfont{\color{arfima}}
\sectionfont{\color{arfima}}  
\subsectionfont{\color{arfima}}

\definecolor{myblue}{rgb}{0.8,0.8,1}
\definecolor{my}{rgb}{1,0.8,0.3}

\usepackage{amsthm}

\newtheoremstyle{plainstyle}  
  {}       
  {}       
  {\itshape}   
  {}       
  {\bfseries}  
  {.}      
  {5pt plus 1pt minus 1pt} 
  {}       

\newtheoremstyle{definitionremarkstyle} 
  {}       
  {}       
  {}       
  {}       
  {\bfseries}  
  {.}      
  {5pt plus 1pt minus 1pt} 
  {}       

\theoremstyle{plainstyle}

\theoremstyle{definitionremarkstyle}

\renewcommand\thefootnote{\textcolor{arfima}{\arabic{footnote}}}
\renewcommand\footnoterule{{\color{arfima}\kern-3pt \hrule width 2in \kern 2.6pt}}

\fancypagestyle{titlepagestyle}{
    \fancyhf{} 
    \fancyfoot[C]{
\footnotesize{This research is part of the project I+D+i
TED2021-131844B-I00, funded by MCIN/ AEI/10.13039/501100011033 and the European
Union NextGeneration EU/PRTR. \includegraphics[height=1cm]{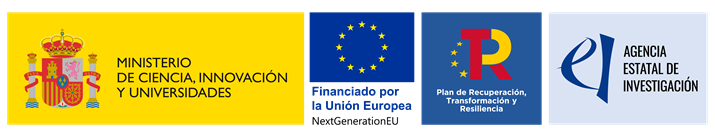}
           }}
    \renewcommand{\headrulewidth}{0pt} 
}

\begin{document}
\newgeometry{top=3cm, bottom=4cm, left=1.5cm, right=1.5cm, footskip=6mm}
\begin{titlepage}
$$\phantom{prueba} \vspace{7cm} $$
    \thispagestyle{titlepagestyle} 
    	\vspace*{-15em}
    \begin{center}
        {\Huge \textbf{Price Discovery in Cryptocurrency Markets}}\\[1cm]
        \vfill
        \textbf{Authors:}\\[0.5cm]
        {\Large Juan Plazuelo Pascual \\ \texttt{jplazuelo@arfimaconsulting.com}}\\[.5cm]
        {\Large Carlos Tardón Rubio \\ \texttt{ctardon@arfimaconsulting.com}}\\[.5cm]
        {\Large Juan Toro Cebada \\ \texttt{jtoro@arfimaconsulting.com}}\\[.5cm]
        {\Large Ángel Hernando Veciana \\ \texttt{ahvecian@eco.uc3m.es}}\\[5cm]
    \end{center}

		
		

	
\fancypagestyle{titlepagestyle}{
    \fancyhf{} 
    \fancyfoot[C]{
        \parbox{0.9\textwidth}{\footnotesize
 This research is part of the project I+D+i TED2021-131844B-I00, funded by MCIN/ AEI/10.13039/501100011033 and the European Union NextGeneration EU/PRTR.
 \\          \begin{center} \includegraphics[height=1cm]{images/logosahv.png} \end{center}
        }
        \vspace{0.5cm}
    }
    \renewcommand{\headrulewidth}{0pt} 
}

\end{titlepage}
\restoregeometry


\pagestyle{fancy}
\fancyhead[R]{\includegraphics[width=0.9cm]{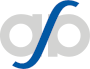}}
\fancyhead[L]{\color{arfima} Crypto}
\fancyfoot[C]{\thepage \phantom{.} de \phantom{.} \pageref{LastPage}}
\renewcommand\thefootnote{\textcolor{arfima}{\arabic{footnote}}}
\renewcommand\footnoterule{{\color{arfima}\kern-3pt \hrule width 2in \kern 2.6pt}}

\newpage
\tableofcontents
\phantom{.}


\chapter*{Executive Summary}

This work is part of the research undertaken within the project \textit{Accomplishing digital finance: Decentralized oracles for a decentralized financial system} (TED2021-131844B-I00) that aims to analyze decentralized price mechanisms. Specifically, in this study, we address the following issues:

\begin{itemize}
    \item \textbf{Evaluation of AMMs as oracles} by comparing them with off-chain data sources. \textbf{In particular, we analyze Uniswap v2} and compare its price formation with that of centralized exchanges. An extensive amount of work on month-end AMMs has been conducted, and an order book has been built to retrieve relevant pricing data for this analysis.
    \item \textbf{Empirical analysis of the ETH price time series} on Binance versus Uniswap is carried out for both a long data span and shorter windows with more granular data. This allows us to investigate differences in price dynamics.
    \item \textbf{Lead-lag relationship estimation} for the assets mentioned above (on-chain versus off-chain) is complemented with Hasbrouck Information Share (IS) analysis and Gonzalo and Granger price decomposition analysis. These measures provide an assessment of price information dynamics between on-chain and off-chain systems.
    \item \textbf{Cointegration analysis and Granger causality tests} are applied to determine price formation in each pricing mechanism. This is performed exhaustively for both long-span and short-window data under stressed liquidity conditions, where gas prices and liquidity constraints play a significant role.
    \item \textbf{Arbitrage opportunity estimation} is conducted by considering price dynamics, mispricing of identical assets, and transaction costs in the form of gas fees.
    \item \textbf{Performance benchmarks for modified AMMs} are established, comparing them to off-chain data sources. Specifically, Uniswap v2 is analyzed in depth to understand its mechanisms.
    \item \textbf{Incorporation of on-chain data} is achieved using the Uniswap blockchain.
    \item \textbf{Benchmarking the performance of modified AMMs} is conducted by comparing Uniswap AMM operations (on-chain) to the same asset's performance on the Binance exchange (off-chain).
\end{itemize}

This document presents an extensive analysis of price discovery in cryptocurrency markets, comparing centralized and decentralized exchanges as well as spot and futures markets. We first conduct the analysis for ETH, followed by a similar approach for BTC.

Chapter 1 introduces the \textit{theoretical framework}, explaining the fundamental differences between centralized exchanges and decentralized finance (DeFi) mechanisms, putting special emphasis on Automated Market Makers (AMMs). We also explain how to construct an order book from a liquidity pool in a decentralized environment, so that it can be compared to the order book data available in centralized exchanges.

Chapter 2 discusses the \textit{methodological approaches} used in this work, including Hasbrouck’s Information Share, Gonzalo and Granger’s Permanent-Transitory decomposition, and the Hayashi-Yoshida methodology. These are applied to assess lead-lag relationships, cointegration, and price discovery, focusing on both centralized and decentralized markets.

Chapter 3 first presents the \textit{empirical analysis of the ETH price time series} on Binance versus Uniswap v2. We analyze data over a one-year period and focus on five specific events in 2024, examining how price discovery occurs across these markets. In this chapter, we also investigate the \textit{spot BTC versus BTC futures market} on the CME, exploring how information flows between these markets. In both cases, we estimate lead-lag relationships, cointegration, and other metrics. We find that centralized markets generally lead in terms of price discovery for Ethereum. Additionally, our analysis indicates that while futures markets tend to lead, moments of high volatility produce mixed results, and the difference between the two is not as significant as in the centralized versus decentralized comparison.

These insights have important implications for traders, investors, and institutions, particularly with regard to liquidity, arbitrage opportunities, and market efficiency. Throughout, \textit{ different metrics serve as performance benchmarks for modified AMMs}, highlighting the interaction between decentralized and centralized market structures.

\chapter{Theoretical framework behind crypto data}

The process of price discovery is central to understanding how information is incorporated into asset prices. It refers to the mechanism through which the price of an asset is determined in the market, reflecting all available information. This document investigates the dynamics of price discovery across different markets, particularly following news events. We focus on understanding how centralized exchanges and decentralized finance platforms are interconnected. Specifically, a natural question in this context is how these centralized and decentralized exchanges interact, especially in terms of information transfer between them. Information asymmetry, or the lack of shared market data, can lead to inefficiencies. Understanding where information is generated (e.g., on-chain activity, liquidity changes, or significant price movements) and how it spreads across both centralized exchanges (CEXs) and decentralized exchanges (DEXs) is important for traders seeking to arbitrage price discrepancies or gauge market sentiment.

Moreover, the rise of decentralized exchanges prompts further questions about how data—such as order flow, pricing, and liquidity—moves between centralized and decentralized ecosystems. For instance, how do pricing and volume data from DEXs influence the behavior of traders on CEXs, and vice versa? The relationship between these two types of exchanges is intricate, as they operate under different mechanisms but are linked through arbitrage and cross-market trading.

To analyze the functioning of price discovery in this environment, we will apply several methodologies, specifically the Hasbrouck Information Share, the Gonzalo-Granger Permanent-Transitory decomposition, and the Hayashi-Yoshida estimator. These methods will be discussed in detail in Chapter 2. Before proceeding, we must develop a deep understanding of both centralized cryptocurrency markets, such as Binance, and decentralized platforms, like Uniswap. Centralized exchanges use a traditional order book model, and there is extensive research on how to extract and work with time-series data from these platforms. However, decentralized platforms have been less thoroughly explored.

Our goal in this chapter is to explain, in detail, how to construct an order book from a liquidity pool in a decentralized environment, so that it can be compared to the order book data available in centralized exchanges. We will begin with basic definitions and work through the technical details of how Uniswap v2 operates.

\section{Digital markets}

Cryptocurrencies are digital or virtual currencies that use cryptography for security and operate on decentralized networks based on blockchain technology. Since the introduction of Bitcoin in 2009, cryptocurrencies have significantly evolved, leading to the development of numerous blockchain platforms and applications.  Blockchain technology, a distributed ledger system, ensures transparency, security, and immutability of data, making it a foundational element for various decentralized applications. Decentralized Finance (DeFi) represents a paradigm shift in the financial sector, aiming to create an open and permissionless financial system that operates without the need for centralized intermediaries like banks and financial institutions. One of the significant innovations within the DeFi ecosystem is the development of Automated Market Makers (AMMs). In AMM-based systems, liquidity providers (LPs) contribute funds to liquidity pools, which are used to execute trades. Examples of popular AMMs include Uniswap, Balancer, and Curve. We will focus on Uniswap in the remainder of the work. On the other hand, digital assets and their derivatives are traded on centralized exchange markets, which use order books to facilitate matching between buyers and sellers. Centralized exchanges (CEXs) typically provide a more controlled and regulated environment, with liquidity concentrated in a few major platforms, allowing for fast execution and advanced trading features. Despite the increasing prominence of decentralized finance (DeFi) and automated market makers (AMMs), a large portion of trading activity still occurs on centralized exchanges, particularly for large institutional trades.

However, the role of DeFi exchanges, which operate without intermediaries and rely on smart contracts, is growing rapidly. These decentralized exchanges (DEXs) provide traders with greater transparency, user autonomy, and access to a wider range of tokenized assets. As the technology and adoption of DeFi platforms advance, we are witnessing an important shift in how digital assets are traded, with DeFi becoming a major player in global markets.

\subsection{Literature review}

The literature on decentralized finance (DeFi) has particularly focused on Automated Market Makers (AMMs) and their role in comparison to traditional centralized exchanges (CEX). DeFi, a blockchain-based financial system, operates without intermediaries by using smart contracts to execute trades. AMMs, which use algorithms instead of traditional limit order books, are central to decentralized exchanges (DEXs) as highlighted by Schär (2021) \cite{RePEc:fip:fedlrv:91428} and John, Kogan, and Saleh (2022) \cite{ssrn4222528}.

Early research explores the role of liquidity providers in DEXs and the coexistence of DEXs with centralized limit order books, as highlighted by Lehar, Parlour, and Zoican (2024) \cite{lehar2024fragmentationoptimalliquiditysupply}. Studies have also examined the empirical differences between DEX and CEX markets, notably by Han, Huang, and Zhong (2021) \cite{han2021trustdefi}, and Heimbach, Wang, and Wattenhofer (2021) \cite{heimbach2021behaviorliquidityprovidersdecentralized}. Some models suggest that AMMs can be more favorable than limit order books in low-volatility environments. Factors such as volatility, transaction costs, and gas fees play significant roles in liquidity provision on DEXs. For instance, high transaction costs can limit arbitrage opportunities and impact price efficiency, as noted by Barbon and Ranaldo (2021) \cite{barbon2021arbitrage}.

Informed trading and liquidity provision in DEXs have been widely studied. Research shows that increased liquidity in DEXs attracts informed traders, reducing adverse selection in CEXs and impacting bid-ask spreads, as noted by Aoyagi and Ito (2021) \cite{aoyagi2021coexisting}. Trades on DEXs that incur higher costs, like gas fees, are often associated with higher informational content, as informed traders are willing to pay more for prompt execution, as shown by Capponi, Jia, and Yu (2024) \cite{capponi2024price}. Arbitrage in DEXs, while prevalent, is constrained by transaction costs, which can hinder price convergence between DEX and CEX markets, as described by Daian et al. (2019) \cite{daian2019flash}. Liquidity providers in DEXs often strategically re-balance their positions based on market expectations, either burning liquidity near the current price for potential gains or withdrawing liquidity to avoid anticipated losses, as discussed by Lehar, Parlour, and Zoican (2024) \cite{lehar2024fragmentationoptimalliquiditysupply}.

Research comparing DEX and CEX markets highlights key differences in transaction costs, price efficiency, and liquidity provider behavior, with DEXs generally facing higher costs and inefficiencies due to gas fees (Barbon and Ranaldo 2021 \cite{barbon2021arbitrage}; Capponi, Jia, and Yu 2024 \cite{capponi2024price}).

In this paper, we investigate the relationship between centralized and decentralized exchanges. The paper is organized as follows. In the first chapter, we introduce the concept of an Automated Market Maker (AMM) and define the mechanism of a canonical AMM, Uniswap. We examine two of the most recent protocols and explain how prices are defined, the granularity of prices is obtained, and how the order books can be reconstructed from available information. This is all necessary to recover prices and order books for the AMMs we will use in a later stage.

In Chapter Two, we define the tools that will be used to investigate information transmission across markets. We explore three standard and well-established methodologies: Hasbrouck’s information share, Gonzalo and Granger’s common trends test, and a methodology for analyzing lead-lag relationships using tick-by-tick data.

Both of these chapters provide the instruments and tools to address the relevant questions regarding information transmission. We first investigate the price dynamics between centralized exchanges (CEX) and decentralized finance (DeFi) markets, focusing on Ethereum in Binance and Uniswap v3 as the assets analyzed in our research. We initially explore these questions using a large data sample.

Next, we examine whether the same information transmission patterns hold true during periods of market stress. In highly stressed markets where liquidity is low, we investigate and revisit the same question: Which market originates the first relevant information, and how is this information distributed?

This chapter  explores the mechanics of Uniswap v2 and Uniswap v3, two prominent AMM protocols, and provides a detailed analysis of their operations and impact on the DeFi landscape. Through this exploration, we aim to provide understanding of both their advantages and limitations, how prices are defined and how liquidity is measured.

\subsection{Blockchain and Smart Contracts}
A blockchain is a decentralized, distributed ledger that records transactions in a secure, transparent, and immutable manner. A ledger is a database that is consensually shared and synchronized across multiple sites, institutions, or geographies, ensuring that records are consistently updated and accessible to all participants.

Smart contracts are self-executing programs on the blockchain with terms directly written into code. They automatically execute and enforce agreements when predefined conditions are met, eliminating the need for intermediaries.

Ethereum is a leading blockchain platform supporting smart contracts, primarily coded in a programming language called Solidity. This facilitates the creation of decentralized applications (DApps) and various DeFi solutions.

\subsection{Evaluation of AMMs as oracles: decentralized finance (DeFi), liquidity pools, and automated market makers—differences with the order book model}

Traditional centralized exchanges like Binance operate using an order book model, where buy and sell orders are listed with specific bid and ask prices. Market makers facilitate trading by constantly providing buy and sell orders, thus maintaining liquidity and ensuring trades can occur efficiently.

In decentralized finance (DeFi), relying on traditional market makers contradicts the goal of decentralization. Instead, liquidity pools are used. A liquidity pool consists of two tokens. When created, it starts with no tokens. The first liquidity provider sets the initial price by depositing both tokens in the correct ratio. If the ratio is incorrect, arbitrage opportunities may arise. When many liquidity participants join, the price tends to align with those in centralized markets. One of the goals of this thesis is to discover which mechanism ultimately leads the process of price discovery.

Liquidity providers receive special tokens called LP (Liquidity Provider) tokens in proportion to their contribution to the pool. To retrieve their liquidity, they must burn these LP tokens. Burning tokens means permanently removing them from circulation by sending them to an irretrievable address. This mechanism, known as an Automated Market Maker (AMM), allows for continuous and decentralized trading.

Uniswap v2 employs a constant product formula to determine prices and allocate tokens to liquidity providers, which we will detail later.

\subsection{Uniswap}
Uniswap is a decentralized exchange protocol that allows users to trade tokens directly from their wallets. It has evolved from Uniswap v2 to Uniswap v3, each offering unique features and improvements. In the following, we will delve into the ins and outs of two different AMM protocols: Uniswap v2 and Uniswap v3. We will show and explain how prices are determined in each protocol and how the granularity of prices is defined. We will also detail how order books can be reconstructed from the rules that define the AMM. Lastly, we will demonstrate how liquidity snapshots, burns, mints, etc., help retrieve the necessary information to recover AMM order books. All these tasks are essential for retrieving the relevant information from the AMM book.

Uniswap v2 requires liquidity providers to deposit their liquidity in a 50-50 ratio, meaning they must provide equal amounts of both tokens in the pool. When liquidity providers add liquidity, the pool size increases, which lowers the slippage of a given trade.

Uniswap v3 introduces a more flexible approach, allowing liquidity providers to concentrate their liquidity within specific price ranges. This means they can allocate their assets to narrower price intervals, improving capital efficiency. When opening a new liquidity position in Uniswap v3, providers specify a price range \([p_a, p_b]\) within which their position is active. A position only needs to maintain enough reserves to support trading within its range and can act like a constant product pool with larger (virtual) reserves within that range.

Some key observations are:

- \textbf{Observation 1}: In Uniswap v2, liquidity providers are required to deposit their liquidity in a \textbf{50-50 ratio}. If liquidity providers add liquidity, the pool size increases, which lowers the slippage of a given trade. In Uniswap v3, the \textbf{required reserves depend on the selected price range relative to the market price}, with larger reserves of token Y required if the price range is skewed to prices lower than the market price, \(p_{\text{mkt}}\).

- \textbf{Observation 2}: Uniswap v3 improves liquidity by allowing liquidity providers to concentrate their liquidity on smaller price ranges. When they open a new liquidity position, they have to specify a price range \textbf{\([p_a, p_b]\) } at which their position is active.

A position only needs to maintain enough reserves to support trading within its range and can therefore act like a constant product pool with larger (virtual) reserves within that range.

\subsubsection{How Trades are Performed in Uniswap v2}

In Uniswap v2, trades are executed using a constant product formula involving liquidity pools. Each trading pair (e.g., ETH/DAI) has its own liquidity pool consisting of two assets. Liquidity providers deposit an equal value of both assets into the pool. The total liquidity is represented by two amounts: \( x \) and \( y \), where \( x \) is the quantity of token A and \( y \) is the quantity of token B.

\subsubsection*{Constant Product Formula}

Uniswap v2 uses a constant product market maker model, which ensures that the product of the quantities of the two assets remains constant. This can be expressed mathematically as:

\[
k = x \cdot y
\]

where \( k \) is a constant. When a trade is performed, the trader provides one token and receives another. Here is the step-by-step process:

\begin{enumerate}
    \item \textbf{Determine Input and Output Amounts:}
    \begin{itemize}
        \item Let’s say a trader wants to swap \( \Delta x \) of token A for token B.
        \item The trader provides \( \Delta x \) of token A to the pool, increasing the pool’s token A balance from \( x \) to \( x + \Delta x \).
    \end{itemize}

    \item \textbf{Calculate Output Using Constant Product Formula:}
    \begin{itemize}
        \item The new amount of token B in the pool must satisfy the constant product formula:
        \[
        (x + \Delta x) \cdot (y - \Delta y) = k
        \]
        \item Solving for \( \Delta y \) (the amount of token B to be received by the trader):
        \[
        \Delta y = y - \frac{k}{x + \Delta x}
        \]
        The pool’s new balance of token B becomes \( y - \Delta y \).
    \end{itemize}

    \item \textbf{Include the Trading Fee:}
    \begin{itemize}
        \item Uniswap v2 charges a 0.30\% fee on each trade, which is added to the liquidity pool. Therefore, the actual amount of tokens provided by the trader, accounting for the fee, is \( \Delta x \) adjusted by the fee.
        \item The effective amount added to the pool is \( \Delta x \cdot (1 - 0.003) = \Delta x \cdot 0.997 \).
    \end{itemize}
\end{enumerate}

\subsubsection*{Example Calculation}

Suppose the pool initially has 1000 ETH (token A) and 4000 DAI (token B). Therefore,

\[
k = 1000 \cdot 4000 = 4,000,000
\]

A trader wants to swap 10 ETH (token A) for DAI (token B).

The effective amount of ETH added to the pool is \( 10 \cdot 0.997 = 9.97 \) ETH.

New ETH balance:

\[
1000 + 9.97 = 1009.97
\]

Using the constant product formula:

\[
1009.97 \cdot (4000 - \Delta y) = 4,000,000
\]

Solving for \( \Delta y \):

\[
\Delta y = 4000 - \frac{4,000,000}{1009.97}
\]

\[
\Delta y \approx 4000 - 3961.07
\]

\[
\Delta y \approx 38.93
\]

So, the trader receives approximately 38.93 DAI.

New DAI balance:

\[
4000 - 38.93 = 3961.07
\]

New ETH balance:

\[
1009.97
\]

\subsubsection{Example for Slippage in Uniswap v2}

Let's understand the concept of slippage with the following example. We are not going to account for fees to isolate the effect of slippage.

\textbf{Initial Parameters:}
\begin{itemize}
    \item Initial liquidity: 10 ETH and 20,000 USDC
    \item Expected price for 1 ETH: 2,000 USDC
\end{itemize}

First, we calculate \( k \):

\[
k = 10 \, \text{ETH} \cdot 20,000 \, \text{USDC} = 200,000
\]

A trader wants to buy 1 ETH. 

\textbf{Expected Price:}

The expected price is 2,000 USDC for 1 ETH. However, due to slippage, the real price may differ.

\textbf{Real Price Calculation:}

\begin{align*}
    &\text{Initial Pool Balances:} \\
    &x = 10 \, \text{ETH}, \quad y = 20,000 \, \text{USDC} \\
    &\text{New Pool Balance after the Trade:} \\
    &x_{\text{new}} = 10 - 1 = 9 \, \text{ETH} \\
    &9 \cdot (20,000 + \Delta y) = 200,000 \\
    &20,000 + \Delta y = \frac{200,000}{9} \\
    &20,000 + \Delta y = 22,222.22 \\
    &\Delta y = 22,222.22 - 20,000 \\
    &\Delta y = 2,222.22 \, \text{USDC}
\end{align*}

So, the real price to buy 1 ETH is 2,222.22 USDC, which is higher than the expected price of 2,000 USDC due to slippage. The real price is higher than the expected price due to the impact of the trade on the liquidity pool.

This motivates using a system where liquidity can be "concentrated". This is how Uniswap v3 was born.

In practice, Uniswap v2 is also effective if the liquidity pool has sufficient liquidity.

\subsection{Trade Data Fields Explanation}

When analyzing trade data in the context of AMMs on decentralized exchanges (DEXs), there are several key fields that we need to pay attention to. We provide a brief explanation of each field:

\begin{itemize}
    \item \textbf{Pool Address:}The unique identifier for the pool where the trade occurred.

    \item \textbf{Fee Tier:} The percentage of the transaction amount taken as a fee by the pool. It indicates the cost of the transaction for the user and can be used to calculate revenue for liquidity providers.

    \item \textbf{Block Number:} The specific block in the blockchain where the transaction was recorded. It provides a timestamp for the transaction.

    \item \textbf{Amounts of Tokens Swapped:} The quantities of the different tokens that were exchanged in the transaction.

    \item \textbf{Pool Price After the Transaction:} The price of the tokens in the pool immediately after the transaction was completed. It is quite important for assessing slippage, as we will notice in the examples.
\end{itemize}

We now provide an example.
\textbf{Example of Trade Data Record}

\begin{verbatim}
{
    "pool_address": "0x1a2b3c4d5e6f7g8h9i0jklmnopqrstuv",
    "fee_tier": "0.3%",
    "block_number": 1234567,
    "amounts_swapped": {
        "tokenA": 100,
        "tokenB": 50
    },
    "pool_price_after_transaction": {
        "tokenA": 2.0,
        "tokenB": 0.5
    }
}
\end{verbatim}

\subsection{DEX Liquidity Data Components}

In decentralized exchanges , it’s also important to understand the different types of events and snapshots that capture the state and changes in liquidity. Here are the key components:

\begin{itemize}
    \item \textbf{Liquidity Events (Mints/Burns):}
    \begin{itemize}
        \item \textbf{Mints:}
        \begin{itemize}
            \item \textbf{Description:} Events where liquidity providers add liquidity to a pool.
            \item \textbf{Fields:}
            \begin{itemize}
                \item \textbf{Pool Address:} The identifier of the liquidity pool.
                \item \textbf{Provider Address:} The address of the user adding liquidity.
                \item \textbf{Amounts Added:} The quantities of each token added to the pool.
                \item \textbf{Timestamp/Block Number:} When the event occurred.
            \end{itemize}
            \item \textbf{Purpose:} Increases the pool’s liquidity, allowing for larger trades with less slippage.
        \end{itemize}
        \item \textbf{Burns:}
        \begin{itemize}
            \item \textbf{Description:} Events where liquidity providers remove liquidity from a pool.
            \item \textbf{Fields:}
            \begin{itemize}
                \item \textbf{Pool Address:} The identifier of the liquidity pool.
                \item \textbf{Provider Address:} The address of the user removing liquidity.
                \item \textbf{Amounts Removed:} The quantities of each token removed from the pool.
                \item \textbf{Timestamp/Block Number:} When the event occurred.
            \end{itemize}
            \item \textbf{Purpose:} Decreases the pool’s liquidity, which can increase slippage for subsequent trades.
        \end{itemize}
    \end{itemize}
    
    \item \textbf{Liquidity Snapshots:}
    \begin{itemize}
        \item \textbf{Description:} Periodic records capturing the state of the liquidity pool at a specific point in time.
        \item \textbf{Fields:}
        \begin{itemize}
            \item \textbf{Pool Address:} The identifier of the liquidity pool.
            \item \textbf{Total Liquidity:} The total amount of each token in the pool.
            \item \textbf{Total Value Locked (TVL):} The total value of assets in the pool, often measured in a common denominator like USD.
            \item \textbf{Price Ratios:} The current price ratio of the tokens in the pool.
            \item \textbf{Timestamp/Block Number:} When the snapshot was taken.
        \end{itemize}
        \item \textbf{Purpose:} Provides a periodic overview of the pool’s state, useful for analyzing trends and changes in liquidity over time.
    \end{itemize}
\end{itemize}

\textbf{Example Records:}

\textit{Mint Event:}

\begin{verbatim}
{
    "event_type": "mint",
    "pool_address": "0x1a2b3c4d5e6f7g8h9i0jklmnopqrstuv",
    "provider_address": "0xa1b2c3d4e5f6g7h8i9j0klmnopqrstuv",
    "amounts_added": {
        "tokenA": 1000,
        "tokenB": 500
    },
    "timestamp": 1650000000,
    "block_number": 1234567
}
\end{verbatim}

\textit{Burn Event:}

\begin{verbatim}
{
    "event_type": "burn",
    "pool_address": "0x1a2b3c4d5e6f7g8h9i0jklmnopqrstuv",
    "provider_address": "0xa1b2c3d4e5f6g7h8i9j0klmnopqrstuv",
    "amounts_removed": {
        "tokenA": 500,
        "tokenB": 250
    },
    "timestamp": 1650000000,
    "block_number": 1234568
}
\end{verbatim}

\textit{Liquidity Snapshot:}

\begin{verbatim}
{
    "snapshot_type": "liquidity",
    "pool_address": "0x1a2b3c4d5e6f7g8h9i0jklmnopqrstuv",
    "total_liquidity": {
        "tokenA": 5000,
        "tokenB": 2500
    },
    "total_value_locked_usd": 10000,
    "price_ratios": {
        "tokenA": 2.0,
        "tokenB": 0.5
    },
    "timestamp": 1650000000,
    "block_number": 1234570
}
\end{verbatim}

\section{Incorporating on-chain data: understanding ticks in Uniswap v3}

In Uniswap v3, liquidity providers (LPs) specify the price range for their liquidity positions. Prices are divided into discrete intervals called "ticks," each corresponding to a price given by:

\[
p_i = 1.0001^i
\]

where \(i\) is an integer.

This way, prices are set at discrete intervals, with each tick representing a 1 basis point (bp) change from its neighbors.
LPs can concentrate their liquidity \textbf{between any two ticks}, creating a specified price range.

\subsection{Example and tick spacing}

\begin{figure}[h]
    \centering
    \includegraphics[width=\textwidth]{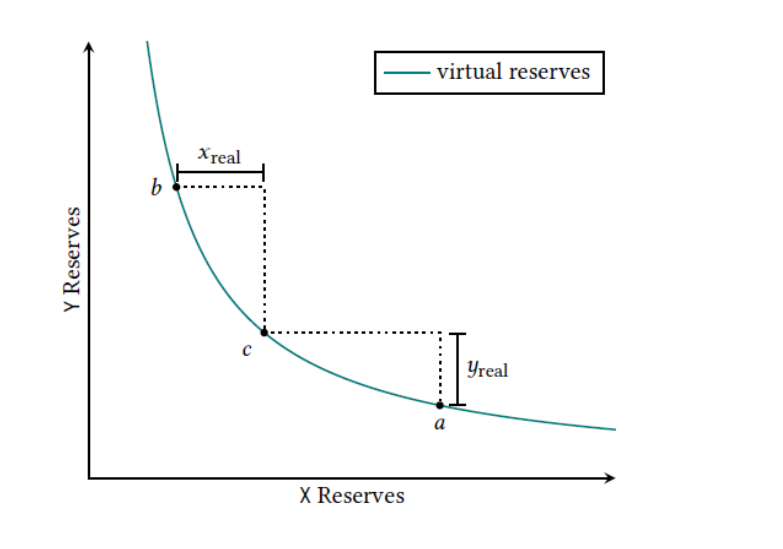} 
    \caption{Dynamics of Price On Uniswap v3}
    \label{fig:mi_imagen}
\end{figure}

\begin{itemize}
    \item \textbf{Tick \(i = 0\)}: 
    \[
    p_0 = 1.0001^0 = 1.0000
    \]
    \item \textbf{Tick \(i = 10\)}: 
    \[
    p_{10} = 1.0001^{10} \approx 1.0010
    \]
    \item \textbf{Tick \(i = 100\)}: 
    \[
    p_{100} = 1.0001^{100} \approx 1.0100
    \]
\end{itemize}

Not all ticks can be initialized but only those divisible by a pre-specified pool parameter, the tick spacing \(s\). For example, the USDC/ETH 0.3\% pool has \(s = 60\). Therefore, only ticks that are divisible by 60 can be initialized, i.e., 

\[
\{ \ldots, -120, -60, 0, 60, 120, \ldots \}
\]

A tick range can then be defined as \([i, i+s]\).

\subsection{Price Range Example}

Suppose an LP wants to provide liquidity in the price range from \(p_a\) to \(p_b\):

\begin{itemize}
    \item \textbf{Lower Bound \(p_a\)} (Tick \(i = 1000\)):
    \[
    p_{1000} = 1.0001^{1000} \approx 1.1052
    \]
    \item \textbf{Upper Bound \(p_b\)} (Tick \(i = 2000\)):
    \[
    p_{2000} = 1.0001^{2000} \approx 1.2214
    \]
\end{itemize}

LPs provide liquidity only when the price is between 1.1052 and 1.2214 USDC/ETH, concentrating their liquidity efficiently within this range.

\section{Token Association with Price Ranges}

Let \(p_i\) denote the price of token X (in units of token Y) that corresponds to tick \(i\). From the Uniswap v3 whitepaper (Adams et al., 2021), we obtain the following relations for \(x_i\), the quantity of tokens X locked in the tick range \([i, i+s]\), and \(y_i\), the quantity of tokens Y locked in the same tick range:

\[
x_i = \frac{L_i}{\sqrt{z_i}} - \frac{L_i}{\sqrt{p_{i+s}}}
\]

\[
y_i = L_i \cdot (\sqrt{z_i} - \sqrt{p_i})
\]

where

\[
z_i =
\begin{cases} 
p_i & \text{if } p_{\text{mkt}} \leq p_i \\
p_{\text{mkt}} & \text{if } p_i < p_{\text{mkt}} < p_{i+s} \\
p_{i+s} & \text{if } p_{i+s} \leq p_{\text{mkt}}
\end{cases}
\]

\subsection{Intuitive Reasoning Behind the Formula}

\begin{itemize}
    \item \textbf{\( p_{\text{mkt}} \leq p_i \)}: The market price is below or at the lower bound of the tick range. All the liquidity is held in token Y, because the market price is too low for any of token X to be in the pool.
    \item \textbf{\( p_i < p_{\text{mkt}} < p_{i+s} \)}: The market price is within the tick range. Both tokens X and Y are present in the liquidity pool, adjusted according to the current market price.
    \item \textbf{\( p_{i+s} \leq p_{\text{mkt}} \)}: The market price is above or at the upper bound of the tick range. All the liquidity is held in token X, because the market price is too high for any of token Y to be in the pool.
\end{itemize}

We can clearly see the difference with Uniswap v2: there, we could only add liquidity in 50-50 proportions.
\subsection{Summary in Intuitive Terms}

A token associated with a price range means that within this price range, the liquidity pool holds and can use that token for trading.

This mechanism ensures that liquidity is efficiently allocated and concentrated around the current market price, optimizing trading and reducing slippage within the specified range.

\begin{figure}[h]
    \centering
    \includegraphics[width=\textwidth]{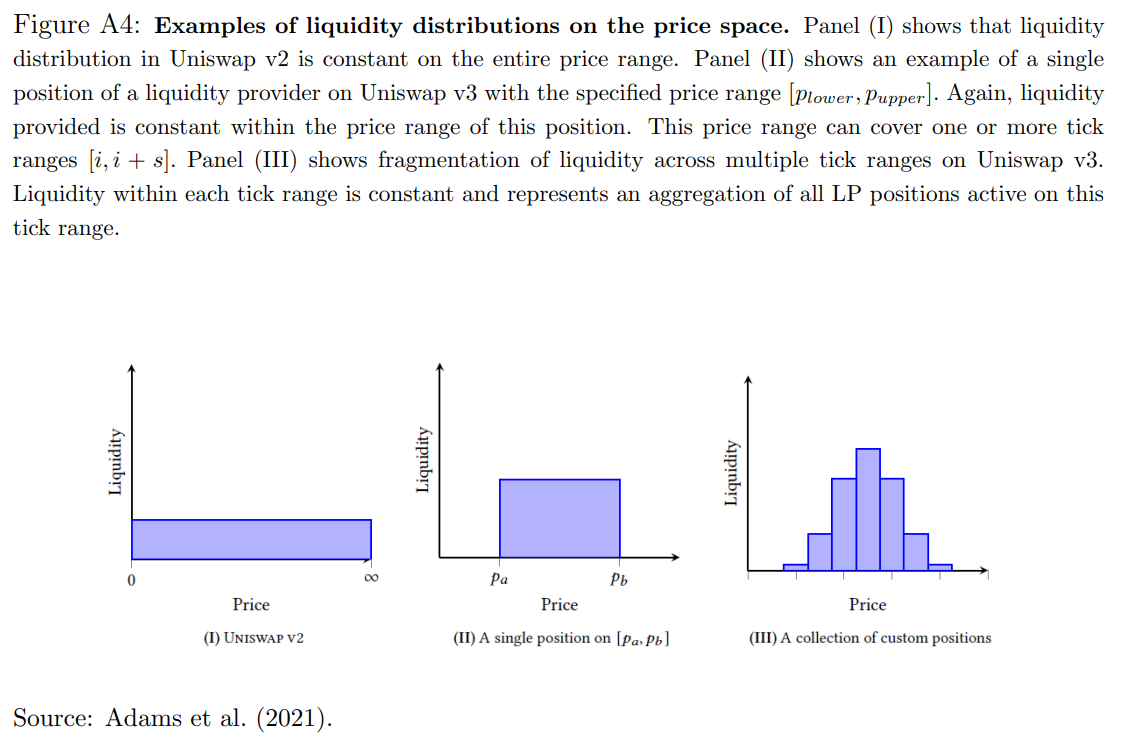} 
    \caption{Comparison of Distributions}
    \label{fig:mi_imagen}
\end{figure}

\section{Numerical example of adding liquidity with a mint in Uniswap v3}

To illustrate this, suppose that an LP would like to add liquidity to USDC/ETH 0.3\% pool (with a \(s\) of 60). Assume the market tick is 200618, such that the market price is:

\[
p_{\text{mkt}} = 1.0001^{200618}
\]

To convert prices to a human-readable format, we have to scale \(p_{\text{mkt}}\) by \(10^{(d_y - d_x)}\), where \(d_y\) is the number of decimals for token Y, and \(d_x\) is the number of decimals for token X. In our example, \(d_y = 18\) for ETH and \(d_x = 6\) for USDC:

\[
p_{\text{mkt, adj}} = \frac{1.0001^{200618}}{10^{(18-6)}} = 0.00051558 \text{ ETH per USDC}
\]

which corresponds to \(1/0.00051558 = 1939.56\) USDC per ETH.

Suppose an LP would like to add 50 ETH (token Y) to the price range \([p_{\text{lower}}, p_{\text{upper}}]\) that corresponds to tick range \([i_{\text{lower}}, i_{\text{upper}}] = [200520, 200640]\), illustrated as Mint 1 on Panel A of Figure A5. Note that this range includes two corresponding elementary tick ranges: \([200520, 200580]\) and \([200580, 200640]\). How many USDC (token X) does the LP have to add to complete their liquidity position?

To compute the required quantity of tokens X, we first infer the liquidity \(L_{\text{pos}}\) of this position from Equation (8):

\[
L_{\text{pos}} = \frac{y_{\text{pos}}}{\sqrt{p_{\text{mkt}}} - \sqrt{p_{\text{lower}}}} = \frac{50 \cdot 10^{18}}{\sqrt{1.0001^{200618}} - \sqrt{1.0001^{200520}}} = 4.505 \cdot 10^{17}
\]

with \(z_p = p_{\text{mkt}}\), because the position includes the market tick, \(p_{\text{lower}} < p_{\text{mkt}} < p_{\text{upper}}\). Note again to convert to "human readable" format, we have to scale \(x_{\text{pos}}\) by \(10^{4}\), i.e., \(x_{\text{pos, adj}} = x_{\text{pos}} / 10^{6} = 21812\) USDC.

Suppose there are two additional mints to the pool, as illustrated on Panel A of Table 1. Table 2 below summarizes all three liquidity positions in the pool:

\begin{table}[h]
\centering
\caption{Adding liquidity to USDC/ETH 0.3\% pool}
\begin{tabular}{ccccc}
\hline
\textbf{\(i_{\text{lower}}\)} & \textbf{\(i_{\text{upper}}\)} & \textbf{\(x_{\text{pos}}\)} & \textbf{\(y_{\text{pos}}\)} & \textbf{\(L_{\text{pos}}\)} \\
\hline
200520 & 200640 & 21812 & 50 & 4.505e17 \\
200580 & 200640 & 44934 & 100 & 9.281e17 \\
200580 & 200700 & 250848 & 60 & 1.392e18 \\
\hline
\end{tabular}
\end{table}

To compute aggregated liquidity on each tick range, \(L_i\), we have to add up liquidity of all active liquidity positions, \(L_{\text{pos}}\), on a given tick range. Panel B of Figure A5 shows distribution of aggregated liquidity by tick range. For example, there is only one active liquidity position on tick range \([200520, 200580]\). Therefore, \(L_i = L_{\text{pos}} = 4.505e17\). On the next tick range, \([200580, 200640]\), all three liquidity positions are active, such that \(L_i = \sum_{j=1}^{3} L_{\text{pos, j}} = 2.77e18\). From Equations (7) and (8), we can then compute aggregated \(x_i\) and \(y_i\) liquidity reserves for each tick range \([i, i+s]\). Table 3 below summarizes distribution of liquidity and reserves by tick range:

\begin{table}[h]
\centering
\caption{Liquidity distribution by tick range}
\begin{tabular}{cccc}
\hline
\textbf{Tick Range} & \textbf{\(L_i\)} & \textbf{\(x_i\) (USDC)} & \textbf{\(y_i\) (ETH)} \\
\hline
200520, 200580 & 4.505e17 & 0 & 30.58 \\
200580, 200640 & 2.77e18 & 134159 & 119.42 \\
200640, 200700 & 1.392e18 & 183427 & 0 \\
\hline
\end{tabular}
\end{table}

Note that tick ranges that do not include the market tick, \(i_{\text{mkt}} = 200618\), have only reserves in one of the tokens, either X (if \(i > i_{\text{mkt}}\)), or Y (if \(i + s < i_{\text{mkt}}\)).

\section{Processing the Data: The Importance of Liquidity Snapshots and How to Build Them}

To understand and analyze the state of these liquidity pools over time, it is essential to create and update liquidity snapshots. These snapshots provide a clear and concise representation of the pool's status at any given point, including the amounts of tokens held and the total liquidity. This information is vital for liquidity providers, traders, and researchers to evaluate the performance and dynamics of the pool.

This section outlines the process of building and updating liquidity snapshots. It begins with the initial liquidity snapshot, which establishes the baseline state of the pool. We then explore how to update the snapshot following various events, such as liquidity provision, liquidity burn, and trade events. By systematically capturing these changes, we can maintain an accurate and up-to-date view of the pool's liquidity.
\subsection{Initial liquidity snapshot}
\begin{figure}[h]
    \centering
    \includegraphics[width=1\textwidth]{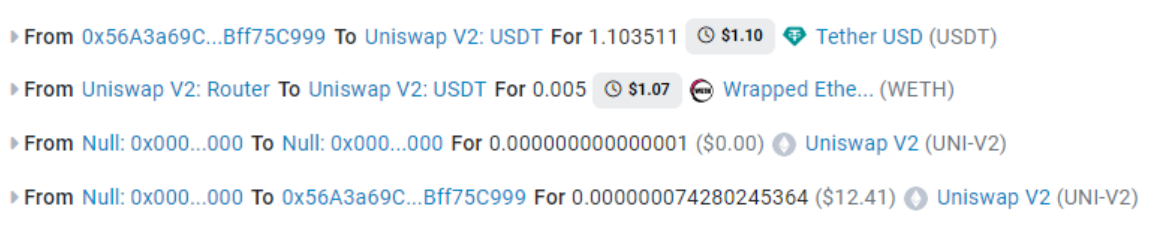} %
    \caption{Example of Initial Liquidity Providing Unsiwap}
    \label{fig:mi_imagen}
\end{figure}
We can construct an initial snapshot with a real-world example. We work with the following notation: $x_t$ and $y_t$ are the amount of tokens the liquidity pool has at time $t$, and $L_t$ the amount of liquidity tokens.

\[
x_0 = 0.5 \times 10^6
\]
\[
y_0 = 1.103511 \times 10^6
\]

The total liquidity tokens $L_t$ can be calculated using the constant product formula, which ensures that the product of the quantities of the two assets remains constant. The formula is given by:

\begin{equation}
\begin{aligned}
L_t &= \text{Total liquidity tokens} = \sqrt{k} \\
    &= \sqrt{(0.5 \times 10^6) \times (1.103511 \times 10^6)} = 74280246364
\end{aligned}
\end{equation}

We can see in Figure 3 that there are also 1000 UNI-v2 tokens that are being burned. In the end, liquidity providers get
Total - burnt = 74280246364 - 1000 = 74280245364 liquidity UNI-v2 tokens.

\subsection{How to build the liquidity snapshot at \( t + 1 \)}

Assume that we have the liquidity snapshot \( t \). That is, we have \( x_t \), \( y_t \), and \( L_t \). How can we build the liquidity snapshot at \( t + 1 \), after a liquidity provision or trade event has happened? Depends on the event we encounter.

\subsubsection{Liquidity provision event}
In this case, a new participant wants to add amounts \( ax_t \) and \( ay_t \) to the liquidity pool, in order to receive \( l_t \) liquidity tokens. In this case:

\[
x_{t+1} = x_t + ax_t
\]
\[
y_{t+1} = y_t + ay_t
\]
\[
l_t = \min \left( \frac{ax_t \cdot L_t}{x_t}, \frac{ay_t \cdot L_t}{y_t} \right)
\]
\[
L_{t+1} = L_t + l_t \quad (2)
\]
\newpage
\subsubsection{Liquidity burn event}
In this case, a participant wants to burn their \( l_t \) liquidity tokens, to receive amounts \( ax_t \) and \( ay_t \). In this case:

\[
ax_t = \frac{l_t \cdot x_t}{L_t}
\]
\[
ay_t = \frac{l_t \cdot y_t}{L_t}
\]
\[
x_{t+1} = x_t - ax_t
\]
\[
y_{t+1} = y_t - ay_t
\]
\[
L_{t+1} = L_t - l_t \quad (3)
\]

\subsubsection{Trade event}
In this case, the invariant has to be unchanged, where \( f \) are the fees (set to 0.003), and \( a_x \) and \( a_y \) are the inflow/outflow of tokens \( x \) and \( y \):

\[
(x_{t+1} - f \cdot a_x) \cdot (y_{t+1} - f \cdot a_y) \geq x_t \cdot y_t
\]

There are two options here, depending on what data we have:

1. First case: we have the total input and output of each token that comes in/out of the contract. That is, we know \( ax_t \) and \( ay_t \). This case simplifies the calculation, which becomes:

\[
x_{t+1} = x_t + ax_t
\]
\[
y_{t+1} = y_t + ay_t
\]
\[
L_{t+1} = L_t \quad (4)
\]

2. Second case: we don’t know one of the inflows/outflows. Let’s say we only know \( a_x \). This case is more difficult since we have to make an assumption: the trader does not donate tokens to the pool. In this case, the invariant becomes a strict equality:

\[
(x_{t+1} - f \cdot a_x) \cdot (y_{t+1} - f \cdot a_y) = x_t \cdot y_t
\]

\newpage

Therefore, we have:

\[
x_{t+1} = x_t + a_x
\]
\[
a_y = f^{-1} \left( y_{t+1} - \frac{x_t \cdot y_t}{(x_{t+1} - f \cdot a_x)} \right)
\]
\[
y_{t+1} = y_t + a_y
\]
\[
L_{t+1} = L_t \quad (5)
\]

\section{Conclusions}
In this chapter, we explored the theoretical framework behind price discovery in cryptocurrency markets, focusing on the comparison between centralized exchanges and decentralized platforms. We examined how Automated Market Makers (AMMs) like Uniswap operate, providing a detailed analysis of their mechanics, pricing models, and liquidity management. We have understood the differences between order book models and liquidity pool systems. This sets the foundation for further analysis of how price discovery functions in these contrasting environments.

\chapter{The methodologies used for the empirical analysis}
The process of price discovery is central to understanding how information is incorporated into asset prices. It refers to the mechanism through which the price of an asset is determined in the market, reflecting all available information. This document investigates the dynamics of price discovery across different markets, particularly following news events. We employ three main price discovery techniques: Hasbrouck's information share (1995)\cite{hasbrouck1995one}, Gonzalo and Granger's decomposition (1995)\cite{Gonzalo}, and lead-lag metrics discussed by Huth \cite{huth2014high}.

Price discovery has been extensively analyzed in the literature through various methodologies. Hasbrouck introduced his Information Share (IS) measure, which quantifies the contribution of each market to the efficient price. Gonzalo and Granger's proposed decomposition offers insights into the long-term price discovery process. Additionally, lead-lag relationships highlight the timing differences in the incorporation of information between markets.

Lead-lag analysis investigates the temporal relationship between price movements in different markets. A leading market reflects new information before others, while a lagging market adjusts its prices subsequently. This approach, however, faces challenges of model misspecification. Hasbrouck (1995) pointed out that improper model specifications could lead to incorrect inferences about the information shares of markets. Therefore, careful consideration of model assumptions is crucial for accurate lead-lag analysis.

Hasbrouck's IS measure evaluates the proportional contribution of each market's innovations to the variance of the efficient price. It provides a way to quantify the relative importance of different markets in the price discovery process. This measure has been widely applied in studies examining the role of various trading venues in reflecting new information.

The Gonzalo-Granger decomposition separates the permanent and transitory components of asset prices, focusing on the long-term equilibrium relationship. This methodology allows researchers to identify the underlying drivers of price changes and distinguish between persistent and temporary movements. It is particularly useful in analyzing cointegrated markets, where long-term relationships exist despite short-term deviations. We will provide precise definitions of the cointegration concept.

In this chapter, we provide an overview of the prior research relevant to the problem of price discovery and cointegration in financial markets. We begin with a general introduction to the key concepts, followed by a review of the primary methodologies that will be applied in the analysis: the Hasbrouck Information Share, the Gonzalo-Granger Permanent-Transitory decomposition, and the Hayashi-Yoshida methodology.

\section{Previous work in price discovery}

\subsection{Price discovery in financial markets}

Price discovery is the process by which the true value of an asset is determined as market participants gather and interpret news across different markets. Intermarket arbitrage ensures that the prices of the same asset in different markets do not drift apart significantly. Statistically, this is shown as cointegrated I(1) variables, meaning they share one or more common stochastic factors. The shared factor is often referred to as the "implicit efficient price," reflecting the asset's fundamental value. This price is driven by news and is common across all markets trading the asset.
There are two dominant approaches to analyzing price discovery:
\begin{itemize}
    \item Hasbrouck’s Information Shares (IS) Model (1995): It measures each market's contribution to the variance of the innovations in the common factor. Each market's contribution to this variance is referred to as its "information share."
    \item Gonzalo and Granger’s Permanent-Transitory (PT) Decomposition (1995): This approach focuses on the error correction process, dealing with permanent shocks that lead to disequilibrium due to different rates at which markets process news. It measures each market's contribution to the common factor based on the markets’ error correction coefficients.
\end{itemize}
Key differences between the models include how they account for contemporaneous correlation. The (IS) model, for instance, addresses this issue by using Cholesky factorization, which requires ordering the prices. This order dependency results in upper and lower bounds of information shares, with the mean of these bounds often used to interpret the results. In contrast, the (PT) model does not account for contemporaneous correlation, which may lead to potentially different outcomes. The IS and PT models tend to produce similar results when the residuals from the (VECM) are uncorrelated. However, when there is substantial contemporaneous correlation between the markets, the two models yield divergent results.

\section{Cointegration in a Vector Autoregression (VAR) model}
\subsection{Vector Autoregression Model(VAR)}
A Vector Autoregression (VAR) model is a statistical model used to capture the linear interdependencies among multiple time series. In a VAR model, each variable is modeled as a linear function of its own lagged values, as well as the lagged values of all other variables in the system.

For a vector of \( n \) time series variables \( X_t = \begin{pmatrix} x_{1t} \\ x_{2t} \\ \vdots \\ x_{nt} \end{pmatrix} \), a VAR(\( p \)) model can be written as:

\[
X_t = A_1 X_{t-1} + A_2 X_{t-2} + \dots + A_p X_{t-p} + u_t
\]

where:
\begin{itemize}
    \item \( A_i \) are \( n \times n \) coefficient matrices for \( i = 1, \dots, p \).
    \item \( u_t \) is an \( n \times 1 \) vector of white noise error terms, which are assumed to be normally distributed with zero mean and constant covariance matrix \( \Sigma_u \).
\end{itemize}

\subsection{Cointegration in a VAR Model}
Intuitively, cointegration refers to a situation where multiple time series are individually non-stationary, but a linear combination of them is stationary. In other words, even though the individual series may wander widely over time, there exists a long-term equilibrium relationship between them.

In the context of a VAR model, cointegration can be analyzed using the concept of a Vector Error Correction Model (VECM), which is a reparameterization of the VAR model that explicitly incorporates the cointegrating relationships.

\subsection{Vector Error Correction Model (VECM)}
The VECM representation for a VAR(\( p \)) model is given by:

\[
\Delta X_t = \Pi X_{t-1} + \sum_{i=1}^{p-1} \Gamma_i \Delta X_{t-i} + u_t
\]

where:
\begin{itemize}
    \item \( \Delta X_t = X_t - X_{t-1} \) represents the first difference of the series.
    \item \( \Pi = \sum_{i=1}^{p} A_i - I \) is an \( n \times n \) matrix that contains information about the long-run relationships between the variables.
    \item \( \Gamma_i \) are \( n \times n \) matrices capturing short-run dynamics.
    \item \( u_t \) is the vector of error terms.
\end{itemize}

The key component of the VECM that indicates cointegration is the matrix \( \Pi \). If the rank of \( \Pi \) is reduced (i.e., \( 0 < \text{rank}(\Pi) = r < n \)), then there are \( r \) cointegrating relationships among the \( n \) variables in \( X_t \).

\begin{itemize}
    \item If \( \Pi \) has full rank (i.e., \( r = n \)), all variables are stationary.
    \item If \( \Pi \) has zero rank (i.e., \( r = 0 \)), there are no cointegrating relationships, and the VAR model is in differences (a standard VAR model).
\end{itemize}

When \( 0 < r < n \), the matrix \( \Pi \) can be decomposed as:

\[
\Pi = \alpha \beta'
\]

where:
\begin{itemize}

    \item \( \beta \) is an \( n \times r \) matrix of cointegrating vectors, representing the long-term equilibrium relationships between the variables.
    
    \item \( \alpha \) is an \( n \times r \) matrix representing the adjustment coefficients, indicating how the variables adjust towards the long-term equilibrium.
    
\end{itemize}

As we will see, the methodologies we will study place special importance on the \( \alpha \) coefficient.

\subsection{Simple example: bivariate case}

For a bivariate system \( X_t = \begin{pmatrix} x_{1t} \\ x_{2t} \end{pmatrix} \), suppose \( r = 1 \) (i.e., there is one cointegrating relationship). Then:

\[
\Pi = \alpha \beta' = \begin{pmatrix} \alpha_1 \\ \alpha_2 \end{pmatrix} \begin{pmatrix} \beta_1 & \beta_2 \end{pmatrix}
\]

This implies:

\[
\Pi X_{t-1} = \alpha_1 (\beta_1 x_{1,t-1} + \beta_2 x_{2,t-1}) + \alpha_2 (\beta_1 x_{1,t-1} + \beta_2 x_{2,t-1})
\]

The expression \( \beta_1 x_{1,t-1} + \beta_2 x_{2,t-1} \) represents the cointegrating relationship (the long-term equilibrium), and \( \alpha_1 \) and \( \alpha_2 \) represent how each variable adjusts to deviations from this equilibrium.

\section{Cointegration in the Moving Average (MA) representation}
Better insight into the concept of cointegration can be grasped if we move to the MA, and express the process as function of innovations that drives it. 
\subsection{Moving Average representation(VMA)}
For a vector of \( n \) time series variables \( X_t = \begin{pmatrix} x_{1t} \\ x_{2t} \\ \vdots \\ x_{nt} \end{pmatrix} \), the Moving Average (MA) representation can be expressed as:

\[
X_t = C(L) u_t
\]

where:
\begin{itemize}
    \item \( C(L) \) is a matrix polynomial in the lag operator \( L \).
    \item \( u_t \) is an \( n \times 1 \) vector of white noise errors (innovations).
\end{itemize}

This can be rewritten to its integrated form as:

\[
X_t = \mu + C(1) \sum_{i=1}^{t} u_i + C^*(L) u_t
\]

where:
\begin{itemize}
    \item \( \mu \) is a deterministic term (possibly including a drift).
    \item \( C(1) \) is the long-run impact matrix, representing the cumulative effect of shocks on the levels of the series. It is the sum of the moving average coefficients.
    \item \( C^*(L) \) represents the transitory components, capturing the effects that dissipate over time. \( C^*(L) \) is a matrix polynomial in the lag operator L.
\end{itemize}
\subsection{Common stochastic trends}

In the MA representation, the term \( C(1) \sum_{i=1}^{t} u_i \) represents the common stochastic trends in the system. These trends are responsible for the long-term non-stationarity in the individual series.

If \( X_t \) is cointegrated, the rank of the matrix \( C(1) \) will be less than \( n \). Specifically, if there are \( r \) cointegrating relationships, the rank of \( C(1) \) will be \( n - r \). This implies that there are \( n - r \) common stochastic trends driving the non-stationarity in \( X_t \), and \( r \) linearly independent cointegrating vectors that form stationary combinations.

Thus, cointegration in the Moving Average representation indicates that there are fewer common stochastic trends than the number of series in the system, and that these series are bound together by long-term equilibrium relationships. We will later discuss the decomposition of the cointegrated process into permanent and transitory components, but first, we will briefly examine the implications of cointegrated processes in financial prices and Hasbrouck's methodology.

\section{Implications of cointegrated processes in financial prices}

Cointegration in financial prices is crucial when dealing with related assets like spot and futures contracts, stocks of related companies, or currency pairs. It implies a long-term equilibrium relationship, even though individual price series may show short-term non-stationary behavior.

This equilibrium relationship is maintained by economic forces such as arbitrage, which prevent prices from diverging indefinitely. The concept is closely tied to the Error Correction Mechanism (ECM), which explains how deviations from equilibrium are corrected over time.

For risk management, cointegration provides confidence in portfolio stability, as related assets are less likely to experience large price divergences. It also highlights the limitations of traditional lead-lag analysis, which can be misleading when assets are cointegrated.

Finally, cointegration has implications for price discovery and market efficiency, as it ensures that prices across different markets or instruments for the same asset reflect the same underlying information, maintaining market efficiency.

\section{Lead-lag analysis in the presence of cointegration:}

In the following, we demonstrate how lead-lag analysis in the presence of cointegration might lead to spurious conclusions. We illustrate this with a simple process.
Assume we have two cointegrated price series \( P_t \) and \( Q_t \), representing the prices of the same asset in two different markets. The Vector Error Correction Model (VECM) for these two series is given by:

\[
\Delta P_t = \alpha_1 (Q_{t-1} - P_{t-1}) + \epsilon_{1t} \tag{1a}
\]
\[
\Delta Q_t = \alpha_2 (Q_{t-1} - P_{t-1}) + \epsilon_{2t} \tag{1b}
\]

where:
\begin{itemize}
    \item \( \Delta P_t = P_t - P_{t-1} \) and \( \Delta Q_t = Q_t - Q_{t-1} \) are the changes in the prices.
    \item \( Q_{t-1} - P_{t-1} \) is the cointegrating relationship (i.e., the error correction term).
    \item \( \alpha_1 \) and \( \alpha_2 \) are the adjustment coefficients for \( P_t \) and \( Q_t \), respectively, which describe how each price adjusts to deviations from the long-term equilibrium.
    \item \( \epsilon_{1t} \) and \( \epsilon_{2t} \) are white noise error terms.
\end{itemize}
\subsection{Misspecification of models}
To understand how this leads to a misleading interpretation of lead-lag relationships, we apply recursive substitution to equation \( (1b) \), which describes \( \Delta Q_t \).

First, recall equation \( (1b) \):

\[
\Delta Q_t = \alpha_2 (Q_{t-1} - P_{t-1}) + \epsilon_{2t}
\]

Now, expand \( Q_{t-1} \) and \( P_{t-1} \) in terms of their lagged values:

\[
Q_{t-1} = Q_{t-2} + \Delta Q_{t-1}
\]
\[
P_{t-1} = P_{t-2} + \Delta P_{t-1}
\]

Substituting these into the error correction term \( (Q_{t-1} - P_{t-1}) \) gives:

\[
Q_{t-1} - P_{t-1} = (Q_{t-2} + \Delta Q_{t-1}) - (P_{t-2} + \Delta P_{t-1})
\]

This simplifies to:

\[
Q_{t-1} - P_{t-1} = (Q_{t-2} - P_{t-2}) + (\Delta Q_{t-1} - \Delta P_{t-1})
\]

Now substitute this back into the expression for \( \Delta Q_t \):

\[
\Delta Q_t = \alpha_2 [(Q_{t-2} - P_{t-2}) + (\Delta Q_{t-1} - \Delta P_{t-1})] + \epsilon_{2t}
\]

Expanding this:

\[
\Delta Q_t = \alpha_2 (Q_{t-2} - P_{t-2}) + \alpha_2 (\Delta Q_{t-1} - \Delta P_{t-1}) + \epsilon_{2t}
\]

This process can be continued by expanding \( Q_{t-2} - P_{t-2} \) in terms of \( Q_{t-3} \) and \( P_{t-3} \):

\[
Q_{t-2} - P_{t-2} = (Q_{t-3} + \Delta Q_{t-2}) - (P_{t-3} + \Delta P_{t-2})
\]

Substituting this back into our expression for \( \Delta Q_t \):

\[
\Delta Q_t = \alpha_2 [(Q_{t-3} - P_{t-3}) + (\Delta Q_{t-2} - \Delta P_{t-2})] + \alpha_2 (\Delta Q_{t-1} - \Delta P_{t-1}) + \epsilon_{2t}
\]

Simplifying:

\[
\Delta Q_t = \alpha_2 (Q_{t-3} - P_{t-3}) + \alpha_2 (\Delta Q_{t-2} - \Delta P_{t-2}) + \alpha_2 (\Delta Q_{t-1} - \Delta P_{t-1}) + \epsilon_{2t}
\]

This shows that lead-lag analysis is misleading. The recursive substitution shows that \( \Delta Q_t \) is influenced by past differences between \( Q_t \) and \( P_t \), as well as past changes in \( Q_t \) and \( P_t \). However, these influences are not just one-way (i.e., \( P_t \) leading \( Q_t \))—they are driven by the cointegrating relationship. In other words, \( \Delta Q_t \) is responding to the disequilibrium between \( Q_{t-1} \) and \( P_{t-1} \), and the adjustments involve terms that appear to "lag" or "lead" but are in fact part of a joint adjustment process to restore equilibrium.

This finite-lag model assumes convergent representations where none exist. We know that only lags up to order 2 are relevant in the true model, but in the current representation, many more lags appear. This can lead to spurious correlations among the coefficients, making some past lags seem more relevant than they actually are. Accurate modeling requires considering infinite lags, which are impractical for empirical analysis but essential for correct specification.

To summarize, addressing misspecification requires recognizing cointegration and using VECMs to incorporate both short-term and long-term relationships. This approach avoids misleading inferences and provides a more accurate understanding of price dynamics in multiple markets.
\subsection{Hasbrouck's framework}
Hasbrouck’s argument is that in the presence of cointegration, using autoregressive models to assess lead-lag relationships can be misleading. The recursive expansion of the VECM shows that the price changes are influenced by multiple past values of both series, reflecting their interdependence due to the cointegrating relationship. This undermines the simple interpretation of causality that lead-lag analysis might suggest. Instead, the changes in both series are better understood as part of a system where both are moving together to maintain a long-term equilibrium relationship.

Hasbrouck's analysis of price discovery in cointegrated markets focuses on understanding how much each market contributes to the formation of the efficient price, which is the price implied by the long-term equilibrium relationship.

\begin{itemize}
    \item \textbf{Information Share (IS)}: Hasbrouck's Information Share (IS) metric quantifies the contribution of each market to the variance of the innovations in the common factor (the efficient price).
    \item \textbf{Permanent-Transitory Decomposition}: The efficient price can be seen as the permanent component of the price process, while deviations from this price are transitory and tend to revert over time.
\end{itemize}
\section{Permanent-Transitory decomposition \mbox{using} the integrated Moving Average form}

Starting from the Vector Error Correction Model (VECM):

\[
\Delta X_t = \alpha \beta' X_{t-1} + \sum_{i=1}^{k} \Gamma_i \Delta X_{t-i} + u_t
\]

This VECM can be transformed into its Vector Moving Average (VMA) representation:

\[
\Delta X_t = C(L) u_t
\]

where \( C(L) \) is a matrix polynomial in the lag operator \( L \). By summing the moving average coefficients, the integrated form of this moving average representation (which accumulates the shocks) is given by:

\begin{equation}
X_t = C(1) \sum_{i=1}^{t} u_i + C^*(L) u_t
\end{equation}

Here:
\begin{itemize}
    \item \( C(1) \) is the long-run impact matrix, which captures the cumulative effect of shocks on the system.
    \item \( C^*(L) \) represents the transitory components, which have only a temporary impact on \( X_t \).
\end{itemize}

The equation can be interpreted as follows: The permanent component, \( W_t = C(1) \sum_{i=1}^{t} u_i \), represents the cumulative effect of past shocks on the system, which causes permanent (non-reverting) changes in the price series \( X_t \). This component corresponds to the long-term equilibrium that the system will converge to, and it is equivalent to the efficient price or common factor in Hasbrouck's framework. On the other hand, the transitory component, \( G_t = C^*(L) u_t \), represents the short-term deviations from the equilibrium that dissipate over time. These effects are temporary and do not contribute to the long-term level of the price series.

Now, Johansen \cite{Johansen91} proves that

\begin{equation}
C(1) = \beta_\bot \left( \alpha'_\bot \left( I - \sum_{i=1}^k \Gamma_i \right) \beta_\bot \right)^{-1} \alpha'_\bot.
\end{equation}

So equation 2.1 can be rewritten as 

\begin{equation}
X_t = \beta_\bot C \left( \sum_{i=1}^t u_i \right) + C^*(L)u_t,
\end{equation}

with

\begin{equation}
\psi = \left( \alpha'_\bot \left( I - \sum_{i=1}^k \Gamma_i \right) \beta_\bot \right)^{-1} \alpha'_\bot.
\end{equation}

The final step in calculating the information share (IS) involves removing the contemporaneous correlation in \(u_t\). This is achieved by generating a new set of errors:

\begin{equation}
u_t = Fe_t,
\end{equation}

where \( \Omega \) is the covariance matrix of the innovations \( e_t \) and \(F\) is the lower triangular matrix such that \(\Omega = FF'\). The proportion of the innovation variance attributed to \(e_j\) is calculated as:

\begin{equation}
IS_j = \frac{([\psi F]_j)^2}{\psi \Omega \psi'},
\end{equation}

where \([\psi F]_j\) represents the \(j\)th element of the row matrix \(\psi F\).
Denote

\[
F = \begin{pmatrix} 
f_{11} & 0 \\ 
f_{12} & f_{22} 
\end{pmatrix} 
= \begin{pmatrix} 
\sigma_1 & 0 \\ 
\rho \sigma_2 & \sigma_2 (1 - \rho^2)^{1/2} 
\end{pmatrix}.
\]

Simplifying, we obtain:

\[
IS_1 = \frac{(\gamma_1 f_{11} + \gamma_2 f_{12})^2}{(\gamma_1 f_{11} + \gamma_2 f_{12})^2 + (\gamma_2 f_{22})^2},
\]

\[IS_2= \frac{(\gamma_2 f_{22})^2}{(\gamma_1 f_{11} + \gamma_2 f_{12})^2 + (\gamma_2 f_{22})^2}.
\]

where \(\gamma_1\) and \(\gamma_2\) are components of \(\alpha_\bot\), and \(f_{ij}\) are elements of the matrix \(F\).

Intuitively, the information share measures the extent to which innovations in a particular market contribute to the variance of the efficient price. A higher information share indicates that the market plays a more significant role in incorporating new information into the security price.

As we can see, the information share for each market can then be computed using the elements of \( \Omega \) and \(\alpha\). In the original paper (1995) \cite{hasbrouck1995one}, Hasbrouck does not estimate the metric with the VECM model, but with its corresponding Vector Moving Average (VMA) representation. As Baillie et al. (2002) \cite{BAILLIE2002309} point out, it is easier to estimate the metric with the VECM representation, since in the end we only need the vector $\alpha_\bot$ and the matrix of error correlations.

A significant advantage of Hasbrouck's approach is that it does not require the direct estimation of the VMA, which can sometimes be challenging. All necessary calculations can be performed by estimating the VECM. With the vector \(\alpha_\bot\) and the matrix of correlation of errors, we are well-equipped to calculate the information share.

In summary, Hasbrouck's information share provides a metric for understanding the contributions of different markets to price discovery. By quantifying the proportion of the efficient price variance attributable to each market, this measure tries to identify where new information is most rapidly and accurately reflected in security prices.

\section{Permanent-Transitory decomposition: The Gonzalo and Granger methodology}
The permanent component \(W_t\), as presented by Gonzalo and Granger (1995) \cite{Gonzalo}, captures the long-term equilibrium price that is driven by the common stochastic trends in the system. This component is defined as:

\[
W_t = \alpha'_\perp X_t \tag{34}
\]

where:
\begin{itemize}
    \item \(X_t\) is the vector of prices, e.g., \(X_t = (P_{1t}, P_{2t}, \dots, P_{nt})'\).
    \item \(\alpha_\perp\) is the orthogonal complement of the adjustment coefficient vector \(\alpha\).
\end{itemize}
The PT decomposition splits the price vector \(X_t\) into its permanent and transitory components:

\[
X_t = A_1 \alpha'_\perp X_t + A_2 \beta' X_t \tag{35}
\]

where:
\begin{itemize}
    \item \(A_1 = \beta_\perp (\alpha'_\perp \beta_\perp)^{-1}\), which maps the permanent component \(W_t\) into \(X_t\), indicating how the permanent component influences the overall price vector. \( \beta_\perp \) is the orthogonal complement of the cointegrating vector \( \beta \).
     \item \(A_2 = \alpha (\beta' \alpha)^{-1}\), which is related to the transitory component. It captures the short-term deviations (transitory component) that influence the price vector \( X_t \).\footnote{Stock and Watson's model represents a time series \( Y_t \) as \( Y_t = f_t + G_t \), where \( f_t \) is the common trend representing the permanent component, and \( G_t \) is the transitory component. The Permanent Component \( W_t = \alpha'_\perp X_t \) in the PT decomposition is analogous to the common trend \( f_t \) in the Stock and Watson model, capturing the common stochastic trend driving long-term movements in \( X_t \). Similarly, the Transitory Component \( A_2 \beta' X_t \) in the PT decomposition corresponds to \( G_t \) in the Stock and Watson model, reflecting the short-term fluctuations expected to dissipate over time.}
\end{itemize}

Substituting, we get:

\[
X_t = \beta_\perp (\alpha'_\perp \beta_\perp)^{-1} \alpha'_\perp X_t + \alpha (\beta' \alpha)^{-1} \beta' X_t.
\]

Since \(\alpha'_\perp \alpha = 0\) and \(\beta'_\perp \beta = 0\), the terms simplify to:

\[
X_t = \beta_\perp \alpha'_\perp X_t + \alpha \beta' X_t.
\]

Thus, equation (35) holds true, as the permanent and transitory components combine to give \(X_t\).

We will place special emphasis on the meaning of $\alpha$ and $\beta$. $\beta$ is the cointegrating vector, and we can interpret it as the vector that models the long-term relationships between the time series, as it contains the coefficients of the linear stationary combinations. However, sometimes one series will deviate from the other due to noise. $\alpha$ is the error correction vector, and it indicates how quickly the series adjust to these deviations. A more formal treatment justifying this can be found on pages 39-42 in Johansen (1996) \cite{johansen1996}.

In general, particularizing to a bivariate case, the permanent component \( W_t \) is driven by the series that adjusts the least to deviations from the equilibrium. For instance, if $\alpha_2$ is much smaller than $\alpha_1$, then \( x_{2t} \) contributes more to the permanent component.

Tests can be performed to analyze the value of \(\alpha\) estimated in the VECM. This way, we can obtain the value for \(\alpha_\bot\) and then perform two tests:
\begin{itemize}
    \item One where the null hypothesis is \(H_0: (1,0)\) is \(\alpha_\bot\).
    \item One where the null hypothesis is \(H_0: (0,1)\) is \(\alpha_\bot\).
\end{itemize}
These tests can be performed directly on \(\alpha\) via a likelihood ratio test. The distribution of these likelihood ratio tests follows a chi-squared (\(\chi^2\)) distribution with one degree of freedom, as we are testing a single restriction on \(\alpha_\bot\).

\begin{figure}[h]
    \centering
    \includegraphics[width=0.6\textwidth]{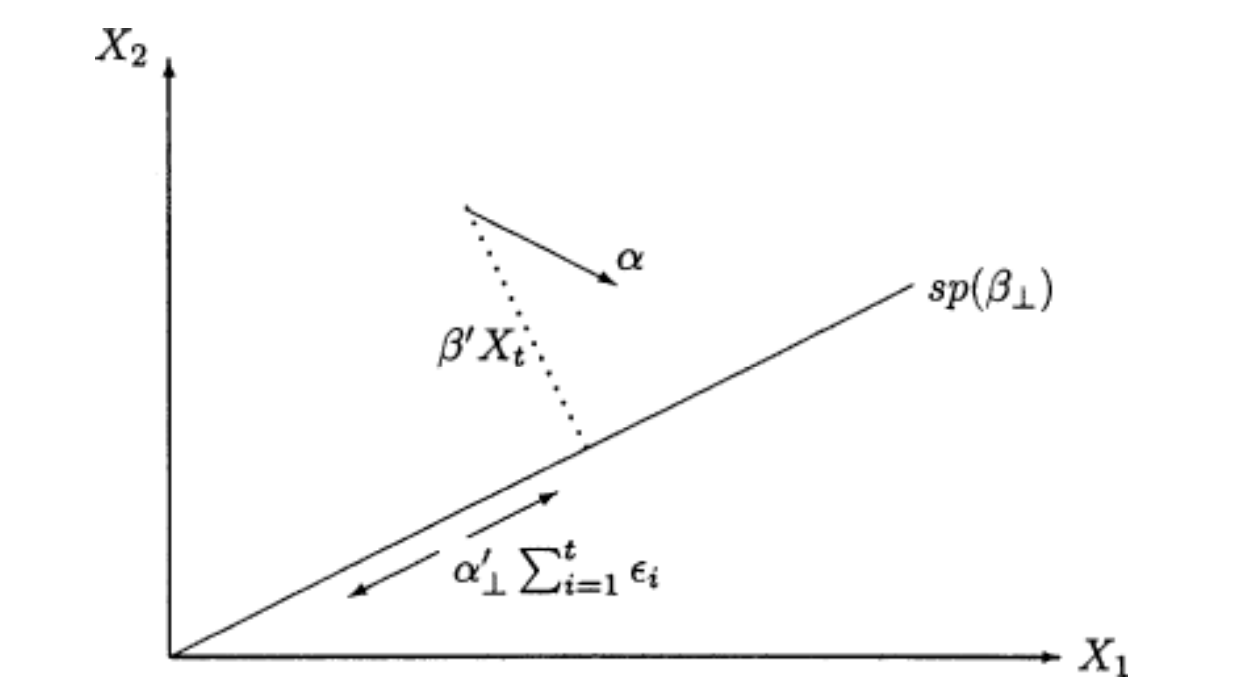}
    \caption{Illustration of the relationship between $\beta$ and $\alpha$ in a cointegrated system.}
\end{figure}

Figure 2.1 illustrates the dynamics of the cointegrated VECM system. The common trend moves the variables along the orthogonal $\beta$ space. Any error or disequilibrium in the cointegrated vector is corrected back to the equilibrium relationship by the $\alpha$ coefficients. The specific weight of each $\alpha$ determines which market is correcting the most. In a similar way, the orthogonal $\alpha$ reveals which variable represents the common stochastic trend that carries the most weight in the overall dynamics.

\section{Hayashi-Yoshida metric for analyzing lead-lag relationships}

A lead-lag relationship refers to the temporal dynamic between two instruments, where the price movements of one instrument (the leader) precede and potentially predict the movements of another instrument (the lagger). The liquidity of the market can play a significant role here.
The Hayashi-Yoshida metric and the extensions that we are going to discuss do not require the data to be synchronous, which makes it particularly interesting for high-frequency data.
The main resource used here has been the paper by Huth and Abergel \cite{huth2014high}. We have seen in previous sections that both the Hasbrouck and the Gonzalo and Granger methodologies require synchronous time series. This is restrictive, especially if we have more granular data like tick-to-tick (T2T) data, which is asynchronous for different markets.

Understanding lead-lag relationships is important for price discovery because they reveal the sequence in which information is incorporated into the prices of different instruments. If one instrument consistently leads another, it suggests that the leading instrument is quicker to reflect new information, thereby playing a more dominant role in the price discovery process.

\subsection{Definition of the metrics used}

In a formal setting, using stochastic calculus, we define the metrics for two Itô processes \(X, Y\) such that
\[
\begin{aligned}
dX_t &= \mu_t^X dt + \sigma_t^X dW_t^X, \\
dY_t &= \mu_t^Y dt + \sigma_t^Y dW_t^Y
\end{aligned}
\]
\[
d\langle W^X, W^Y \rangle_t = \rho_t dt,
\]
and independent observation times \(0 = t_0 \leq t_1 \leq \ldots \leq t_{n-1} \leq t_n = T\) for \(X\) and \(0 = s_0 \leq s_1 \leq \ldots \leq s_{m-1} \leq s_m = T\) for \(Y\), the original paper \cite{hayashiori} shows that  
\[
\sum_{i,j} r_i^X r_j^Y 1_{\{O_{ij} \neq \emptyset\}},
\]
where
\[
\begin{aligned}
O_{ij} &= ]t_{i-1}, t_i] \cap ]s_{j-1}, s_j], \\
r_i^X &= X_{t_i} - X_{t_{i-1}}, \\
r_j^Y &= Y_{s_j} - Y_{s_{j-1}},
\end{aligned}
\]
is an unbiased and consistent estimator of \(\int_0^T \sigma_t^X \sigma_t^Y \rho_t dt\) as the observation intervals become finer. In practice, this involves summing each product of increments whenever they share any time overlap. For constant volatilities and correlation, this method yields a consistent estimator for the correlation
\[
\hat{\rho} = \frac{\sum_{i,j} r_i^X r_j^Y 1_{\{O_{ij} \neq \emptyset\}}}{\sqrt{\sum_i (r_i^X)^2 \sum_j (r_j^Y)^2}}.
\]

It is possible and more useful to consider a lagged version for the correlation

\[
\hat{\rho}(\ell) = \frac{\sum_{i,j} r_i^X r_j^Y 1_{\{O_{ij}^\ell \neq \emptyset\}}}{\sqrt{\sum_i (r_i^X)^2 \sum_j (r_j^Y)^2}}
\]

where 
\begin{align}
O_{ij}^\ell &= ]t_{i-1}, t_i] \cap ]s_{j-1} - \ell, s_j - \ell]
\end{align}
This is the key statistic from which the derived metrics, such as the Lead-Lag Ratio (LLR) and the maximum lag, will be calculated.

The approach is related to Granger causality, but it offers additional insights. While Granger causality indicates which asset is leading the other in a given pair, these metrics go further by considering the strength and characteristic timing of the lead/lag relationship. In this context, the maximum level of the cross-correlation function and the specific lag at which it occurs are crucial factors to take into account.

Specifically, the Hayashi-Yoshida and the Lead-Lag Ratio (LLR), both measured with observation times of trades without conditioning by types of movements, provide key insights into the lead-lag dynamics:

 \begin{itemize}
        \item \textbf{HY\_lead\_lag}: The lag (in milliseconds) that maximizes the absolute value of the lead-lag estimator, where the estimator is given by:
\begin{align}
    HY\_lead\_lag:=\arg\max_{\ell}|\hat{\rho}(\ell)|
\end{align}

        \item \textbf{LLR}: The formula is 
        \[
\text{LLR} := \frac{\sum_{i=1}^{p} \rho^2(\ell_i)}{\sum_{i=1}^{p} \rho^2(-\ell_i)}
\]

where index \(i\) refers to a discrete grid of lags we are arbitrarily choosing.
    \end{itemize}
The LLR output is always positive: \(X\) leads \(Y\) if \(LLR > 1\), and \(Y\) leads \(X\) if \(LLR < 1\).

The LLR compares the squared correlations at positive lags (where \(X\) leads \(Y\)) with those at negative lags (where \(Y\) leads \(X\)). Positive lags capture how much earlier price changes in \(X\) predict price changes in \(Y\), while negative lags capture the reverse. The ratio summarizes this lead-lag behavior, with values greater than 1 indicating that \(X\) consistently leads \(Y\), and values less than 1 indicating that \(Y\) leads \(X\).

It’s important to note that LLR alone does not capture the timing or strength of the lead-lag relationship. To fully understand these dynamics, the maximum cross-correlation value and the lag at which it occurs must also be considered. In practice, the LLR can be noisier, especially with high-frequency data, but it still provides valuable information about the directional relationship between assets.

\subsection*{Relationships between the various statistics}
In this section we present a summary of the statistics used and the key properties of each.
\begin{displaymath}
\begin{array}{|c|c|c|c|}
\hline
 Object   & Formula   & Input     & Output\\ \hline
\text{HY function} & 
\hat{\rho} = \frac{\sum_{i,j} r_i^X r_j^Y 1_{\{O_{ij} \neq \emptyset\}}}{\sqrt{\sum_i (r_i^X)^2 \sum_j (r_j^Y)^2}}
 & X,Y & \text{correlation} \\ \hline
\text{HY lagged function} & \hat{\rho}(\ell) = \frac{\sum_{i,j} r_i^X r_j^Y 1_{\{O_{ij}^\ell \neq \emptyset\}}}{\sqrt{\sum_i (r_i^X)^2 \sum_j (r_j^Y)^2}} & X,Y,\ell & \text{correlation of $X$,${Ylag}$} \\ \hline
\text{HY lead/lag time} &\arg\max_{\ell}|\hat{\rho}(\ell)| & X,Y     & \text{$\rho$-maximising lag time}\\ \hline
LLR & \frac{\sum_{i=1}^{p} \rho^2(\ell_i)}{\sum_{i=1}^{p} \rho^2(-\ell_i)} &   \rho, X,Y,\{\ell_1,...,\ell_p\} & \text{lead/lag indicator} \\ \hline
\end{array}
\end{displaymath}
\begin{itemize}
    \item The two time-series $ T_X,T_Y$ are not necessarily synchronous. The HY function is a way of measuring correlation between two asynchronous time series, and it can be generalised to the lagged HY metric. This lagged metric can be considered as shifting the asset Y and then applying HY to the shifted intervals.
    \item Intuitively, the most heavily traded assets tend to incorporate information into prices faster than others so they should lead. In general:
    \begin{center}
        Asset X more liquid $\implies$ Asset X leads$\implies$ $LLR>1$
    \end{center}
        However, this rule does not always hold and should be carefully verified in each specific case.

\end{itemize}
\subsection{Limitations of trade time}

In the formal definition, we stated that \( C \) and \( Y \) are two Itô processes with independent observation times, but we have not yet specified what these times represent in practice. There are two logical alternatives:
\begin{itemize}
    \item \textbf{Trade time}: consider a new time point whenever there is a new trade.
    \item \textbf{Tick time}: consider a new time point whenever there is a variation in the midquote.
\end{itemize}

These do not necessarily coincide: there can be variations in the midquote without trades actually occurring, and trades can happen without the midquote changing.
\newpage

Therefore, we can consider two different time indices for the analysis: one that changes every time there is a trade (we call this trade time) and one that changes every time there is a non-null variation in the midquote (we call this tick time).

\begin{figure}[h]
\centering
\includegraphics[width=0.9\textwidth]{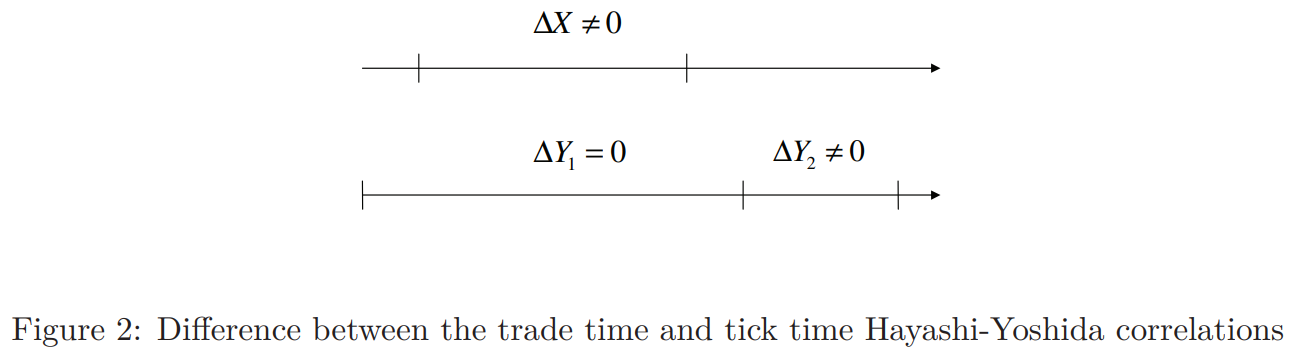}
\caption{A possible visualization of the differences between trade time and tick time \cite{huth2014high}}
\label{fig:trade_tick_time}
\end{figure}

We first construct an idealized example where the correlation \(\hat{\rho}(\ell)\) for an asset \( Y \) differs when measured with tick time versus trade time. The example data is shown in Table 1, where the returns are already computed. On the right, we compute the returns only every time a trade occurs. If a trade does not occur, the return is zero. If a trade occurs, it accumulates the virtual returns of the contemporaneous time and all the virtual returns that occurred when there were no trades.

We can identify the virtual returns with tick time, that is, those are the returns measured with all the midquotes whenever there is a change in the bid or the ask.

As both columns are differently, it is clear that if we consider the HY metric that we introduce later in this article, we will not get the same results. But we are going to give a concrete example to see it more clearly.
\subsection*{Asset X}
\begin{table}[h!]
\centering
\begin{tabular}{|c|c|c|}
\hline
\textbf{Period} & \textbf{\( r_{it} \) (Tick Returns)} & \textbf{\( r_{it}^0 \) (Trade Returns)} \\
\hline
1 & 0.01 & 0.01 \\
2 & 0.02 & 0.02 \\
3 & 0.03 & 0 \\
4 & 0.04 & 0 \\
5 & 0.05 & 0.12 \\
\hline
\end{tabular}
\caption{Tick Returns (\( r_{it} \)) and Trade Returns (\( r_{it}^0 \)) for Asset X}
\label{tab:returns_X}
\end{table}
\newpage
\subsection*{Asset Y}
\begin{table}[h!]
\centering
\begin{tabular}{|c|c|c|}
\hline
\textbf{Period} & \textbf{\( r_{jt} \) (Tick Returns)} & \textbf{\( r_{jt}^0 \) (Trade Returns)} \\
\hline
1 & 0.02 & 0 \\
2 & -0.01 & -0.01 \\
3 & 0.01 & 0.01 \\
4 & -0.02 & 0 \\
5 & 0.03 & 0.01 \\
\hline
\end{tabular}
\caption{Tick Returns (\( r_{jt} \)) and Trade Returns (\( r_{jt}^0 \)) for Asset Y}
\label{tab:returns_Y}
\end{table}
\subsection*{Hayashi-Yoshida correlation calculations}
For simplicity, in this case we consider the both assets to be synchronous.
\subsubsection*{With Tick Time (Virtual Returns)}
\small
\[
\begin{aligned}
\hat{\rho}_{\text{tick}} &= \frac{\sum_{i,j} r_i^X r_j^Y 1_{\{O_{ij} \neq \emptyset\}}}{\sqrt{\sum_i (r_i^X)^2 \sum_j (r_j^Y)^2}} \\
&= \frac{(0.01 \cdot 0.02) + (0.02 \cdot -0.01) + (0.03 \cdot 0.01) + (0.04 \cdot -0.02) + (0.05 \cdot 0.03)}{\sqrt{(0.01^2 + 0.02^2 + 0.03^2 + 0.04^2 + 0.05^2) \cdot (0.02^2 + (-0.01)^2 + 0.01^2 + (-0.02)^2 + 0.03^2)}} \\
&= \frac{0.001}{\sqrt{0.0055 \cdot 0.0019}}= 0.309
\end{aligned}
\]
\normalsize

\subsubsection*{With Trade Time (Observed Returns)}

\[
\begin{aligned}
\hat{\rho}_{\text{trade}} &= \frac{\sum_{i,j} r_i^X r_j^Y 1_{\{O_{ij} \neq \emptyset\}}}{\sqrt{\sum_i (r_i^X)^2 \sum_j (r_j^Y)^2}} \\
&= \frac{(0.01 \cdot 0) + (0.02 \cdot -0.01) + (0 \cdot 0.01) + (0 \cdot 0) + (0.12 \cdot 0.01)}{\sqrt{(0.01^2 + 0.02^2 + 0^2 + 0^2 + 0.12^2) \cdot (0 + 0.0001 + 0.0001 + 0 + 0.0001)}} \\
&= \frac{0 - 0.0002 + 0 + 0 + 0.0012}{\sqrt{0.0149 \cdot 0.0003}}= \frac{0.001}{\sqrt{0.00000447}}= 0.474
\end{aligned}
\]
We can observe that the correlation measured in trade time is bigger and introduces some spurious correlation.

\subsubsection{Challenges associated with trade time and summary} Intuitively, it is evident that tick time (understood as the clock that increments each time there is a non-zero variation in the midquote between two trades) should be more precise than trade time. For a more detailed discussion of this intuition, refer to pages 84 through 98 of \textit{The Econometrics of Financial Markets} by Campbell, Lo, and MacKinlay \cite{Lo}, which provides a rigorous examination of these issues in a more abstract setting.

In summary, we will analyze both the Lead-Lag Ratio (LLR) and the maximum lag using datasets indexed by tick time, as this provides a more precise measure of market activity. This approach allows for a clearer understanding of the lead-lag dynamics between assets, avoiding potential distortions associated with trade time.

\chapter{Empirical analysis}
In the previous chapters, we explored several methodologies and techniques for studying price discovery across different markets. These included the Hasbrouck methodology, the Gonzalo-Granger methodology, and the Hayashi-Yoshida methodology. In this chapter, we apply these methodologies to analyze price discovery in cryptocurrency markets.

We will focus on two compelling case studies: the first compares spot ETH on a centralized exchange (Binance) with ETH in a decentralized liquidity pool on Uniswap v2. We directly incorporate on-chain data from Uniswap, capturing real-time transactions, liquidity events, and price movements within the pool to evaluate how decentralized market dynamics compare to centralized markets. The second case examines the relationship between BTC futures prices on CME and its spot price on Binance.
In the first case, our goal is to understand the connections between traditional centralized finance and the emerging decentralized finance (DeFi) ecosystem, and to determine whether information flows more quickly to one market than the other. In the second case, we aim to assess a similar relationship, this time between a spot cryptocurrency and its futures counterpart. This analysis will provide valuable insights into which market incorporates the most information during key events.

The structure of the analysis will be as follows: first, we will analyze the series over an extended period, with the aim of understanding the long-term relationships between the assets. Then, we will focus on five specific dates characterized by special conditions, such as significant price drops, large gains, or notable events. This will help us determine whether the dynamics observed in the long-term analysis are still present during these particular events.

To accurately assess the performance of decentralized AMMs in comparison to centralized exchanges, it is essential to establish a set of benchmarks. Rather than relying on a single benchmark, which could introduce noise into the results, we will analyze various metrics explained in the first part of this document. These multiple benchmarks will provide a robust framework for evaluating the efficiency, accuracy, and liquidity behavior of AMMs. By comparing the modified AMMs to off-chain data, such as from Binance, we can objectively measure their ability to reflect market prices and discover new information.

For each specific date, we will conduct two distinct analyses: the first will focus on centralized versus decentralized markets, while the second will compare the spot market with the futures market. In both cases, we will focus on specific 15-minute and 20-minute time frames during the day, particularly during periods of heightened volatility and price movements. This allows us to determine which market incorporates information more quickly and effectively during critical events.

We will analyze the same five sample dates for both cases, each occurring in a different month, to examine the underlying dynamics under varying conditions.

\section{Empirical Analysis of ETH Time Series on Binance vs Uniswap v2}
For the centralized market data, we sourced it from Binance \cite{binance}, one of the largest and most liquid cryptocurrency exchanges globally. The data includes detailed price and volume information, recorded at one-second intervals, which allows for high-frequency analysis of market activity.\footnote{The data has been processed into a database for analysis, and an example of the data retrieval process can be provided upon request.}

For decentralized data, we have used Etherscan. In this case, we retrieved all transactions from the API that occurred in the relevant blocks since May 19, 2020, when the Uniswap v2 pool was created. This data includes every interaction with the liquidity pool, such as swaps, liquidity additions, and removals, providing a comprehensive view of the decentralized trading activity.

As mentioned earlier, we will begin with a long-term analysis, covering the period from September 2023 to 2024, to understand the relationships between both markets over time. Following that, we will conduct the analysis for five specific recent events. All of these events took place in 2024, with each date selected from different months, starting in March, to ensure the broadest sample possible. Below, we briefly explain the reasons for selecting each date.

\textbf{March 5th} saw a significant increase in trading volume in both centralized and decentralized markets. This period was highly volatile due to uncertainty regarding the SEC’s potential approval of Ethereum ETFs. Gas prices were also at their highest for 2024.

\textbf{April 30th} was marked by a sharp decline in cryptocurrency prices alongside rising gas prices. Ethereum ETFs were launched in Hong Kong, but they experienced losses, contributing to downward pressure on prices.

\textbf{May 20th} was characterized by a significant rise in ETH prices after reports suggested that the approval of Ethereum ETFs had suddenly become much more likely.

\textbf{August 5th} witnessed a major decline in equities markets, which also affected the cryptocurrency ecosystem.

Finally, \textbf{September 6th} was similar to August, with large outflows in equity markets that had a ripple effect on cryptocurrencies.

For each of these days, we will focus on specific 15-minute and 20-minute time frames, which we believe are particularly interesting because they include significant price movements and potentially more price-relevant information.

To provide a clear understanding of market conditions during the specific events, we will first  present several key descriptive statistics for the specific events we will analyze. The \textbf{percent price change during the event (\%)} indicates the percentage change in Bitcoin's price during the specific event window, where a negative value represents a price decline and a positive value represents an increase. The \textbf{total daily volume} refers to the total amount traded on that day, capturing the overall market activity for the full trading day and providing context for the volume during the specific event. The \textbf{daily maximum price (USD)} reflects the highest price reached by Bitcoin over the course of the trading day, offering insight into the price peak, while the \textbf{daily minimum price (USD)} represents the lowest price reached during the trading day, helping to understand the full extent of price volatility. These descriptive statistics offer a quantitative view of market behavior, helping us understand the intensity of trading activity, the magnitude of price changes, and overall volatility during key events.

In the following analysis, one of the main goals is evaluating AMMs as oracles by comparing them with off-chain data sources. This comparison will help assess the reliability and effectiveness of AMMs in providing accurate price information relative to external, centralized platforms.

In the following analysis, one of the primary objectives is to evaluate AMMs as oracles by comparing their performance with off-chain data sources. This comparison aims to determine the extent to which AMMs can serve as reliable and autonomous providers of real-time market data, similar to traditional oracles. By analyzing the consistency and accuracy of on-chain price discovery relative to external data from centralized exchanges, like Binance, we can assess the effectiveness of AMMs in reflecting true market conditions and identifying potential inefficiencies or arbitrage opportunities.

\subsection{Long-term relationships analysis: one-year interval, analysis from September 2023 to September 2024}

We will conduct a yearly analysis to understand the long-term relationships between the decentralized and centralized markets. They are both driven by a common stochastic trend. For this exercise, we will use the two time series representing both markets and estimate a Vector Error Correction Model (VECM) of order 1, meaning we include one lag. The two variables, decentralized and centralized markets, are both non-stationary, which justifies the use of the VECM framework.

After estimating the model, we will apply the following three methodologies to analyze the price discovery process: Hasbrouck’s Information Share, Gonzalo-Granger’s Permanent-Transitory Decomposition, and the Hayashi-Yoshida method.

We begin by presenting the estimated coefficients for the VECM, using one lag and considering two dimensions: one for the centralized market and one for the decentralized market. The variables are non-stationary, where \(X_1\) denotes the price in the centralized market, and \(X_2\) denotes the price in the decentralized market.

The \textbf{VECM} of order 1 can be expressed as:

\[
\Delta \mathbf{X}_t = \alpha \beta' \mathbf{X}_{t-1} + \Gamma_1 \Delta \mathbf{X}_{t-1} + \mathbf{u}_t
\]

Where \(\Delta \mathbf{X}_t\) were the first differences of the variables \(X_1\) (centralized) and \(X_2\) (decentralized), \(\alpha\) was the adjustment coefficient matrix, \(\beta\) the cointegrating vector, \(\Gamma_1\) the coefficient for the lagged differences, and \(\mathbf{u}_t\) the error term.

Using the values from the VECM estimation, the explicit system of equations is:

\[
\Delta X_1(t) = -0.026 \cdot (X_1(t-1) - 0.999 \cdot X_2(t-1)) + u_{t,1}
\]

\[
\Delta X_2(t) = 0.958 \cdot (X_1(t-1) - 0.999 \cdot X_2(t-1)) + u_{t,2}
\]

Where \(u_{t,1}\) and \(u_{t,2}\) are the residuals for each equation.

We first investigate the residual autocorrelations in order to check that we have found a description of the data
consistent with the assumption of white noise errors.

In order to ensure that the estimated VECM model is adequate, we perform several diagnostic tests to validate its performance. Our model has two dimensions: centralized and decentralized. We use only one lag in our cointegrated model, and we need to confirm that it is well specified. We first investigate the residual autocorrelations to check whether the model produces white noise errors, which would indicate that the model has captured the dynamics of the data appropriately. Below, we present the results of these diagnostics based on the residuals and their properties.

\begin{figure}[h]
    \centering
    \includegraphics[width=0.65\textwidth]{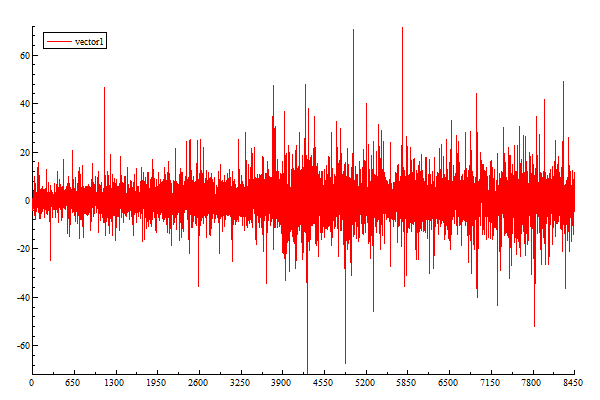}
    \caption{Residuals of the decentralized price over time}
    \label{fig:residual_defi}
\end{figure}

First, we examine the residuals over time for the decentralized price series, Price\_defi, as shown in Figure \ref{fig:residual_defi}. The residuals fluctuate around zero and do not exhibit any visible trend, which indicates that the model captures the underlying structure well. The spikes present are random and do not suggest any systemic issues in the model.

\begin{figure}[h]
    \centering
    \includegraphics[width=0.7\textwidth]{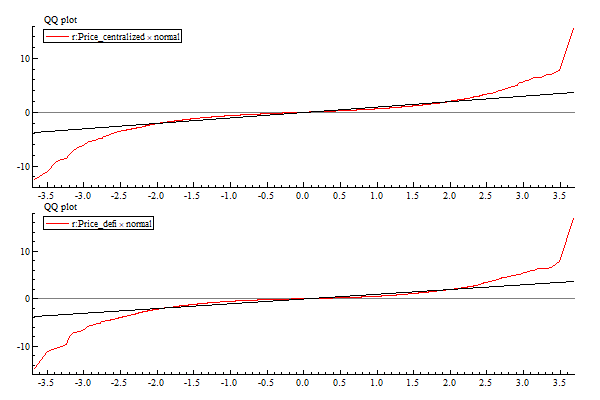}
    \caption{QQ Plots for Residuals of Price\_centralized and Price\_defi}
    \label{fig:qq_plots}
\end{figure}

Next, we evaluate the QQ plots for the residuals of Price\_centralized and Price\_defi, which are depicted in Figure \ref{fig:qq_plots}. These plots assess whether the residuals are normally distributed. Both QQ plots show that the residuals align well with the theoretical quantiles of a normal distribution, except for slight deviations in the tails, which is typical in financial data. These results confirm that the residuals are reasonably close to being normally distributed.

\begin{figure}[h]
    \centering
    \includegraphics[width=1\textwidth]{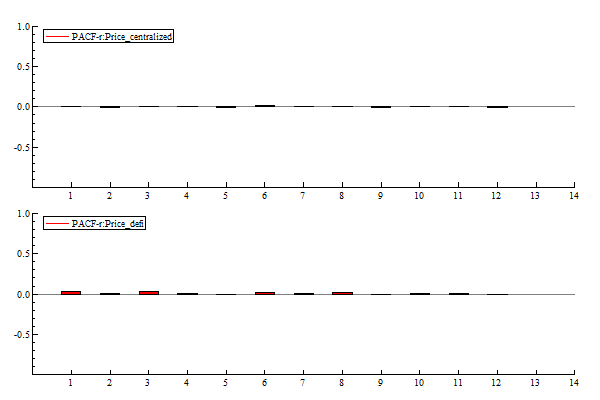}
    \caption{ACF and PACF for Residuals of centralized and decentralized prices.}
    \label{fig:acf}
\end{figure}

We also assess the autocorrelation structure of the residuals using ACF and PACF plots. Figures \ref{fig:acf} and \ref{fig:pacf} present the autocorrelation function (ACF) and partial autocorrelation function (PACF) for both Price\_centralized and Price\_defi. These plots show no significant spikes, indicating that the residuals are uncorrelated, which is an essential feature of a well-specified model. The lack of serial correlation in the residuals supports the adequacy of the VECM specification.

In conclusion, the diagnostic checks demonstrate that the VECM model is well-specified. The residuals show no significant autocorrelation, and the QQ plots confirm that they are reasonably normally distributed. These results provide confidence in the validity of the model, which successfully captures the dynamics between the decentralized and centralized markets.

After conducting diagnostic checks, we are confident that our model is well-specified and confirm that the estimated coefficients are as follows:

\begin{itemize}
    \item \textbf{Beta (Cointegrating Vector):}
    \[
    \beta = \begin{bmatrix} 
    1.000 \\ 
    -0.999 
    \end{bmatrix}
    \]
    \item \textbf{Alpha (Adjustment Coefficients):}
    \[
    \alpha = \begin{bmatrix} 
    -0.026 \\ 
    0.958 
    \end{bmatrix}
    \]
    \item \textbf{Standard Errors of Alpha:}
    \[
    \text{Errors} = \begin{bmatrix} 
    0.029 \\ 
    0.026 
    \end{bmatrix}
    \]
\end{itemize}
\newpage
\subsubsection{Cointegration analysis and Granger causality tests}
The Johansen Cointegration Test was used to confirm whether the time series are cointegrated over the long run.

\begin{table}[h]
\centering
\caption{Johansen Cointegration Test using Trace Test Statistic (10\% Significance Level)}
\begin{tabular}{cccc}
\hline
\textbf{Rank (r)} & \textbf{Cointegrating Relations} & \textbf{Test Statistic} & \textbf{Critical Value} \\
\hline
0 & 2 & 1856 & 10.47 \\
1 & 2 & 4.059e-05 & 2.976 \\
\hline
\end{tabular}
\end{table}

 The test statistic for rank 0 was 1856, which significantly exceeds the critical value of 10.47, leading us to reject the null hypothesis of no cointegration. This indicates that there is at least one cointegrating relationship between the two markets. For rank 1, the test statistic was 4.059e-05, which is lower than the critical value of 2.976. Therefore, we fail to reject the null hypothesis for rank 1, confirming the existence of exactly one cointegrating relationship between the centralized and decentralized markets.

We have thus confirmed that the VECM model we estimated is well-specified and adequate for describing our data.
Following the cointegration confirmation, we apply Hasbrouck’s Information Share methodology to assess which market incorporates more information over the long term.

\begin{table}[h]
\centering
\caption{Hasbrouck's Information Share}
\begin{tabular}{|c|c|c|}
\hline
\textbf{Market}          &  &  \\ \hline
Centralized Market       & 0.999   & 0.982 \\ \hline
Decentralized Market     & 0.001   & 0.018 \\ \hline
\end{tabular}
\label{table:hasbrouck_info_share_corrected}
\end{table}

We compute the metric twice because Hasbrouck's information share can yield different results depending on the order in which we consider the markets. Specifically, whether we treat \(X_1\) as the centralized market or the decentralized market changes the outcome. This difference arises from the way the information share is calculated, reflecting the varying contributions of each market to price discovery. As a result, we calculate the metric in both orderings to capture this asymmetry.
The results indicate that the centralized market overwhelmingly dominates, with an information share of 0.9996 for both orders, while the decentralized market has a share of 4.2071 $\times 10^{-4}$. The interpretation is as follows: the closer the number is to one, the more information the centralized market contributes. This shows that the centralized market plays a leading role in price discovery.

Next, the Gonzalo-Granger methodology was employed to further investigate the dynamics between the markets. The alpha coefficients obtained from the VECM were \(\alpha_1 = -0.0259\) and \(\alpha_2 = 0.958\). The likelihood ratio test for \(\alpha = [1,0]\) produced a \(\chi^2(1) = 1293.5\) with a p-value of 0, which strongly rejects the null hypothesis. Conversely, for \(\alpha = [0,1]\), we get \(\chi^2(1) = 0.84287\) with a p-value of 0.3586, meaning that we fail to reject the null hypothesis in this case. So in this case we have that \(\alpha = [0,1]\).

The result \(\alpha = [0,1]\) suggests that the decentralized market responds more to price discrepancies, while the centralized market drives the overall price changes. This aligns with the interpretation of the VECM system illustrated in Figure 2.1, where the common trend is influenced by the orthogonal vector \(\alpha^\perp = [1,0]\). In this context, \(\alpha^\perp\) represents the influence of the first market, which is the centralized market, indicating that it leads price discovery.
Thus, the orthogonal alpha vector \(\alpha^\perp = [1,0]\) confirms that the centralized market leads, as it is the primary contributor to the common stochastic trend. This means the centralized market influences the price dynamics, while the decentralized market adjusts to correct any disequilibrium.

Next, we apply the Granger causality test to assess the predictive relationship between the decentralized and centralized markets. The Granger causality test helps to determine whether the past values of the centralized market prices can predict the future values of the decentralized market prices and vice versa.

The test was performed with a maximum of 5 lags, and the results indicate a strong predictive relationship between the centralized and decentralized markets. Specifically, the centralized market Granger-causes the decentralized market, as indicated by a highly significant F-test p-value of \( 3.69 \times 10^{-283} \) for lag 1, with similar significance across lags 2 to 5.

In summary, the Granger causality test shows that the centralized market has a strong predictive influence on the decentralized market. This further supports the earlier findings from the VECM analysis, where the centralized market was shown to lead in price discovery.
\subsubsection{Lead-lag relationships estimation.}
The lead-lag methodology is, in some sense, a more refined and interesting version of the Granger causality test, as it tests similar concepts but provides richer output. Therefore, in the rest of the analysis, we have decided to focus on the lead-lag methodology and do not report the results of the Granger causality test. Finally, using the Hayashi-Yoshida Lead-Lag methodology, we computed the lead-lag ratio (LLR), which was found to be 1.20. The lead-lag time was calculated to be 0 seconds, which is consistent with the fact that the analysis was not conducted using tick-by-tick data. Despite the absence of detectable lead-lag time, the LLR result clearly indicates that the centralized market leads the decentralized market over the long run, because the number is bigger than 1.

In conclusion, the yearly analysis supports the previous findings from the date-specific analyses. The centralized market consistently leads in terms of price discovery due to its higher liquidity and faster incorporation of information. All three methodologies – Hasbrouck, Gonzalo-Granger, and Hayashi-Yoshida – align in demonstrating that the centralized market dominates the decentralized market in long-term price relationships. This concludes the long-term analysis.

Next, we shift our focus to specific events, analyzing various descriptive statistics in addition to the metrics already introduced. To provide a clearer understanding of market conditions during these events, we first present several key descriptive statistics that are crucial for interpreting market behavior. The percent price change during the event (\%) indicates the percentage change in Bitcoin's price over the specific event window. A negative value represents a price decline, while a positive value reflects an increase in price. This metric helps gauge the direction and magnitude of price movements triggered by the event.

The total daily volume represents the total amount traded throughout the trading day, which provides an understanding of overall market activity and the intensity of trading during the event. It contextualizes the event's impact in relation to the full day's activity. Additionally, the daily maximum price (USD) reflects the highest price reached by Bitcoin during the trading day, offering insight into the peak price during the event window. On the other hand, the daily minimum price (USD) captures the lowest price reached, providing a measure of the extent of price volatility within the same period.

These descriptive statistics offer a quantitative view of market behavior, helping to understand the intensity of trading activity, the magnitude of price changes, and overall volatility during key events. This event-based analysis complements the long-term findings by focusing on short-term dynamics under specific market conditions.

\newpage
\subsection{Benchmarking the performance of AMMs: analysis for March 5th, 2024}
We selected this date for various reasons. There was a clear increase in volume in the Uniswap v2 pool, with 53.644 million US dollars, more than double the volume of the previous day. Ethereum was in a bullish trend and reached a new high on this day, but ended up rejecting the critical level of 3800 and experienced a significant decline in the latter part of the day. Gas prices also rose considerably, with an average of 98.68 Gwei throughout the day, the highest average in all of 2024. To put this in perspective, gas prices typically fluctuate between 20 to 50 Gwei during normal market conditions, so this was a significant increase, indicating heightened network congestion and trading activity. Additionally, there were ongoing uncertainties regarding whether the SEC would approve Ethereum ETFs, which added to the volatility and market activity.

We are going to study which market incorporated the most information during one of the most volatile periods of the day, from 19:30 to 19:45.
\begin{table}[H]
\centering
\begin{tabular}{|l|r|r|}
\hline
\textbf{Statistic}                     & \textbf{Value Uniswap v2} & \textbf{Value Binance}  \\ \hline
Price Change During Event (\%)         & -2.76                     & -3.01                   \\ \hline
Event Maximum Price(USD)                & 3567.91
 & 3580.88                 \\ \hline
Event Minimum Price (USD)              & 3458.38                   & 3442.15                 \\ \hline
\end{tabular}
\caption{Descriptive statistics for the event on March 5th: Uniswap v2 pool WETH/USDT and ETH on Binance}
\label{tab:descriptive_event_march}
\end{table}

We have 81 observations from the Uniswap v2 pool and 901 observations from Binance. We next plot the evolution of prices in both centralized and decentralized exchanges in Figure 3.1 .
\begin{figure}[h]
    \centering
    \includegraphics[width=0.9\textwidth]{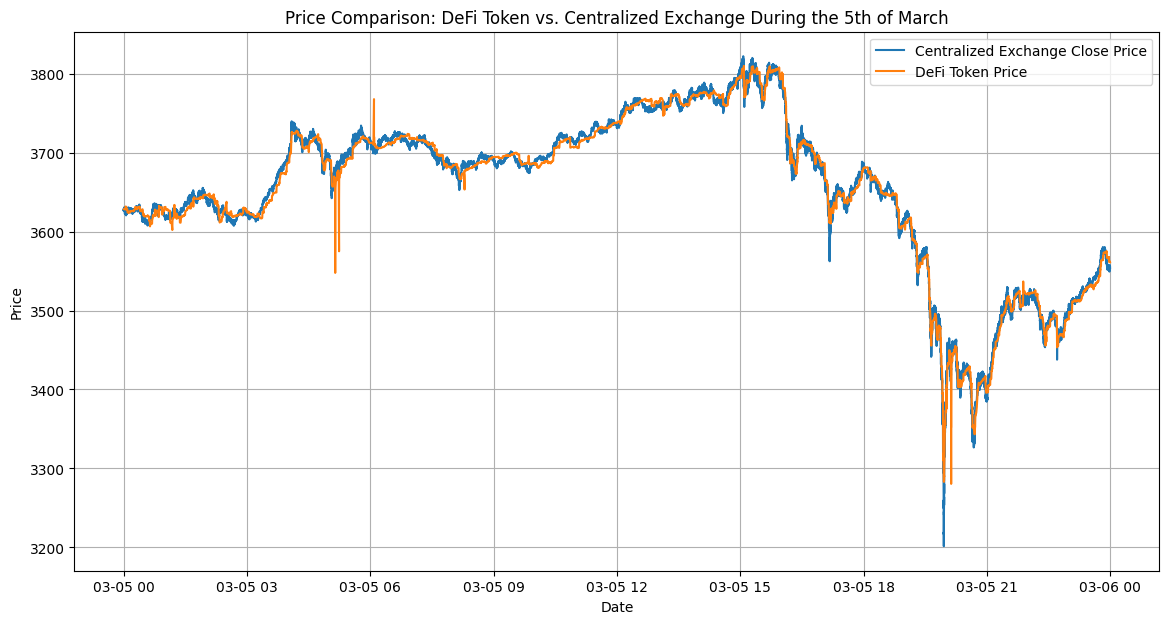}
    \caption{Plot of prices in centralized and decentralized markets on March 5th 2024.}
\end{figure}

We observe that there is a significant divergence between both prices from what we would call a common efficient price around the moments of most volatility , going back to normal when it is less volatile. We observe a significant divergence during the first 15 minutes of the most volatile period too. We  take a closer look at this period.
\begin{figure}[h]
    \centering
    \includegraphics[width=0.9\textwidth]{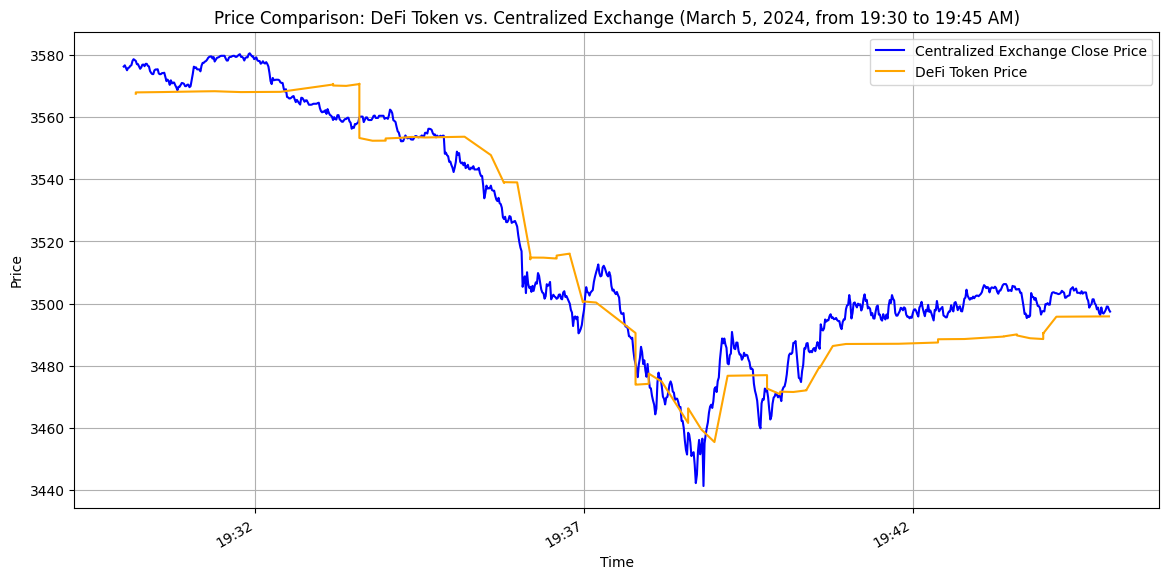}
    \caption{Plot of prices in centralized and decentralized markets during the 15-minute event in March.}
\end{figure}
We observe that the centralized data incorporates more information at first glance.

We begin by estimating a VECM of order 1, following the same procedure outlined in Section A. To ensure the robustness of the model, we analyze its congruence through a series of diagnostic tests, including tests for residual normality, absence of autocorrelation, and normality. These tests confirm that the residuals behave appropriately, ensuring the model is well-specified and fit for further analysis. 

Given that the series are clearly non-stationary, we perform the Johansen Cointegration test at the 10 percent significance level to assess the extent of cointegration relationships between the series. The test confirms that both series are cointegrated, allowing us to move forward with a more detailed analysis of the system's long-term equilibrium dynamics. 

For reasons of space, we will focus on presenting the \(\alpha\) coefficient, as it plays a crucial role in understanding the speed of adjustment toward equilibrium and is the key parameter for the price discovery metrics. With a well-specified and congruent model in place, we can confidently proceed with a battery of price dynamics analyses that further investigate the relationships between the markets.

\begin{table}[h]
\centering
\caption{Johansen Cointegration Test using Trace Test Statistic (10\% Significance Level)}
\begin{tabular}{cccc}
\hline
\textbf{Rank (r)} & \textbf{Cointegrating Relations} & \textbf{Test Statistic} & \textbf{Critical Value} \\
\hline
0 & 2 & 50.29 & 13.43 \\
1 & 2 & 2.113 & 2.705 \\
\hline
\end{tabular}
\end{table}
The results show that the null hypothesis of no cointegration (rank 0) is rejected, as the test statistic (50.29) exceeds the critical value (13.43), indicating at least one cointegrating relationship. Furthermore, the null hypothesis of at most one cointegrating relationship (rank 1) is also rejected, as the test statistic (2.113) exceeds the critical value (2.705), confirming exactly two cointegrating relationships. 
After estimating the model, we will apply the following three methodologies to analyze the price discovery process: Hasbrouck’s Information Share, Gonzalo-Granger’s Permanent-Transitory Decomposition, and the Hayashi-Yoshida method.

We start with the Hasbrouck methodology. The output of the information shares is as follows:

\begin{table}[h]
\centering
\caption{Hasbrouck's Information Share}
\begin{tabular}{|c|c|c|}
\hline
\textbf{Market}          & \textbf{} & \textbf{} \\ \hline
Centralized Market       & 0.959        & 0.966        \\ \hline
Decentralized Market     & 0.041        & 0.034        \\ \hline
\end{tabular}
\label{table:hasbrouck_info_share_corrected}
\end{table}

These results indicate that the market which explains the most variance is clearly the first one, the centralized market.

In the Gonzalo-Granger methodology, when estimating the VECM, we obtain the vector \(\alpha = [\alpha_1, \alpha_2] = [0.008, 0.038]\). This result is consistent with the previous analyses: since \(\alpha_2 > \alpha_1\), it indicates that the second market (the decentralized one) reacts more to discrepancies, so the first market (the centralized one) leads.

We performed a likelihood ratio test for \(\alpha = [1,0]\) and obtained \(\chi^2(1) = 43.399\) with a p-value of \(0\), so we reject this hypothesis. Similarly, for \(\alpha = [0,1]\), we obtained \(\chi^2(1) = 1.931\) with a p-value of \(0.164\). Therefore, we do not have evidence to reject the null hypothesis that \(\alpha = [0,1]\), or equivalently, \(\alpha^{\perp} = [1,0]\). According to the Gonzalo and Granger methodology, this means that the centralized market leads the decentralized one, because the second market (the decentralized one) reacts more to discrepancies, while the first market (the centralized one) drives the price changes.

We continue with the Hayashi-Yoshida methodology. We obtain a lead-lag ratio (LLR) of \(1.86\). The lead-lag time is calculated as \(0\) seconds, which is expected given that the analysis was not performed with tick-by-tick data..

The LLR of \(1.86\) indicates again in this case that the centralized exchange is leading the decentralized market. This conclusion is in agreement with the oth\textbf{}er methodologies, where the centralized market typically leads due to its higher liquidity. 
\newpage
\subsubsection{Gas Fee Calculation}
To evaluate the feasibility of this strategy, we must first account for gas fees. On March 5th, 2024, the average gas price was 98.68 Gwei, as sourced from Etherscan. The gas fee for a typical ERC-20 smart contract transaction can be calculated using a gas limit of 50,000, which is sufficient for the interactions we analyze. Given that:
\[
1 \text{ Gwei} = 0.000000001 \text{ ETH},
\]
the gas fee for a single transaction is:
\[
\text{Gas Fee (in ETH)} = 98.68 \times 50,000 \times 10^{-9} = 0.004934 \text{ ETH}.
\]
Since any arbitrage opportunity involves both buying and selling, we must account for gas fees twice. Thus, the total gas fees for each arbitrage attempt are approximately \$30.58, assuming an ETH price of around \$3,100.

\subsubsection{Arbitrage Analysis}
We computed the potential arbitrage by comparing the prices on Uniswap v2 (decentralized) and Binance (centralized). For each timestamp, we subtracted the centralized price from the decentralized price and deducted the gas fees, which were applied twice to account for both entry and exit transactions.

Additionally, we calculated the gas fee as a percentage of the price deviation between the two markets. For example, with a price difference of \$35.04 at 19:36:08, a gas fee of \$30.58 represents approximately:
\[
\frac{30.58}{35.04} \times 100 \approx 87.3\%
\]
of the price deviation. This high percentage indicates that the gas fee alone consumes most of the potential arbitrage profit.

In Table \ref{table:arbitrage_potential}, we present the 5 most profitable arbitrage opportunities. However, even these opportunities are not large enough to offset the gas fees. Additionally, we computed the mean loss from attempting this arbitrage strategy across the entire period, further illustrating the difficulty of finding profitable opportunities.

\begin{table}[h]
\centering
\caption{Top 5 Arbitrage Potential Cases (Gas Fees Considered)}
\begin{tabular}{|c|c|c|c|c|}
\hline
\textbf{Timestamp} & \textbf{Price (Centralized)} & \textbf{Price (Defi)} & \textbf{Gas Fee (USD)} & \textbf{Arbitrage Potential (USD)} \\
\hline
19:36:08 & 3503.95 & 3538.99 & 34.92 & 0.11 \\
19:36:05 & 3505.42 & 3538.99 & 34.92 & -1.36 \\
19:36:10 & 3506.40 & 3538.99 & 34.92 & -2.34 \\
19:36:06 & 3507.64 & 3538.99 & 34.92 & -3.58 \\
19:36:07 & 3508.80 & 3538.99 & 34.92 & -4.74 \\
\hline
\end{tabular}
\label{table:arbitrage_potential}
\end{table}

Across all timestamps analyzed, the mean loss from attempting this arbitrage strategy (after accounting for gas fees) is approximately \$45.30. This further demonstrates that, given the high transaction fees and network congestion, arbitrage opportunities were not profitable during this period.

\subsubsection{Conclusion on Arbitrage Opportunities}
These results suggest that during periods of high network activity, such as March 5th, 2024, arbitrage opportunities are scarce due to elevated gas fees. Our analysis shows that even when price divergences occur, the cost of executing trades on the Ethereum network effectively cancels out any potential profit. It is also important to note that we have not accounted for slippage, which would further reduce the profitability of such strategies. In conclusion, arbitrage opportunities in the Ethereum market for ETH are difficult to exploit in high-fee environments.
\subsection{Benchmarking the performance of AMMs: analysis for April 30th, 2024}
We selected April 30th in our analysis due to the significant decline observed across various cryptocurrencies, including Ethereum. Ethereum had risen significantly in the preceding days, and on this date, we observed notable profit-taking. Additionally, average gas prices rose from 11.27 Gwei to 17.41 Gwei, reflecting increased network congestion. The trading volume for our Ethereum pool on April 30th reached 9.837 million dollars, marking a substantial increase compared to both the preceding and following days.
This decline was largely due to more investors cashing out profits, particularly those who entered the market during the downturns of 2022 and 2023, as well as ETF investors who saw a strong rise in share prices after buying in early 2024.

\begin{table}[H]
\centering
\begin{tabular}{|l|r|r|}
\hline
\textbf{Statistic}                     & \textbf{Value Uniswap v2} & \textbf{Value Binance}  \\ \hline
Price Change During Event (\%)         & -1.41                     & -1.35                   \\ \hline
Event Maximum Price(USD)                & 3114.91
 & 3115.48                 \\ \hline
Event Minimum Price (USD)              & 3070.67                   & 3056.85                 \\ \hline
\end{tabular}
\caption{Descriptive statistics for the event on April 30th: Uniswap v2 pool WETH/USDT and ETH on Binance}
\label{tab:descriptive_event_april}
\end{table}

We have 48 data observations from the Uniswap v2 pool and 1201 observations from Binance, with one data point recorded every second.

\begin{figure}[h]
    \centering
    \includegraphics[width=0.8\textwidth]{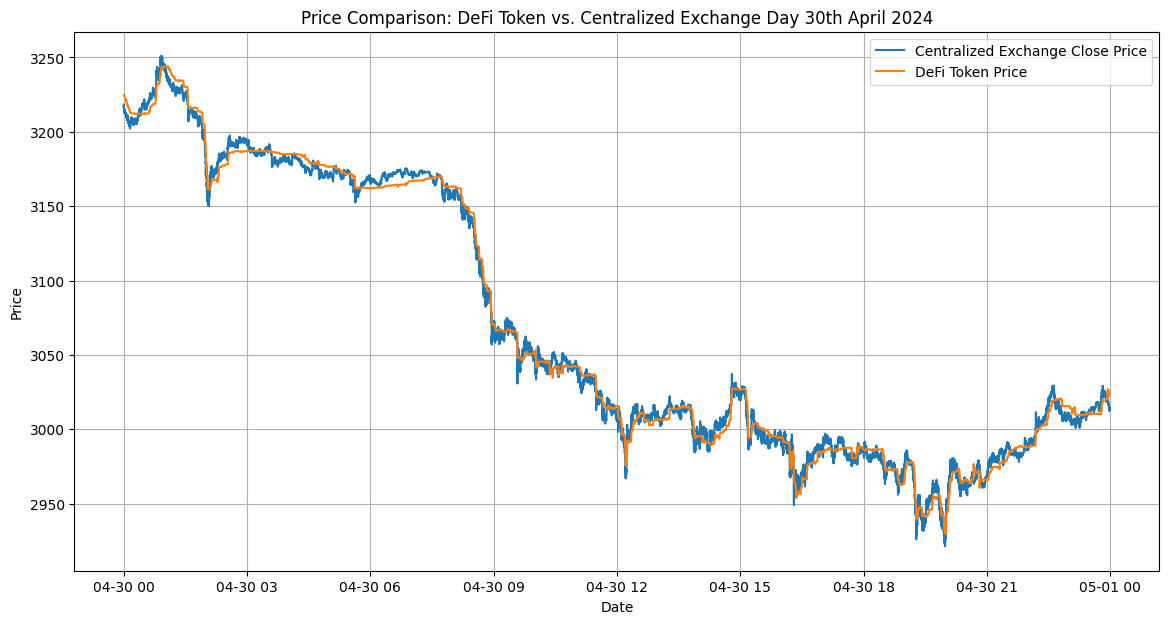}
    \caption{Plot of Ethereum prices on April 30th 2024 on both Uniswap v2 and the centralized exchange Binance.}
\end{figure}

\begin{figure}[h]
    \centering
    \includegraphics[width=0.8\textwidth]{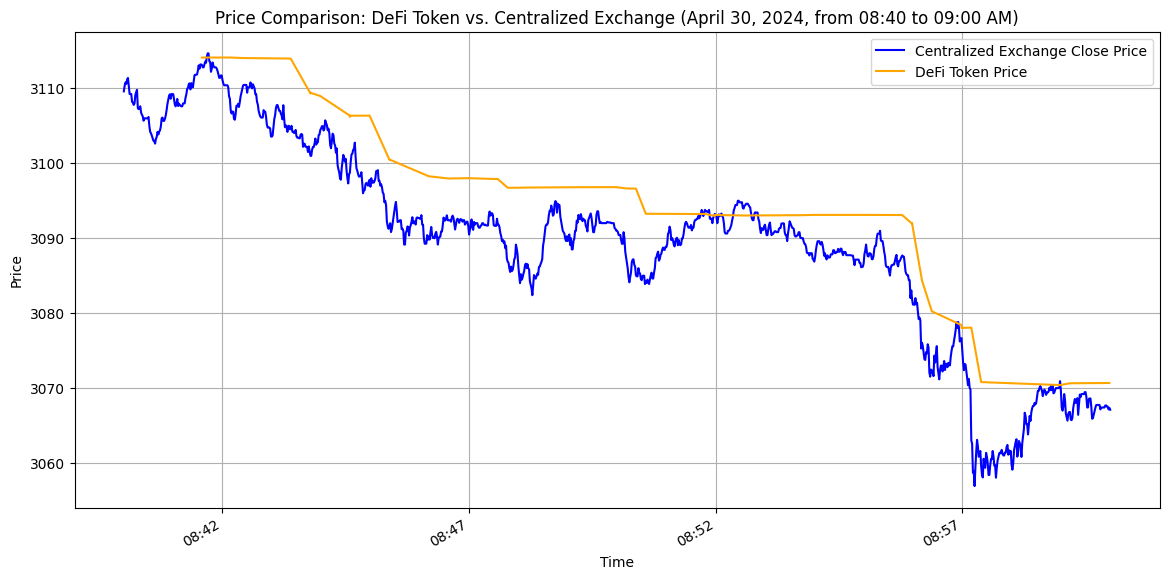}
    \caption{Plot of prices during the 20 minute event on April 30th 2024.}
\end{figure}

We observe that there is a significant divergence between both prices from what we would call a common efficient price around the moments of most volatility , and then they go back to normal.

We see that during the first 20 minutes of the most volatile period there is a significant divergence. We look more closer into this period.

We begin by estimating a VECM, as in the long-run example. For reasons of space, we will only present the \(\alpha\) coefficient, as it is the most important one for the metrics.

The series are clearly non-stationary. First, we compute the Johansen Cointegration test with a 10 percent significance level to confirm that both series are cointegrated.

\begin{table}[h]
\centering
\caption{Johansen Cointegration Test using Trace Test Statistic (10\% Significance Level)}
\begin{tabular}{cccc}
\hline
\textbf{Rank (r)} & \textbf{Cointegrating Relations} & \textbf{Test Statistic} & \textbf{Critical Value} \\
\hline
0 & 2 & 32.10 & 13.43 \\
1 & 2 & 0.1839 & 2.705 \\
\hline
\end{tabular}
\end{table}

Results show that the null hypothesis of no cointegration (rank 0) is rejected, as the test statistic (32.10) exceeds the critical value (13.43), indicating at least one cointegrating relationship. However, the null hypothesis of at most one cointegrating relationship (rank 1) cannot be rejected, as the test statistic (0.1839) is below the critical value (2.705), , confirming exactly one cointegrating relationship.

After estimating the model, we will apply the following three methodologies to analyze the price discovery process: Hasbrouck’s Information Share, Gonzalo-Granger’s Permanent-Transitory Decomposition, and the Hayashi-Yoshida method.

We start with the Hasbrouck methodology. The output of the information shares is as follows:

\begin{table}[h]
\centering
\caption{Hasbrouck's Information Share}
\begin{tabular}{|c|c|c|}
\hline
\textbf{Market}          &  &  \\ \hline
Centralized Market       & 0.965  & 0.923 \\ \hline
Decentralized Market     & 0.076  & 0.003 \\ \hline
\end{tabular}
\label{table:hasbrouck_info_share_corrected}
\end{table}

These results indicate that the market which explains the most variance is clearly the first one, the centralized market. The analysis was conducted by calculating the metric twice, changing the order of the variables each time.

In the Gonzalo-Granger methodology, when estimating the VECM, we obtain the vector \(\alpha = [\alpha_1, \alpha_2] = [-0.0079, 0.013]\). This result is consistent with the previous analyses: since \(\alpha_2 > \alpha_1\), it indicates that the second market (the decentralized one) reacts more to discrepancies, so the first market (the centralized one) leads.

We performed a likelihood ratio test for \(\alpha = [1,0]\) and obtained \(\chi^2(1) = 76.31\) with a p-value of \(0\). Similarly, for \(\alpha = [0,1]\), we obtained \(\chi^2(1) = 1.4287\) with a p-value of \(0.23\). Based on these results, we reject the null hypothesis for \(\alpha = [1,0]\), but we fail to reject the null hypothesis for \(\alpha = [0,1]\). Therefore, we conclude that \(\alpha^{\perp} = [1,0]\), which means that the centralized system leads.

We continue with the Hayashi-Yoshida methodology. We obtain a lead-lag ratio (LLR) of \(1.47\). The lead-lag time is calculated as \(2\) seconds. The decentralized market looked particularly illiquid in this date, so the result is reasonable.

The LLR of \(1.47\) indicates that the centralized exchange is leading the decentralized market, which aligns with our expectations. This conclusion is in agreement with the other methodologies, where the centralized market typically leads due to its higher liquidity. 
\subsection{Benchmarking the performance of AMMs: analysis for May 20th, 2024}
This is another interesting date for Ethereum, as prices rose sharply. During this time, there was ongoing debate about whether the SEC would approve Ethereum ETFs \cite{fxstreet2024}, and news emerged that approval was likely after weeks of uncertainty. On this day, the average gas price increased significantly, more than tripling, from 5.68 Gwei the day before to 18.81 Gwei. This represents a significant relative increase, reflecting heightened network activity and congestion. We will analyze which market incorporated the most information during one of the most volatile periods of the day, from 19:23 to 19:37.

\begin{table}[H]
\centering
\begin{tabular}{|l|r|r|}
\hline
\textbf{Statistic}                     & \textbf{Value Uniswap v2} & \textbf{Value Binance}  \\ \hline
Price Change During Event (\%)         & 4.78                      & 6.78                    \\ \hline
Event Maximum Price(USD)                & 3371.73
 & 3431.15                 \\ \hline
Event Minimum Price (USD)              & 	3217.91                   & 3213.05                 \\ \hline
\end{tabular}
\caption{Descriptive statistics for the event on May 20th: Uniswap v2 pool WETH/USDT and ETH on Binance}
\label{tab:descriptive_event_may}
\end{table}

We have 43 observations from the Uniswap v2 pool and 841 observations from Binance.
\begin{figure}[h]
    \centering
    \includegraphics[width=0.9\textwidth]{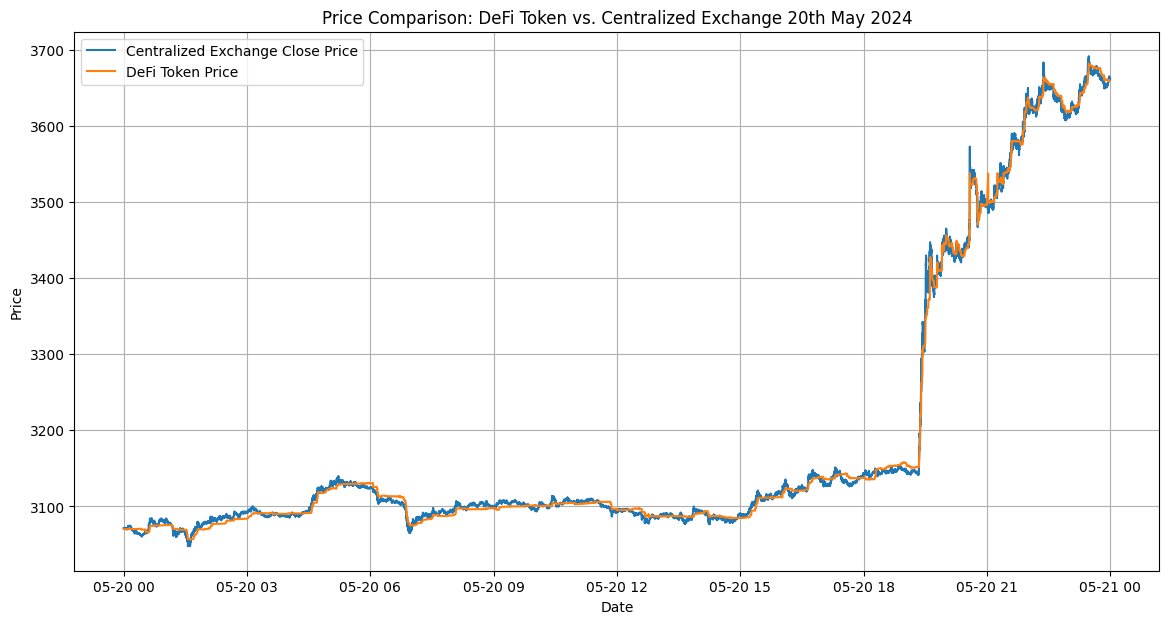}
    \caption{Plot of prices in centralized and decentralized markets during the 20th of May.}
\end{figure}

We observe that there is a significant divergence between both prices from what we would call a common efficient price around the moments of most volatility  again, going back to normal when it is less volatile.

We see that during the first 14 minutes of the most volatile period there is a significant divergence. We look more closer into this period.
\begin{figure}[h]
    \centering
    \includegraphics[width=0.9\textwidth]{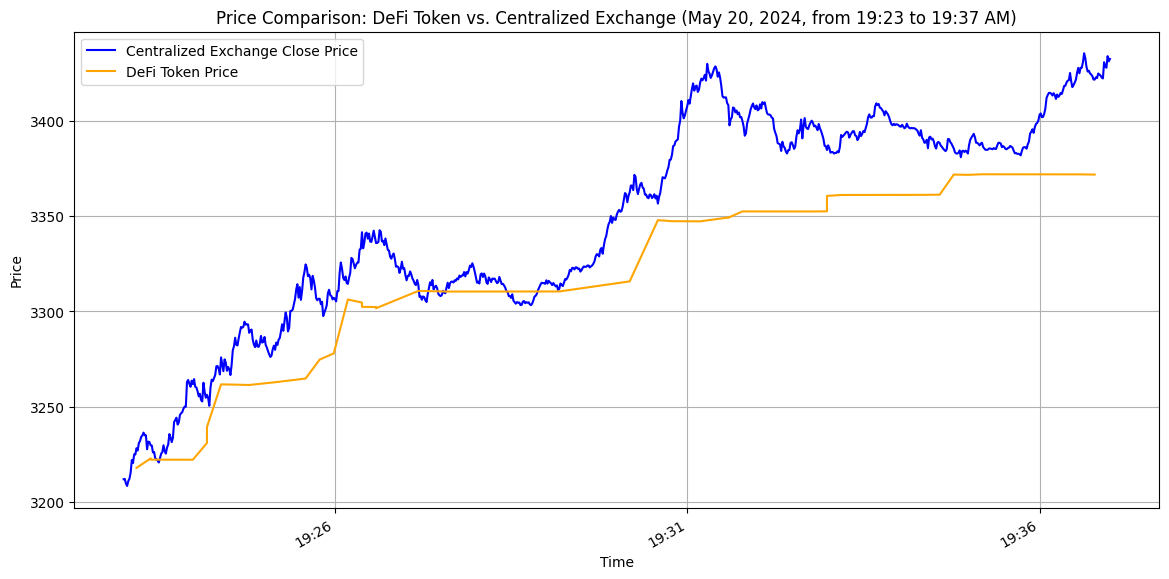}
    \caption{Plot of prices in centralized and decentralized markets during the 14 minute event of 20th May.}
\end{figure}

We observe that the centralized data incorporates more information at first glance, it is more detailed.

We begin by estimating a VECM, as in the long-run example. For reasons of space, we will only present the \(\alpha\) coefficient, as it is the most important one for the metrics.

The series are clearly non-stationary. First, we compute the Johansen Cointegration test with a 10 percent significance level to confirm that both series are cointegrated.

\begin{table}[h]
\centering
\caption{Johansen Cointegration Test using Trace Test Statistic (10\% Significance Level)}
\begin{tabular}{cccc}
\hline
\textbf{Rank (r)} & \textbf{Cointegrating Relations} & \textbf{Test Statistic} & \textbf{Critical Value} \\
\hline
0 & 2 & 16.80 & 13.43 \\
1 & 2 & 6.755 & 2.705 \\
\hline
\end{tabular}
\end{table}
In this case we do not find that both series are cointegrated with this test for this particular case, but both series are clearly cointegrated in the long-run and we decide to perform the analysis anyway.

We will apply the following three methodologies to analyze the price discovery process: Hasbrouck’s Information Share, Gonzalo-Granger’s Permanent-Transitory Decomposition, and the Hayashi-Yoshida method.

We start with the Hasbrouck methodology. The output of the information shares is as follows:

\begin{table}[h]
\centering
\caption{Hasbrouck's Information Share}
\begin{tabular}{|c|c|c|}
\hline
\textbf{Market}          &  &  \\ \hline
Centralized Market       & 0.628   & 0.583 \\ \hline
Decentralized Market     & 0.416   & 0.371 \\ \hline
\end{tabular}
\label{table:hasbrouck_info_share_corrected}
\end{table}

These results indicate that the market which explains the most variance is 
the first one, the centralized market.

In the Gonzalo-Granger methodology, when estimating the VECM, we obtain the vector \(\alpha = [\alpha_1, \alpha_2] = [-0.0048, 0.0084]\).

We performed a likelihood ratio test for \(\alpha = [1,0]\) and obtained \(\chi^2(1) = 34.45\) with a p-value of \(0\), so we reject this hypothesis. Similarly, for \(\alpha = [0,1]\), we obtained \(\chi^2(1) = 0.40\) with a p-value of \(0.52\). Therefore, we do not have evidence to reject the null hypothesis that \(\alpha = [0,1]\), or equivalently, \(\alpha^{\perp} = [1,0]\). According to the Gonzalo and Granger methodology, this means that the centralized market leads the decentralized one, because the second market (the decentralized one) reacts more to discrepancies, while the first market (the centralized one) drives the price changes.

We continue with the Hayashi-Yoshida methodology. We obtain a lead-lag ratio (LLR) of \(1.22\). The lead-lag time is calculated as \(0\) seconds, which is expected given that the analysis was not performed with tick-by-tick data..

The LLR of \(1.22\) indicates again in this case that the centralized exchange is leading the decentralized market. This conclusion is in agreement with the other methodologies, where the centralized market typically leads due to its higher liquidity.

\subsection{Benchmarking the performance of AMMs: analysis for August 5th, 2024}
We selected August 5th as the focal point of our analysis due to the significant decline observed across various cryptocurrencies, including Ethereum.

The beginning of August was particularly volatile for cryptocurrencies, making it an ideal period for this analysis. For instance, during the time window from 00:00:55 to 01:10:00 on August 5th, Ethereum experienced a dramatic dip of over 20\%, dropping from nearly 2700 to around 2200 dollars. Gwei prices averaged 51.68 during the day, a significant increase compared to the previous day's levels. Normal gas prices typically fluctuate between 20 and 40 Gwei during periods of moderate network activity, so this rise indicates heightened network congestion, likely driven by the surge in trading volume and volatility. Additionally, we will look into this specific event to analyze how declines in the equities markets impact cryptocurrency markets, as the beginning of August saw significant outflows from equities markets.

Additionally, the trading volume for our Ethereum pool on August 6th reached 191.855 million dollars, marking a significant increase compared to both the preceding and following days.

We have 178 data observations from the Uniswap v2 pool and 900 observations from Binance, with one data point recorded every second.
The beginning of August was quite a volatile period for cryptocurrencies, making it an ideal time for our analysis. 

\begin{table}[H]
\centering
\begin{tabular}{|l|r|r|}
\hline
\textbf{Statistic}                     & \textbf{Value Uniswap v2} & \textbf{Value Binance}  \\ \hline
Price Change During Event (\%)         & -11.1                     & -11.01                   \\ \hline
Event Maximum Price(USD)                & 2588.67
 & 2586.85                 \\ \hline
Event Minimum Price (USD)              & 	2110.34                   & 2115.37                 \\ \hline
\end{tabular}
\caption{Descriptive statistics for the event on August 5th: Uniswap v2 pool WETH/USDT and ETH on Binance}
\label{tab:descriptive_event_combined}
\end{table}

\begin{figure}[h]
    \centering
    \includegraphics[width=0.8\textwidth]{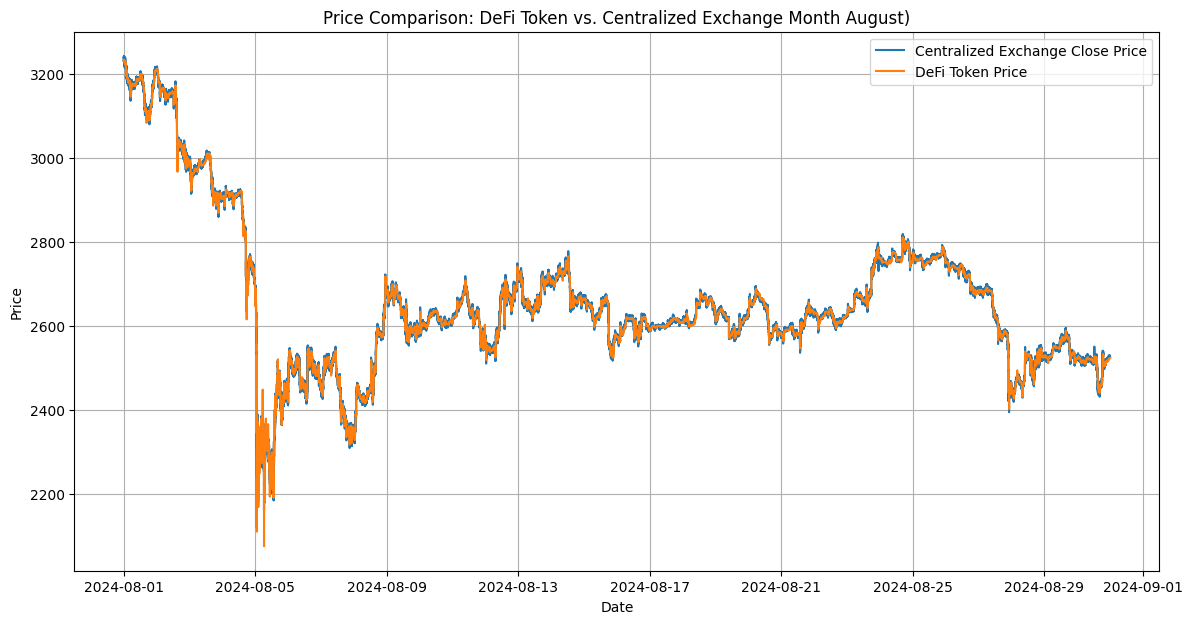}
    \caption{Plot of Ethereum prices in the month of August both on Uniswap v2 and the centralized exchange Binance on August 2024.}
\end{figure}

We observe in Figure 3.1 that prices closely track each other, suggesting a strong cointegration relationship between the two markets.

\begin{figure}[h]
    \centering
    \includegraphics[width=0.8\textwidth]{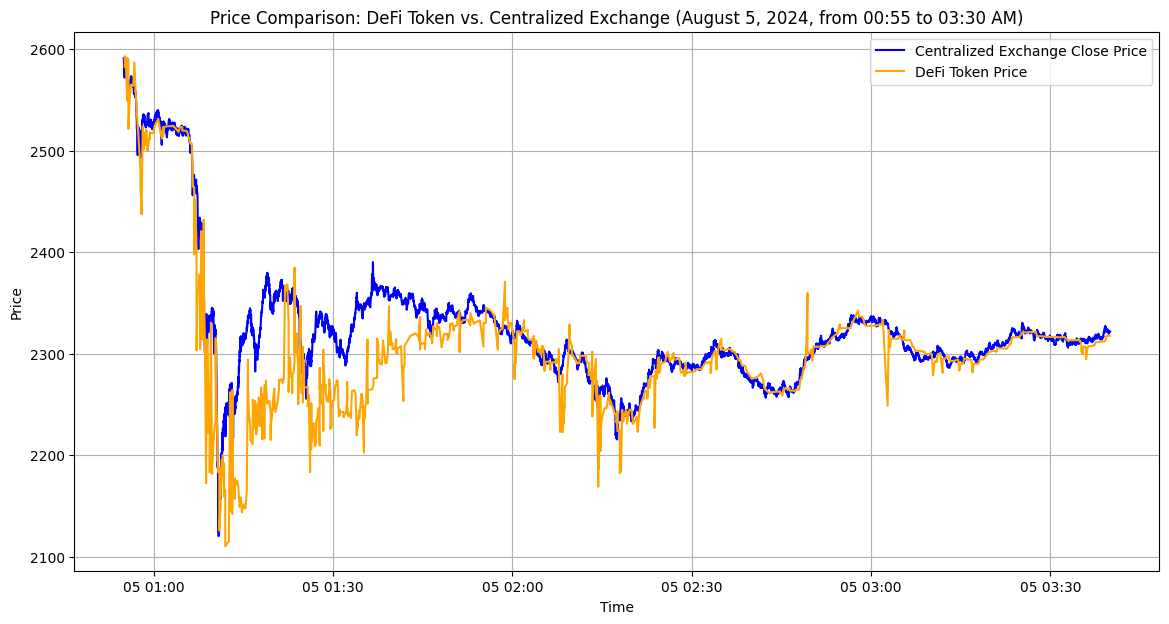}
    \caption{Zoom at the hours near the event on August 5th, 2024.}
\end{figure}
Zooming into the event, we observe that there is a significant divergence between both prices from what we would call a common efficient price around the moments of most volatility , and then they go back to normal.
We see that during the first 15 minutes of the most volatile period there is a significant divergence. We look more closer into this period.
\begin{figure}[h]
    \centering
    \includegraphics[width=0.9\textwidth]{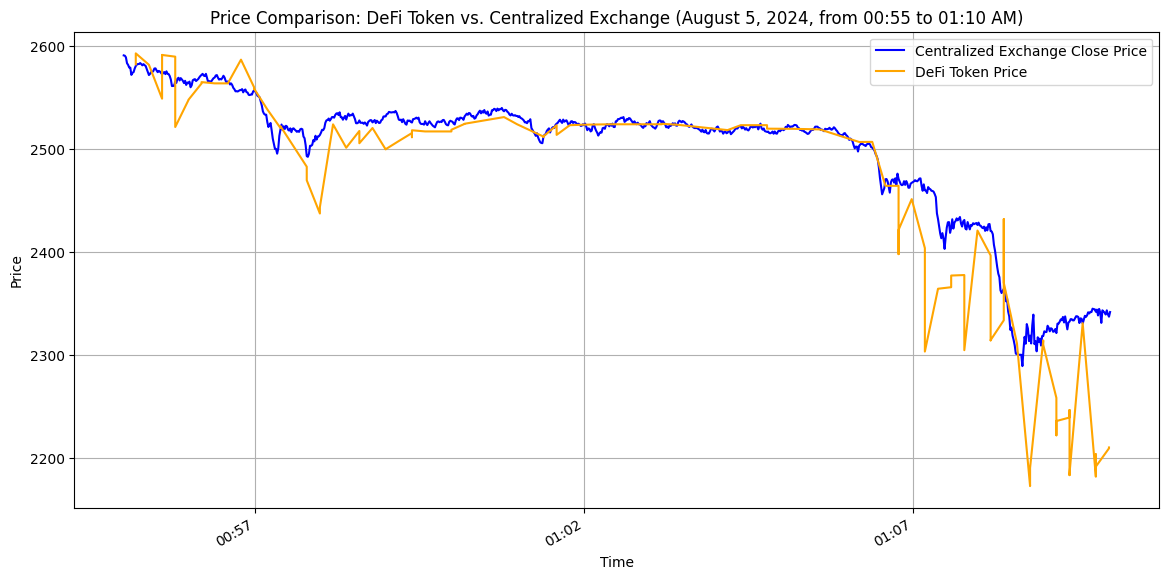}
    \caption{Plot of the evolution of prices in both markets during the 15 minute event on August.}
\end{figure}
The divergence at the start of the price drop is quite significant, and can be interpreted as a deviation from the long-term cointegrating relationship between the two prices.
\newpage
We begin by estimating a VECM, as in the long-run example. For reasons of space, we will only present the \(\alpha\) coefficient, as it is the most important one for the metrics.

The series are clearly non-stationary. First, we compute the Johansen Cointegration test with a 10 percent significance level to confirm that both series are cointegrated.

\begin{table}[h]
\centering
\caption{Johansen Cointegration Test using Trace Test Statistic (10\% Significance Level)}
\begin{tabular}{cccc}
\hline
\textbf{Rank (r)} & \textbf{Cointegrating Relations} & \textbf{Test Statistic} & \textbf{Critical Value} \\
\hline
0 & 2 & 45.54 & 13.43 \\
1 & 2 & 0.2280 & 2.705 \\
\hline
\end{tabular}
\end{table}

The results show that the null hypothesis of no cointegration (rank 0) is rejected, as the test statistic (45.54) exceeds the critical value (13.43), indicating at least one cointegrating relationship. However, the null hypothesis of at most one cointegrating relationship (rank 1) cannot be rejected, as the test statistic (0.2280) is below the critical value (2.705), confirming exactly one cointegrating relationship.

We will apply the following three methodologies to analyze the price discovery process: Hasbrouck’s Information Share, Gonzalo-Granger’s Permanent-Transitory Decomposition, and the Hayashi-Yoshida method.

We start with the Hasbrouck methodology. The output of the information shares is as follows:

\begin{table}[h]
\centering
\caption{Hasbrouck's Information Share}
\begin{tabular}{|c|c|c|}
\hline
\textbf{Market}          &  &  \\ \hline
Centralized Market       & 0.986   & 0.981 \\ \hline
Decentralized Market     & 0.019   & 0.014 \\ \hline
\end{tabular}
\label{table:hasbrouck_info_share_corrected}
\end{table}

These results indicate that the market which explains the most variance is clearly the first one, the centralized market. The analysis was conducted by calculating the metric twice, changing the order of the variables each time.

In the Gonzalo-Granger methodology, when estimating the VECM, we obtain the vector \(\alpha = [\alpha_1, \alpha_2] = [0.008, 0.025]\). This result is consistent with the previous analyses: since \(\alpha_2 > \alpha_1\), it indicates that the second market (the decentralized one) reacts more to discrepancies, so the first market (the centralized one) leads.

We performed a likelihood ratio test for \(\alpha = [1,0]\) and obtained \(\chi^2(1) = 85.141\) with a p-value of \(0.0000\). Similarly, for \(\alpha = [0,1]\), we obtained \(\chi^2(1) = 18.371\) with a p-value of \(0.0001\). These results suggest that the tests are inconclusive. However, by examining the \(\alpha\) coefficients, the evidence also supports the conclusion that the first market, the centralized one, contributes the most to price discovery.

We continue with the Hayashi-Yoshida methodology. We obtain a lead-lag ratio (LLR) of \(1.20\). The lead-lag time is calculated as \(0\) seconds, which is expected given that the analysis was not performed with tick-by-tick data. As a result, no detectable lead-lag time was observed.

The LLR of \(1.20\) indicates that the centralized exchange is leading the decentralized market, which aligns with our expectations. This conclusion is in agreement with the other methodologies, where the centralized market typically leads due to its higher liquidity. 
\newpage
\subsection{Benchmarking the performance of AMMs: analysis for September 6th, 2024}
We selected September 6th for our analysis due to the significant decline observed across various cryptocurrencies, including Ethereum. Average gas prices rose from 7.578 Gwei on September 5th to 10.72 Gwei on September 6th, indicating increased network activity. Additionally, outflows in equities markets may have contributed to the volatility in the cryptocurrency market.

\begin{table}[H]
\centering
\begin{tabular}{|l|r|r|}
\hline
\textbf{Statistic}                     & \textbf{Value Uniswap v2} & \textbf{Value Binance}  \\ \hline
Price Change During Event (\%)         & -0.95                     & -0.83                   \\ \hline
Event Maximum Price(USD)                & 2395.46
 & 2396.94                 \\ \hline
Event Minimum Price (USD)              & 	2364.68                   & 2356.48                 \\ \hline
\end{tabular}
\caption{Descriptive statistics for the event on September 6th 2024: Uniswap v2 pool WETH/USDT and ETH on Binance}
\label{tab:descriptive_event_september_combined}
\end{table}

We have 49 data observations from the Uniswap v2 pool and 901 observations from Binance, with one data point recorded every second.

\begin{figure}[h]
    \centering
    \includegraphics[width=0.8\textwidth]{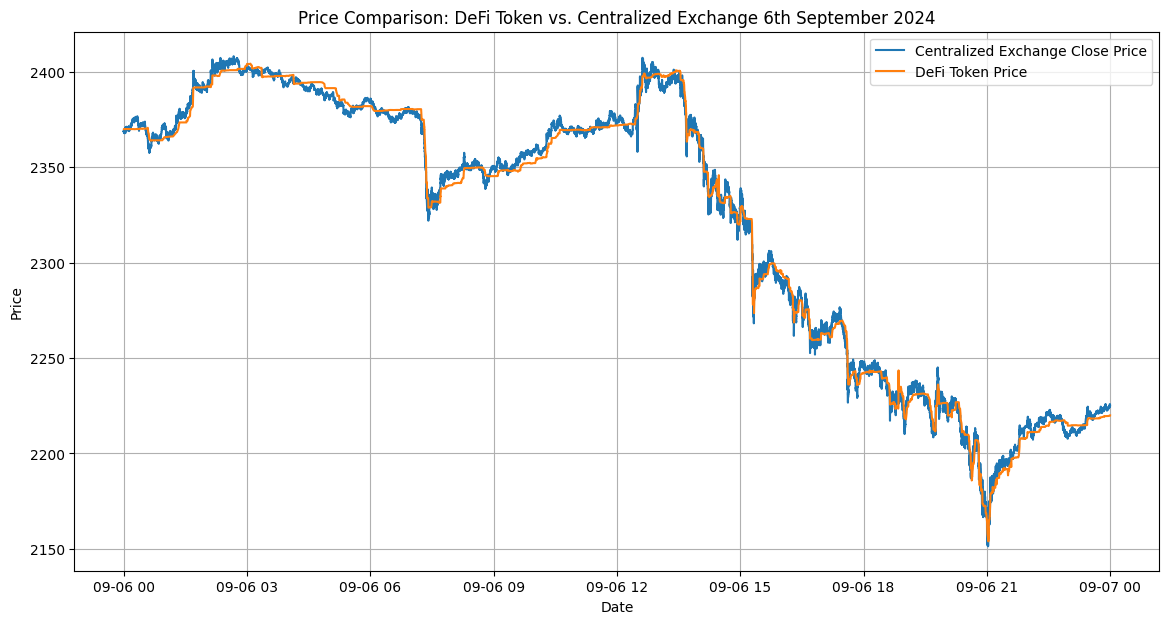}
    \caption{Plot of Ethereum prices in September 6th 2024 on both Uniswap v2 and the centralized exchange Binance.}
\end{figure}

We observe that there is a significant divergence between both prices from what we would call a common efficient price around the moments of most volatility , and then they go back to normal.

We see that during the first 20 minutes of the most volatile period there is a significant divergence. We look more closer into this period.

\begin{figure}[h]
    \centering
    \includegraphics[width=0.8\textwidth]{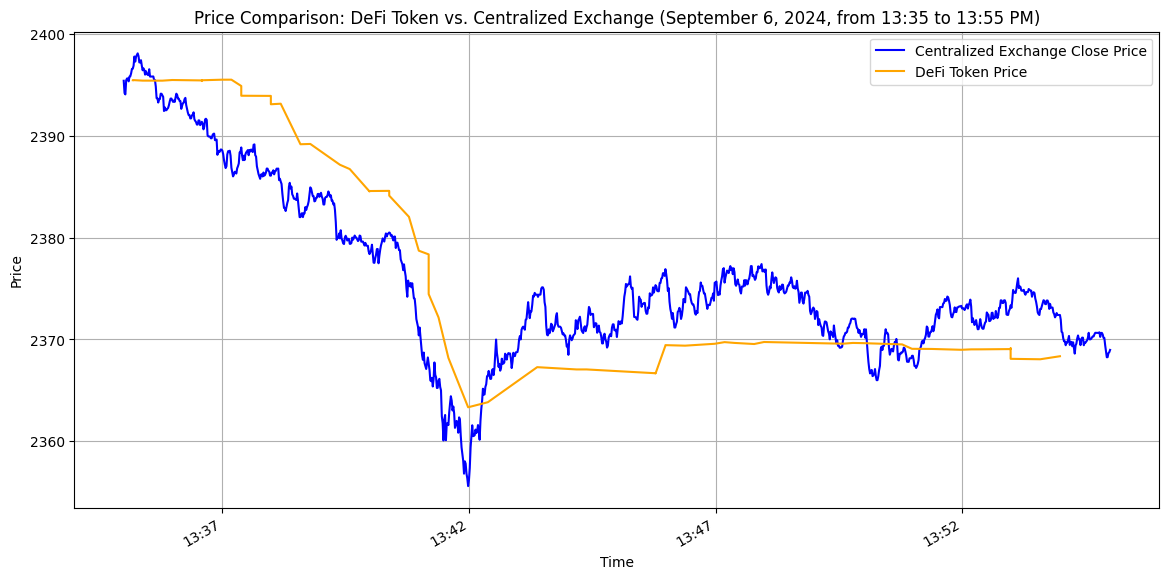}
    \caption{Plot of prices during the 20 minute event on September 6th 2024.}
\end{figure}

We begin by estimating a VECM, as in the long-run example. For reasons of space, we will only present the \(\alpha\) coefficient, as it is the most important one for the metrics.

The series are clearly non-stationary. First, we compute the Johansen Cointegration test with a 10 percent significance level to confirm that both series are cointegrated.

\begin{table}[h]
\centering
\caption{Johansen Cointegration Test using Trace Test Statistic (10\% Significance Level)}
\begin{tabular}{cccc}
\hline
\textbf{Rank (r)} & \textbf{Cointegrating Relations} & \textbf{Test Statistic} & \textbf{Critical Value} \\
\hline
0 & 2 & 51.38 & 13.43 \\
1 & 2 & 2.606 & 2.705 \\
\hline
\end{tabular}
\end{table}

The results show that the hypothesis of no cointegration (rank 0) is rejected, as the test statistic (51.38) exceeds the critical value (13.43), indicating at least one cointegrating relationship. However, the null hypothesis of at most one cointegrating relationship (rank 1) cannot be rejected, as the test statistic (2.606) is below the critical value (2.705), confirming exactly one cointegrating relationship.

We start with the Hasbrouck methodology. The output of the information shares is as follows:

\begin{table}[h]
\centering
\caption{Hasbrouck's Information Share}
\begin{tabular}{|c|c|c|}
\hline
\textbf{Market}          &  &  \\ \hline
Centralized Market       & 0.995   & 0.994 \\ \hline
Decentralized Market     & 0.006   & 0.005 \\ \hline
\end{tabular}
\label{table:hasbrouck_info_share_corrected}
\end{table}

These results indicate that the market which explains the most variance is clearly the first one, the centralized market. The analysis was conducted by calculating the metric twice, changing the order of the variables each time.

In the Gonzalo-Granger methodology, when estimating the VECM, we obtain the vector \(\alpha = [\alpha_1, \alpha_2] = [0.004, 0.019]\). This result is consistent with the previous analyses: since \(\alpha_2 > \alpha_1\), it indicates that the second market (the decentralized one) reacts more to discrepancies, so the first market (the centralized one) leads.

We performed a likelihood ratio test for \(\alpha = [1,0]\) and obtained \(\chi^2(1) = 54.3\) with a p-value of \(0\). So, we reject the null hypothesis. For \(\alpha = [0,1]\), we obtained \(\chi^2(1) = 0.789\) with a p-value of \(0.374\). We reject, so \(\alpha^{\perp} = [1,0]\). According to the Gonzalo and Granger methodology, this means that the centralized market leads the decentralized one, because the second market (the decentralized one) reacts more to discrepancies, while the first market (the centralized one) drives the price changes.

We continue with the Hayashi-Yoshida methodology. We obtain a lead-lag ratio (LLR) of \(1.08\). The lead-lag time is calculated as \(0\) seconds.

The LLR of \(1.08\) indicates that the centralized exchange is leading the decentralized market, which aligns with our expectations. This conclusion is in agreement with the other methodologies, where the centralized market typically leads due to its higher liquidity. 

\section{Arbitrage opportunity estimation. Conclusions of the analysis of centralized markets vs decentralized markets}
We now offer a summary table with the main results for each of the 5 events.

\begin{table}[h]
\centering
\caption{Summary of Results from Different Metrics for 5 Events}

\begin{subtable}[h]{0.45\textwidth}
\centering
\caption{Hasbrouck's Information Share}
\begin{tabular}{|c|c|c|}
\hline
\textbf{Date} & \textbf{Centralized Share} & \textbf{Centralized Leads?} \\ \hline
\textbf{March}     & 0.959  & Yes   \\ \hline
\textbf{April}     & 0.965  & Yes   \\ \hline
\textbf{May}       & 0.628   & Yes   \\ \hline
\textbf{August}    & 0.986  & Yes     \\ \hline
\textbf{September} & 0.995  & Yes    \\ \hline
\end{tabular}
\end{subtable}
\vspace{10pt} 
\centering 

\begin{subtable}[h]{0.45\textwidth}
\centering
\caption{Gonzalo-Granger}
\begin{tabular}{|c|c|}
\hline
\textbf{Date} & \textbf{Centralized Leads?} \\ \hline
\textbf{March}     & Yes \\ \hline
\textbf{April}     & Yes \\ \hline
\textbf{May}       & Yes  \\ \hline
\textbf{August}    & Neither market leads         \\ \hline
\textbf{September} & Yes             \\ \hline
\end{tabular}
\end{subtable}
\vspace{10pt} 
\centering 

\begin{subtable}[h]{0.45\textwidth}
\centering
\caption{Hayashi-Yoshida Results}
\begin{tabular}{|c|c|c|c|}
\hline
\textbf{Date} & \textbf{LLR} & \textbf{Centralized Leads?} \\ \hline
\textbf{March}     & 1.86   & Yes \\ \hline
\textbf{April}     & 1.47   & Yes \\ \hline
\textbf{May}       & 1.22   & Yes \\ \hline
\textbf{August}    & 1.20   & Yes \\ \hline
\textbf{September} & 1.11   & Yes \\ \hline
\end{tabular}
\end{subtable}

\label{table:summary_metrics}
\end{table}
We find that the centralized market consistently leads the decentralized one on every occasion. Only once does the Gonzalo-Granger methodology fail to provide a clear result, but across all examples, it is evident that the centralized market contributes the most to price discovery in ETH. This confirms the intuition we have developed both from the long-run analysis and the specific events. This confirms that arbitrage opportunities do exist between centralized and decentralized markets, offering potential for profit in various scenarios. To maximize these opportunities, a more detailed analysis of transaction costs could provide valuable insights into their impact on profitability.

\section{Analysis for spot cryptocurrency vs futures markets for BTC}

We are also interested in price discovery and how information flows between futures and spot markets. It is logical that both the futures market and the spot market are cointegrated, meaning their prices move together over time, given the arbitrage opportunities that exist between them. Our focus is to understand which market incorporates information faster, shedding light on whether the futures market leads price discovery or if the spot market does.

To achieve this, we will analyze various metrics and conduct a statistical examination, as we did in the decentralized vs. centralized analysis for Ethereum. The aim is to determine the primary driver of price movements, as this could have important implications for traders, investors, and institutions. Our analysis will focus on Bitcoin, given its significance and liquidity in both the spot and futures markets.

The shift from Ethereum to Bitcoin in our analysis was motivated by several factors, including Bitcoin's greater market stability, wider institutional adoption, and enhanced security features. While Ethereum offers flexibility through smart contracts, Bitcoin’s robustness and scalability make it a better fit for our long-term objectives, especially when evaluating market efficiency and price discovery in highly liquid and institutionally relevant markets.

The spot data was sourced from Binance, one of the largest and most liquid cryptocurrency exchanges, using tick-by-tick (t2t) data for the dates under analysis. The futures data come from the CME Micro Bitcoin Futures (MBT), which represent 1/10th the size of a standard Bitcoin futures contract. These futures are designed to offer institutional and retail investors a cost-effective means of managing exposure to Bitcoin with precision. Given that these contracts trade on the CME, a regulated exchange, they serve as a benchmark for institutional-level Bitcoin trading.

We have selected the same dates as those used in the decentralized vs. centralized analysis. In general, cryptocurrencies are highly correlated, so if these dates were interesting for Ethereum, they are likely interesting for Bitcoin as well. Below, we remind the reader of the main reasons these dates were selected.
\newpage
Regarding the data, we also implemented a well-structured database optimized for high-frequency market data. This database design ensures that all analyses are based on timely and accurate information, and the retrieval process leverages efficient query techniques to ensure rapid access to relevant data. These optimizations contribute to the reliability of our results, reinforcing the robustness of our conclusions.

March 5th saw a significant increase in trading volume in both centralized and decentralized markets. This period was highly volatile due to uncertainty regarding the SEC’s potential approval of Ethereum ETFs. Gas prices were also at their highest for 2024. This volatility also impacted Ethereum.

We chose April 30th for several reasons. This date marked significant turbulence in the cryptocurrency market, with notable movements driven by rising Ethereum gas fees, which surged from 11.27 to 17.41 gwei. This increase in fees indicated heightened network congestion and signaled shifts in market sentiment. At the same time, most major cryptocurrencies, including Bitcoin, experienced sharp price declines, making this an ideal period to study market behavior and price discovery across different venues.

Beyond the immediate market movements, broader macroeconomic conditions also played a significant role in shaping market behavior. By April 2024, China was showing signs of economic weakness, with slowing growth and a struggling property sector, contributing to heightened concerns about a global economic slowdown. These concerns were amplified by the ongoing Russia-Ukraine conflict, which continued to destabilize energy markets and weigh on investor sentiment. Global markets, including cryptocurrencies, remained volatile as geopolitical and economic uncertainty persisted.

Bitcoin, in particular, showed a strong correlation with broader equity market volatility during this period. As global equities reacted to concerns over rising interest rates and potential recession risks, Bitcoin’s price movements mirrored those of major stock indices, reflecting its increasing integration into traditional financial markets. This combination of macroeconomic instability, geopolitical tensions, and internal network congestion made April 30th a key period for analyzing the dynamics of price discovery and market sentiment within the cryptocurrency space.

We also selected August 5th for analysis. On this date, a major decline in equities markets occurred, which also impacted the cryptocurrency ecosystem.

On September 6th, there was a notable increase in Ethereum gas prices, rising from 7.698 to 10.72 gwei, reflecting the volatility experienced throughout the day. This surge in gas prices contributed to heightened market activity and provided an ideal opportunity to examine how the CME futures and spot Bitcoin markets reacted to these events.

The sharp decline in cryptocurrencies on that day, particularly Bitcoin, can be attributed to broader macroeconomic pressures. The U.S. stock market experienced a significant downturn, with the SP 500 falling by 1.7\% and the Nasdaq dropping by 2.6\%, led by a selloff in tech stocks. This was driven by a disappointing jobs report that fueled concerns about the strength of the U.S. economy and raised uncertainty around the Federal Reserve’s upcoming interest rate decisions. Investors feared that weak economic data could delay anticipated rate cuts, leading to a risk-off sentiment across financial markets. As Bitcoin and other cryptocurrencies have increasingly become correlated with traditional financial markets, the risk aversion seen in equities spilled over into the crypto space, driving Bitcoin’s price down by more than 3\%.

\subsection{Analysis for March 5th, 2024}
March 5th, 2024, was selected due to the significant market movements observed on this day. Bitcoin experienced a sharp 10\% drop after reaching an all-time high, leading to over \$1 billion in crypto liquidations. This extreme volatility makes it a compelling case for analyzing price discovery between the CME futures market and the spot market on Binance.

We now provide some descriptive statistics for the 15-minute event we have selected for study. Additionally, we plot both time series for the entire day and specifically for the event time window.

\begin{table}[H]
\centering
\begin{tabular}{|l|r|r|}
\hline
\textbf{Statistic}                     & \textbf{Value Futures CME} & \textbf{Value Binance}  \\ \hline
Price Change During Event (\%)         & -1.41                     & -1.54                   \\ \hline
Event Maximum Price(USD)                & 63450
 & 62779.8                 \\ \hline
Event Minimum Price (USD)              & 	62450                   & 61750.01                 \\ \hline
\end{tabular}

\caption{Descriptive statistics for the event on March 5th 2024: Micro BTC futures on CME and spot BTC on Binance}
\label{tab:descriptive_event_september_combined}
\end{table}

\begin{figure}[h]
    \centering
    \includegraphics[width=0.8\textwidth]{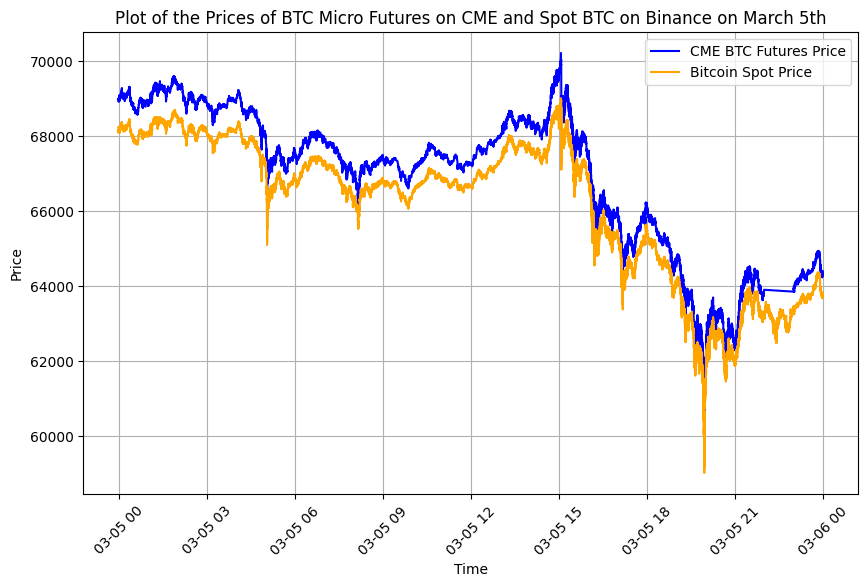}
    \caption{Plot of prices of BTC futures on CME and spot BTC on Binance during the 5th of March}
\end{figure}

We clearly observe that there is a common efficient price, with some divergences occurring during the period of highest volatility.

We notice a significant divergence during the first 15 minutes of this volatile period, which we will examine more closely.
\begin{figure}[h]
    \centering
    \includegraphics[width=0.8\textwidth]{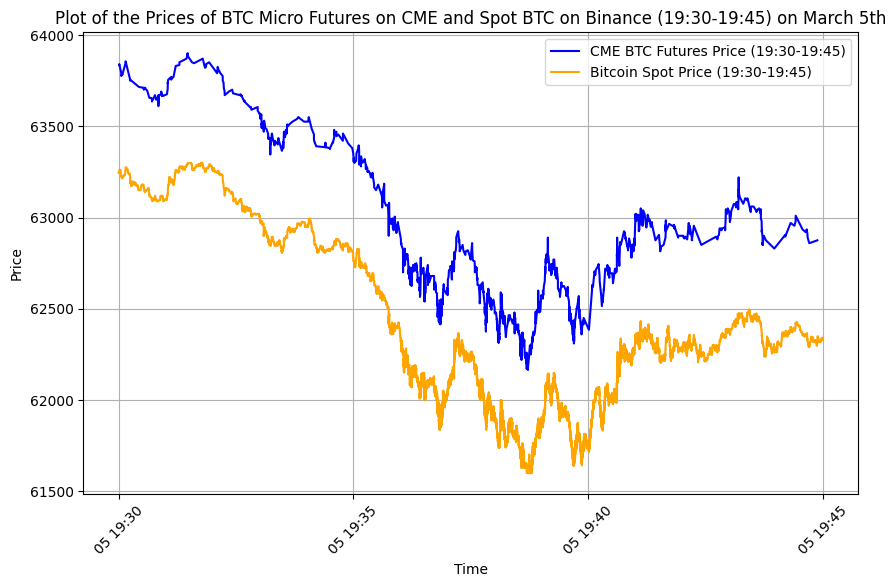}
    \caption{Plot of prices of BTC futures on CME and spot BTC on Binance during the 15-minute event on March 5th}
\end{figure}
In this case, unlike the clear distinction observed in the comparison between centralized and decentralized markets, it is not immediately evident which market incorporates more information. Here, we are comparing a futures market and a spot market, where the primary difference lies in the basis—the spread between the futures price and the spot price.
The basis captures the relationship between the two markets and reflects the cost of carry, as well as potential arbitrage opportunities.

We now begin the statistical analysis. We begin by estimating a VECM, as in the long-run example for the decentralized-centralized case. For reasons of space, we will only present the \(\alpha\) coefficient, as it is the most important one for the metrics.

The series are clearly non-stationary. First, we compute the Johansen Cointegration test with a 10 percent significance level to confirm that both series are cointegrated.

\begin{table}[h]
\centering
\caption{Johansen Cointegration Test using Trace Test Statistic (10\% Significance Level)}
\begin{tabular}{cccc}
\hline
\textbf{Rank (r)} & \textbf{Cointegrating Relations} & \textbf{Test Statistic} & \textbf{Critical Value} \\
\hline
0 & 2 & 28.45 & 13.43 \\
1 & 2 & 0.9432 & 2.705 \\
\hline
\end{tabular}
\end{table}

Results show that the null hypothesis of no cointegration (rank 0) is rejected, as the test statistic (28.45) exceeds the critical value (13.43), indicating at least one cointegrating relationship. However, the null hypothesis of at most one cointegrating relationship (rank 1) cannot be rejected, as the test statistic (0.9432) is below the critical value (2.705), confirming exactly one cointegrating relationship.

Next, we compute the Hasbrouck methodology to begin understanding which market contributes the most to price discovery. The output of the information shares is as follows:

\begin{table}[h]
\centering
\caption{Hasbrouck's Information Share}
\begin{tabular}{|c|c|c|}
\hline
\textbf{Market}          &  &  \\ \hline
Futures (CME)            & 0.563  & 0.562 \\ \hline
Spot (Binance)           & 0.436  & 0.438 \\ \hline
\end{tabular}
\label{table:hasbrouck_info_share_BTC}
\end{table}

These results indicate that the futures market narrowly explains the majority of the variance in both runs of the metric. This is because the information shares for the futures market are slightly higher, suggesting that it incorporates more information compared to the spot market. The analysis was conducted by calculating the metric twice, changing the order of the variables each time, to ensure robustness in the results.

In the Gonzalo-Granger methodology, when estimating the VECM, we obtain the following vector:

\[
\alpha = \begin{bmatrix} 
-0.001 \\ 
-0.302 
\end{bmatrix}
\]

The likelihood ratio test for \(\alpha = [1,0]\) yields \(\chi^2(1) = 203.18\) with a p-value of \(0\), meaning we reject the null hypothesis that the second market (spot) leads.  
The likelihood ratio test for \(\alpha = [0,1]\) yields \(\chi^2(1) = 89.47\) with a p-value of \(0\), meaning we also reject the null hypothesis that the first market (futures) leads.

Therefore, the Gonzalo-Granger methodology is inconclusive in this case.

We continue with the Hayashi-Yoshida methodology. We remind the reader that this methodology is especially relevant for very granular data. Since we have tick-by-tick data for both CME and BTC, this approach is particularly well-suited for our analysis.
We obtain a lead-lag time of \(0.055\) seconds, with a lead-lag ratio (LLR) of \(1.11\). This indicates that the futures market on CME leads the spot BTC market on Binance, with a minimal lag time.

In this case, the methodologies seem to point toward the futures market as the one driving price discovery. Although the Gonzalo-Granger methodology is inconclusive, the Hasbrouck and Hayashi-Yoshida methodologies support the conclusion that the futures market leads in terms of information incorporation.

\newpage
\subsection{Analysis for April 30th, 2024}
April was a bad month for cryptocurrencies, as they went through their worst period since the FTX collapse. We selected the last day of April, which was also a difficult day for cryptocurrencies overall, and Bitcoin was no exception.

We first add some descriptive statistics for the 20 minute event.
\begin{table}[H]
\centering
\begin{tabular}{|l|r|r|}
\hline
\textbf{Statistic}                     & \textbf{Value Futures CME} & \textbf{Value Binance}  \\ \hline
Price Change During Event (\%)         & -0.45                     & -0.51                   \\ \hline
Event Maximum Price(USD)                & 62545
 & 62236.81                 \\ \hline
Event Minimum Price (USD)              & 	62260                   & 61919.34            \\ \hline
\end{tabular}
\caption{Descriptive statistics for the event on April 30th 2024: Micro BTC K24 futures contract on CME and spot BTC on Binance}
\label{tab:descriptive_event_september_combined}
\end{table}

\begin{figure}[h]
    \centering
    \includegraphics[width=0.8\textwidth]{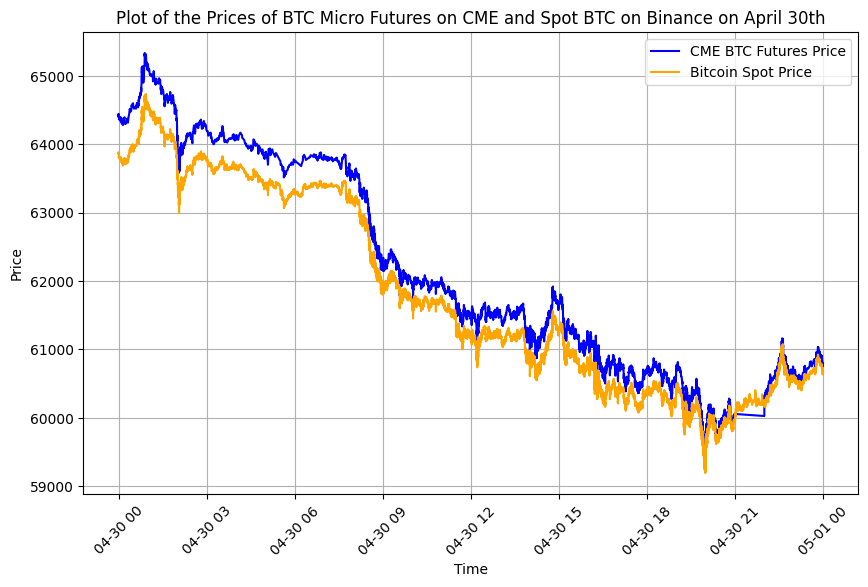}
    \caption{Plot of prices of BTC futures on CME and spot BTC on Binance during the 30th of April}
\end{figure}

We observe there is a common efficient price, with some divergences occurring during the period of highest volatility.

We notice a significant divergence during the first 15 minutes of this volatile period, which we will examine more closely.
\begin{figure}[h]
    \centering
    \includegraphics[width=0.8\textwidth]{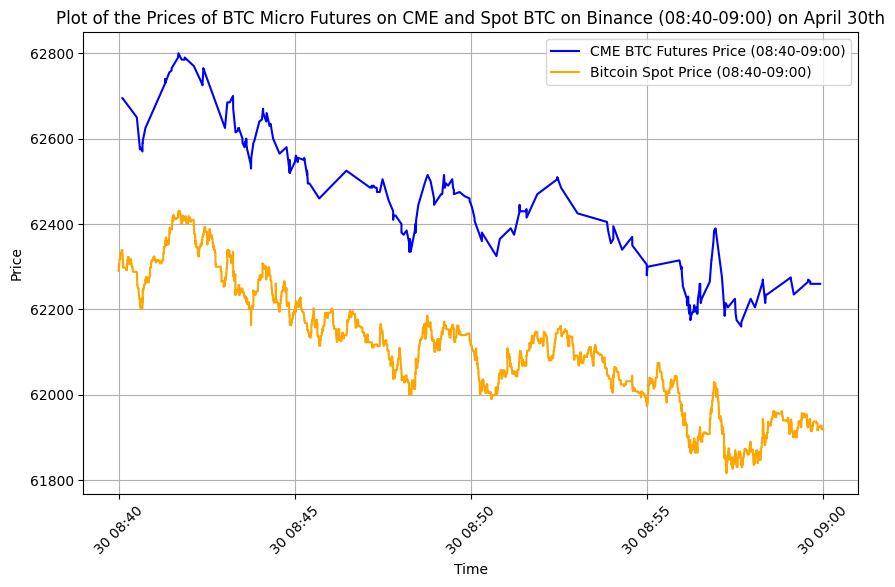}
    \caption{Plot of prices of BTC futures on CME and spot BTC on Binance during the event on April 30th}
\end{figure}
\newpage
In this case, unlike what we observed in the centralized vs decentralized comparison, it is not immediately clear which market incorporates the most information.

We begin by estimating a VECM. For reasons of space, we will only present the \(\alpha\) coefficient, as it is the most important one for the metrics.

The series are clearly non-stationary. First, we compute the Johansen Cointegration test with a 10 percent significance level to confirm that both series are cointegrated.

\begin{table}[h]
\centering
\caption{Johansen Cointegration Test using Trace Test Statistic (10\% Significance Level)}
\begin{tabular}{cccc}
\hline
\textbf{Rank (r)} & \textbf{Cointegrating Relations} & \textbf{Test Statistic} & \textbf{Critical Value} \\
\hline
0 & 2 & 37.36 & 13.43 \\
1 & 2 & 1.8435 & 2.705 \\
\hline
\end{tabular}
\end{table}

Results show that the null hypothesis of no cointegration (rank 0) is rejected, as the test statistic (37.36) exceeds the critical value (13.43), indicating at least one cointegrating relationship. However, the null hypothesis of at most one cointegrating relationship (rank 1) cannot be rejected, as the test statistic (1.8435) is below the critical value (2.705), confirming exactly one cointegrating relationship.
\newpage
We start with the Hasbrouck methodology to understand how the price discovery happens in this event. The output of the information shares is as follows:

\begin{table}[h]
\centering
\caption{Hasbrouck's Information Share}
\begin{tabular}{|c|c|c|}
\hline
\textbf{Market}          &  &  \\ \hline
Futures (CME)            & 0.552  & 0.551 \\ \hline
Spot (Binance)           & 0.448  & 0.449 \\ \hline
\end{tabular}
\label{table:hasbrouck_info_share_BTC}
\end{table}

These results indicate that the futures market explains a slightly larger share of the variance in both runs of the metric. This is because the information shares for the futures market are closer to 1, suggesting that it incorporates marginally more information compared to the spot market. The analysis was conducted by calculating the metric twice, changing the order of the variables each time, to ensure robustness in the results. In reality, both numbers are very close to 0.5, so Hasbrouck's metric says both markets contribute more or less the same to price discovery.

In the Gonzalo-Granger methodology, when estimating the VECM, we obtain the following vector:

\[
\alpha = \begin{bmatrix} 
-0.002 \\ 
-0.0009 
\end{bmatrix}
\]

The likelihood ratio test for \(\alpha = [1,0]\) yields \(\chi^2(1) = 47.033\) with a p-value of \(0\), meaning we reject the null hypothesis that the second market (spot) leads.  
The likelihood ratio test for \(\alpha = [0,1]\) yields \(\chi^2(1) = 28.989\) with a p-value of \(0\), meaning we also reject the null hypothesis that the first market (futures) leads.  
As in the March 5th example, the Gonzalo-Granger methodology is inconclusive in this case.

We continue with the Hayashi-Yoshida methodology. We obtain a lead-lag time of \(0.066\) seconds, with a lead-lag ratio (LLR) of \(1.15\). The positive lag indicates that the first market (futures) is leading, and the lead-lag ratio suggests the same.

The short lead-lag relationship reflects the fact that both markets are highly liquid and frequently arbitraged. In this case, it is unclear which market contributes more to price discovery. Both markets appear to be contributing to price discovery, but the metrics suggest that the futures market is more relevant for price discovery.

\newpage

\subsection{Analysis for May 20th, 2024}
This was a significant date for Ethereum, as the approval of ETFs became much more likely. Bitcoin was also affected by this enthusiasm, though to a lesser extent and not within the same time interval as Ethereum. We first look at some descriptive statistics.
\begin{table}[H]
\centering
\begin{tabular}{|l|r|r|}
\hline
\textbf{Statistic}                     & \textbf{Value BTC Futures CME} & \textbf{Value Binance}  \\ \hline
Price Change During Event (\%)         & -0.58                     & -0.61                   \\ \hline
Event Maximum Price(USD)                & 69825
 & 69481.79                 \\ \hline
Event Minimum Price (USD)              & 	69395                   & 69074.4                 \\ \hline
\end{tabular}
\caption{Descriptive statistics for the event on May 20th 2024: Micro BTC K24 futures contract on CME and spot BTC on Binance}
\label{tab:descriptive_event_september_combined}
\end{table}

\begin{figure}[h]
    \centering
    \includegraphics[width=0.8\textwidth]{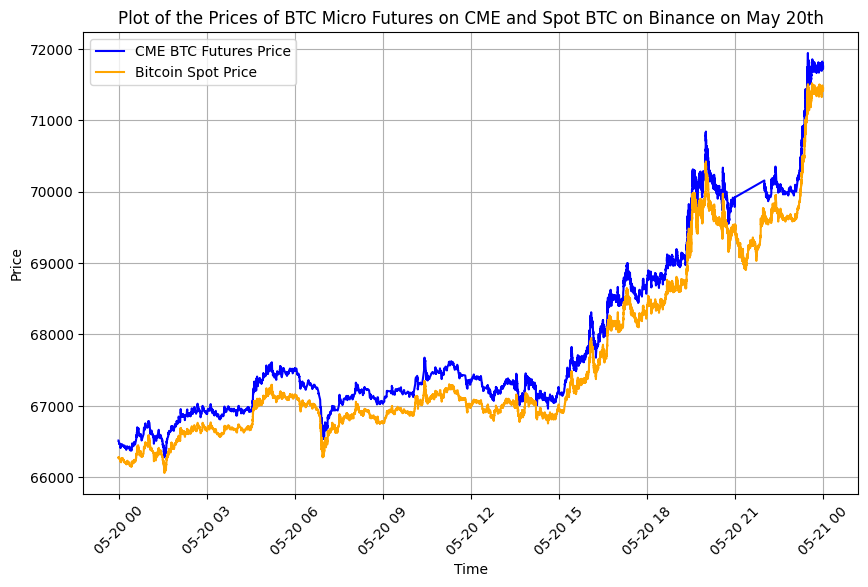}
    \caption{Plot of prices of BTC futures on CME and spot BTC on Binance during the 20th of May}
\end{figure}

There is a common efficient price, with some divergences occurring during the period of highest volatility toward the end of the day.

We notice a significant divergence during the first 15 minutes of this volatile period, which we will examine more closely.
\begin{figure}[h]
    \centering
    \includegraphics[width=0.8\textwidth]{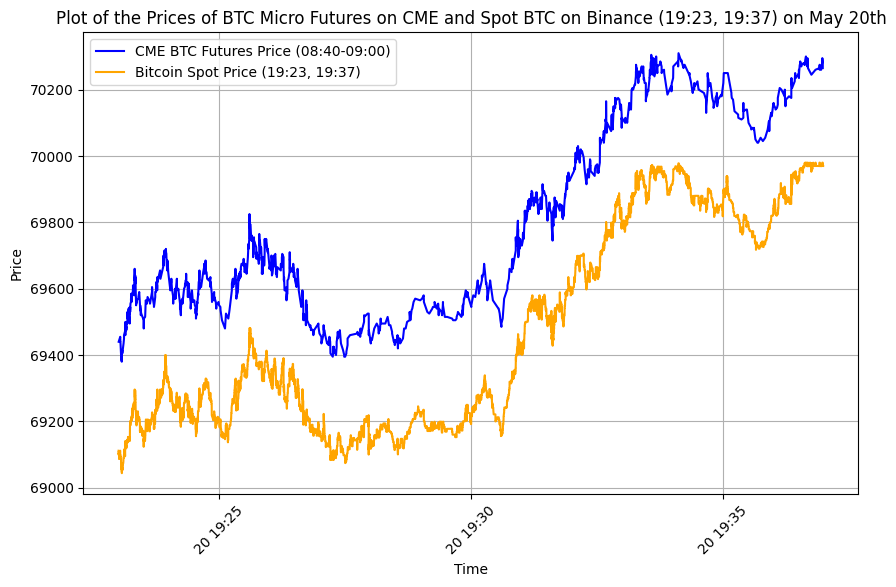}
    \caption{Plot of prices of BTC futures on CME and spot BTC on Binance during the event on May 20th}
\end{figure}
\newpage
In this case, unlike what we observed in the centralized vs decentralized comparison, it is not immediately clear which market incorporates the most information.

We begin by estimating a VECM again. For reasons of space, we will only present the \(\alpha\) coefficient, as it is the most important one for the metrics.

The series are clearly non-stationary. First, we compute the Johansen Cointegration test with a 10 percent significance level to confirm that both series are cointegrated.

\begin{table}[h]
\centering
\caption{Johansen Cointegration Test using Trace Test Statistic (10\% Significance Level)}
\begin{tabular}{cccc}
\hline
\textbf{Rank (r)} & \textbf{Cointegrating Relations} & \textbf{Test Statistic} & \textbf{Critical Value} \\
\hline
0 & 2 & 26.54 & 13.43 \\
1 & 2 & 0.7542 & 2.705 \\
\hline
\end{tabular}
\end{table}

Results show that the null hypothesis of no cointegration (rank 0) is rejected, as the test statistic (26.54) exceeds the critical value (13.43), indicating at least one cointegrating relationship. However, the null hypothesis of at most one cointegrating relationship (rank 1) cannot be rejected, as the test statistic (0.7542) is below the critical value (2.705), confirming exactly one cointegrating relationship.

\newpage

We then continue with the Hasbrouck methodology for analyzing price discovery. The output of the information shares is as follows:

\begin{table}[h]
\centering
\caption{Hasbrouck's Information Share}
\begin{tabular}{|c|c|c|}
\hline
\textbf{Market}          &  &  \\ \hline
Futures (CME)            & 0.52  & 0.51 \\ \hline
Spot (Binance)           & 0.48  & 0.49 \\ \hline
\end{tabular}
\label{table:hasbrouck_info_share_BTC}
\end{table}

In this case, both values are very close to 0.5, so the metric does not provide a clear result, although it narrowly suggests that the futures market contributes slightly more to price discovery.

In the Gonzalo-Granger methodology, when estimating the VECM, we obtain the following vector:

\[
\alpha = \begin{bmatrix} 
-0.003 \\ 
-0.0006 
\end{bmatrix}
\]

The likelihood ratio test for \(\alpha = [1,0]\) yields \(\chi^2(1) = 12.82\) with a p-value of \(0.0003\), meaning we reject the null hypothesis that the second market (spot) leads.  
The likelihood ratio test for \(\alpha = [0,1]\) yields \(\chi^2(1) = 121.74\) with a p-value of \(0\), meaning we also reject the null hypothesis that the first market (futures) leads.  
Thus, the Gonzalo-Granger methodology is inconclusive in this case.

We continue with the Hayashi-Yoshida methodology. We obtain a lead-lag time of \(0.063\) seconds, with a lead-lag ratio (LLR) of \(1.36\). This indicates that the futures market leads the spot market on CME, with a minimal lag time.

In this case, the only methodology that provides a clear result is the Hayashi-Yoshida one, which suggests that the futures market drives price discovery. However, the other methodologies are inconclusive, so this is not a clear-cut case either.

\subsection{Analysis for August 5th, 2024}
The 5th of August was relevant mainly because of the significant fall in equities, which also correlated with cryptocurrencies like Bitcoin.

The following table summarizes the key descriptive statistics for Bitcoin both for CME futures and in Binance during this event:
\begin{table}[H]
\centering
\begin{tabular}{|l|r|r|}
\hline
\textbf{Statistic}                     & \textbf{Value BTC Futures CME} & \textbf{Value Binance}  \\ \hline
Price Change During Event (\%)         & -0.85                     & -0.77                   \\ \hline
Event Maximum Price(USD)                & 56440
 & 56149                 \\ \hline
Event Minimum Price (USD)              & 	56000                   & 55667.3                 \\ \hline
\end{tabular}
\caption{Descriptive statistics for the event on August 5th 2024: Micro BTC Q24 futures contract on CME and spot BTC on Binance}
\label{tab:descriptive_event_september_combined}
\end{table}

\begin{figure}[h]
    \centering
    \includegraphics[width=0.8\textwidth]{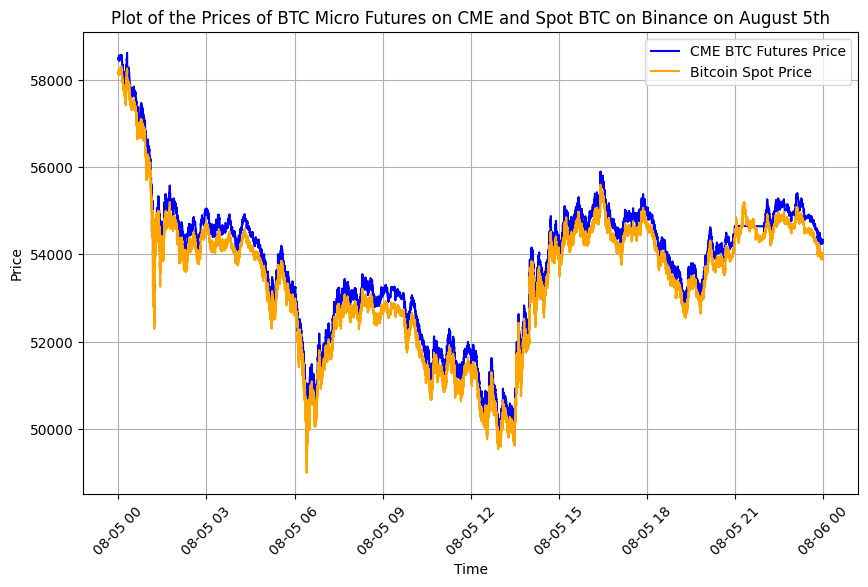}
    \caption{Plot of prices of BTC futures on CME and spot BTC on Binance during the 5th of August}
\end{figure}

We observe that there is a common efficient price, with some divergences occurring during the period of highest volatility.

We notice a significant divergence during the first 15 minutes of this volatile period, which we will examine more closely.
\begin{figure}[h]
    \centering
    \includegraphics[width=0.8\textwidth]{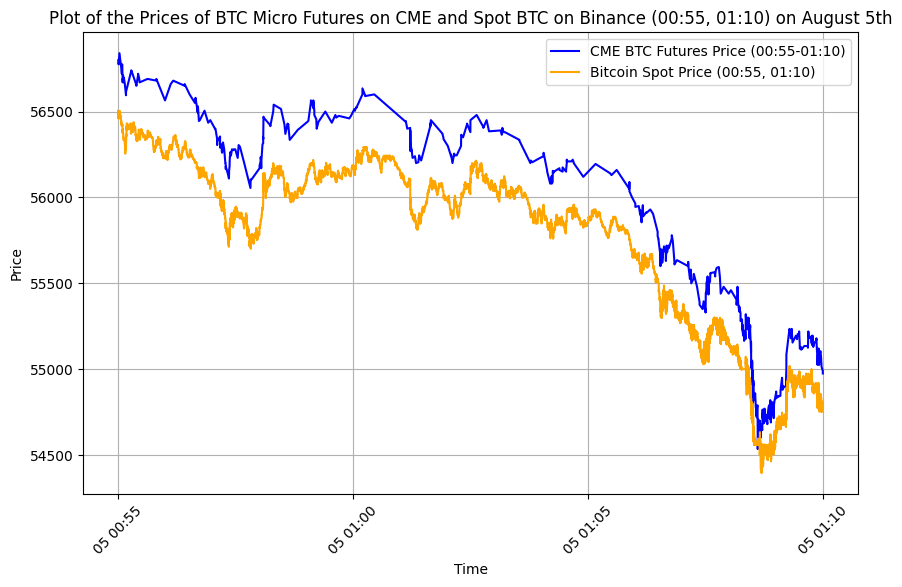}
    \caption{Plot of prices of BTC futures on CME and spot BTC on Binance during event on August 5th}
\end{figure}
In this case, unlike what we observed in the centralized vs decentralized comparison, it is not immediately clear which market incorporates the most information.
\newpage
We begin by estimating a VECM, as in the long-run example for the decentralized-centralized case. For reasons of space, we will only present the \(\alpha\) coefficient, as it is the most important one for the metrics.

The series are clearly non-stationary. First, we compute the Johansen Cointegration test with a 10 percent significance level to confirm that both series are cointegrated.

\begin{table}[h]
\centering
\caption{Johansen Cointegration Test using Trace Test Statistic (10\% Significance Level)}
\begin{tabular}{cccc}
\hline
\textbf{Rank (r)} & \textbf{Cointegrating Relations} & \textbf{Test Statistic} & \textbf{Critical Value} \\
\hline
0 & 2 & 28.77 & 13.43 \\
1 & 2 & 2.012 & 2.705 \\
\hline
\end{tabular}
\end{table}

Results show that the null hypothesis of no cointegration (rank 0) is rejected, as the test statistic (28.77) exceeds the critical value (13.43), indicating at least one cointegrating relationship. However, the null hypothesis of at most one cointegrating relationship (rank 1) cannot be rejected, as the test statistic (2.012) is below the critical value (2.705), confirming exactly one cointegrating relationship.

\newpage
We next compute Hasbrouck's information shares, to understand which market contributes the most to price discovery. The output of the information shares is as follows:

\begin{table}[h]
\centering
\caption{Hasbrouck's Information Share}
\begin{tabular}{|c|c|c|}
\hline
\textbf{Market}          &  &  \\ \hline
Futures (CME)            & 0.15  & 0.13 \\ \hline
Spot (Binance)           & 0.84  & 0.87 \\ \hline
\end{tabular}
\label{table:hasbrouck_info_share_BTC}
\end{table}

These results indicate that the spot market on Binance explains the majority of the variance in both runs of the metric. This is because the information shares for the spot market are much closer to 1, suggesting that it incorporates more information compared to the futures market. The analysis was conducted by calculating the metric twice, changing the order of the variables each time, to ensure robustness in the results.

The results from the Gonzalo-Granger decomposition and cointegration analysis provide insights into the price discovery process.

In the Gonzalo-Granger methodology, when estimating the VECM, we obtain the following vector:

\[
\alpha = \begin{bmatrix} 
0.001 \\ 
-0.007 
\end{bmatrix}
\]

The likelihood ratio test for \(\alpha = [1,0]\) yields \(\chi^2(1) = 0.93\) with a p-value of \(0.334\), meaning we do not have reason to reject the null hypothesis that the second market leads (spot). This supports the conclusion that the spot market on Binance leads, with \(\alpha_2 > \alpha_1\), indicating that the spot market reacts more to discrepancies.

The likelihood ratio test for \(\alpha = [0,1]\) yields \(\chi^2(1) = 52.94\) with a p-value of \(0\), meaning we reject the null hypothesis that the first market leads (futures).
So in conclusion, according to Gonzalo and Granger's methodology, the spot market seems to incorporate more information.

We continue with the Hayashi-Yoshida methodology. We obtain a lead-lag time of \(-0.057\) seconds, with a lead lag ratio (LLR) of \(0.97\). This indicates that the spot market leads the futures market on CME, with a minimal lag time.

The lead-lag relationship being so short reflects the fact that both markets are highly liquid and frequently arbitraged, with all three methodologies agreeing that the spot BTC market incorporates the most information during this event. This conclusion is supported by the Hasbrouck, Gonzalo-Granger, and Hayashi-Yoshida methodologies, all of which point to the spot market leading in terms of information incorporation.

\newpage
\subsection{Analysis for September 6th, 2024}
Similar to the event in August, the capital flight from equities also caused a price decrease in cryptocurrencies, with Bitcoin being affected as well.

We summarize the key descriptive statistics for Bitcoin during this event:

\begin{table}[H]
\centering
\begin{tabular}{|l|r|r|}
\hline
\textbf{Statistic}                     & \textbf{Value BTC Futures CME} & \textbf{Value Binance}  \\ \hline
Price Change During Event (\%)         & -1.00                     & -1.07                   \\ \hline
Event Maximum Price(USD)                & 56935
 & 56668                 \\ \hline
Event Minimum Price (USD)              & 	56325                   & 56101                 \\ \hline
\end{tabular}
\caption{Descriptive statistics for the event on Sept 6th 2024: Micro BTC U24 futures contract on CME and spot BTC on Binance}
\label{tab:descriptive_event_september_combined}
\end{table}

\begin{figure}[h]
    \centering
    \includegraphics[width=0.8\textwidth]{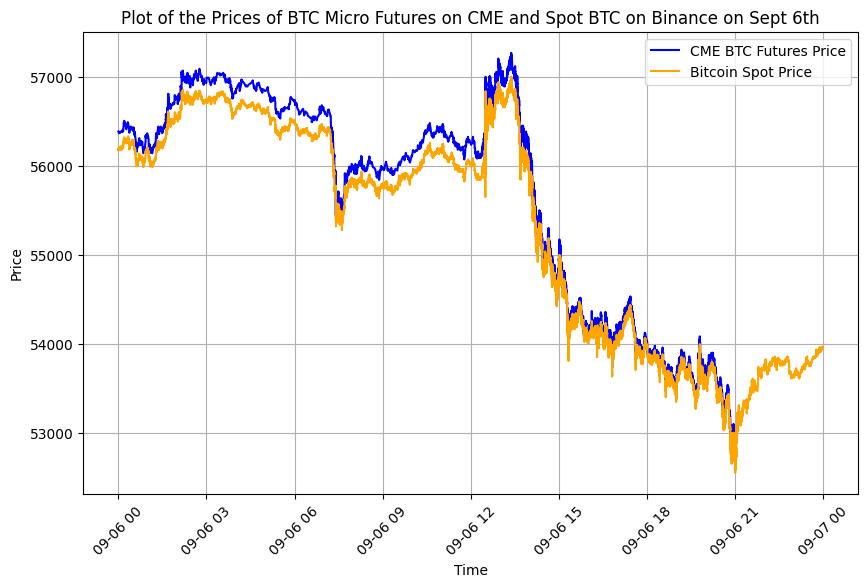}
    \caption{Plot of prices of BTC futures on CME and spot BTC on Binance during the 6th of September}
\end{figure}

We observe that there is a common efficient price, with some divergences occurring during the period of highest volatility.

\begin{figure}[h]
    \centering
    \includegraphics[width=0.8\textwidth]{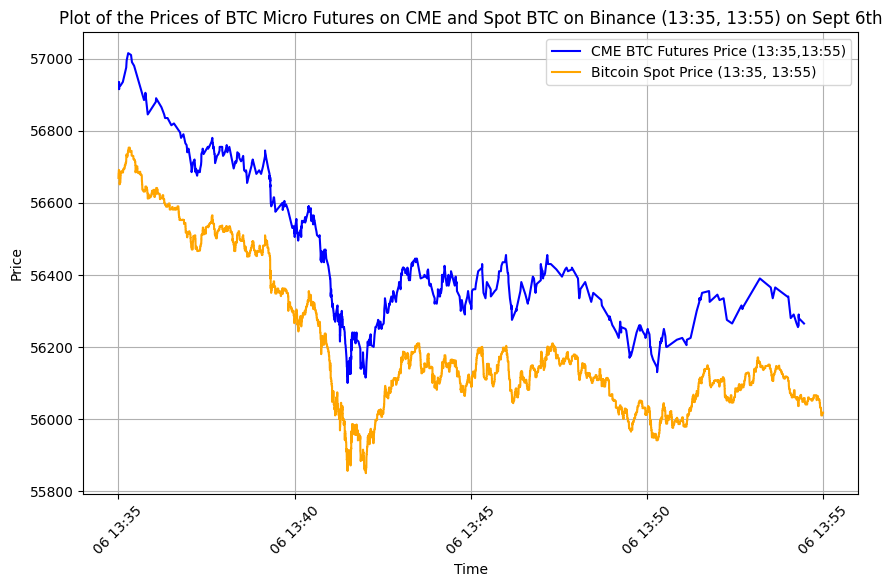}
    \caption{Plot of prices of BTC futures on CME and spot BTC on Binance during the event on 6th of September}
\end{figure}
In this case, unlike what we observed in the centralized vs decentralized comparison, it is not immediately clear which market incorporates the most information.

\newpage
We begin by estimating a VECM, as in the long-run example for the decentralized-centralized case. For reasons of space, we will only present the \(\alpha\) coefficient, as it is the most important one for the metrics.
The series are clearly non-stationary. First, we compute the Johansen Cointegration test with a 10 percent significance level to confirm that both series are cointegrated.

\begin{table}[h]
\centering
\caption{Johansen Cointegration Test using Trace Test Statistic (10\% Significance Level)}
\begin{tabular}{cccc}
\hline
\textbf{Rank (r)} & \textbf{Cointegrating Relations} & \textbf{Test Statistic} & \textbf{Critical Value} \\
\hline
0 & 2 & 18.56 & 13.43 \\
1 & 2 & 0.9834 & 2.705 \\
\hline
\end{tabular}
\end{table}

Results show that the null hypothesis of no cointegration (rank 0) is rejected, as the test statistic (18.56) exceeds the critical value (13.43), indicating at least one cointegrating relationship. However, the null hypothesis of at most one cointegrating relationship (rank 1) cannot be rejected, as the test statistic (0.9834) is below the critical value (2.705), confirming exactly one cointegrating relationship.

\newpage

Hasbrouck's information share estimates provide insights into the contribution of each market to the price discovery process. The first estimate reflects the futures market’s share, and the second reflects the spot market’s share:

\begin{table}[H]
\centering
\begin{tabular}{|l|c|c|}
\hline
\textbf{Market}              &  &  \\ \hline
Futures Market (CME)         & 0.5390   & 0.5403   \\ \hline
Decentralized Spot Market    & 0.4610   & 0.4597   \\ \hline
\end{tabular}
\caption{Hasbrouck Information Share for Futures and Spot Markets of BTC during the event}
\label{tab:hasbrouck}
\end{table}

These figures suggest a significant flow of information between the two markets, without a clear dominance by either.

In the Gonzalo-Granger methodology, when estimating the VECM, we obtain the following vector:

\[
\alpha = \begin{bmatrix} 
-0.002 \\ 
-0.0005 
\end{bmatrix}
\]

The likelihood ratio test for \(\alpha = [1,0]\) yields \(\chi^2(1) = 47.033\) with a p-value of \(0\), meaning we reject the null hypothesis that the second market leads (spot). This indicates that the futures market on CME could be leading, as \(\alpha_1 > \alpha_2\), suggesting the futures market reacts more to discrepancies.

The likelihood ratio test for \(\alpha = [0,1]\) yields \(\chi^2(1) = 28.989\) with a p-value of \(0\), meaning we also reject the null hypothesis that the first market leads (futures). So, in this case, Gonzalo and Granger's methodology is inconclusive in determining which market leads, as both hypotheses are rejected.

We continue with the Hayashi-Yoshida methodology. We obtain a lead-lag time of \(0.15\) seconds, with a lead-lag ratio (LLR) of \(1.94\). This indicates that the futures market on CME leads the spot market on Binance, with a minimal lag time.

In this case, the Hayashi-Yoshida methodology is the only one that provides a clear result, supporting the conclusion that the futures market drives price discovery. However, since Gonzalo-Granger is inconclusive and the Hasbrouck methodology does not clearly indicate a leader, this is not a definitive result.

\section{Conclusions of the analysis of spot BTC vs futures markets
on CME}

\begin{table}[h]
\centering
\caption{Summary of Results from Different Metrics for 5 Events}

\begin{subtable}[h]{0.45\textwidth}
\centering
\caption{Hasbrouck's Information Share}
\begin{tabular}{|c|c|c|}
\hline
\textbf{Date} & \textbf{Futures Share} & \textbf{Futures Leads?} \\ \hline
\textbf{March}     & 0.563  & Yes   \\ \hline
\textbf{April}     & 0.552  & Yes   \\ \hline
\textbf{May}       & 0.521   & Yes   \\ \hline
\textbf{August}    & 0.158  & No     \\ \hline
\textbf{September} & 0.539  & Yes    \\ \hline
\end{tabular}
\end{subtable}
\vspace{10pt} 
\centering 

\begin{subtable}[h]{0.45\textwidth}
\centering
\caption{Gonzalo-Granger}
\begin{tabular}{|c|c|}
\hline
\textbf{Date} & \textbf{Futures Leads?} \\ \hline
\textbf{March}     & Neither market leads \\ \hline
\textbf{April}     & Neither market leads \\ \hline
\textbf{May}       & Neither market leads  \\ \hline
\textbf{August}    & Spot         \\ \hline
\textbf{September} & Neither market leads             \\ \hline
\end{tabular}
\end{subtable}
\vspace{10pt} 
\centering 

\begin{subtable}[h]{0.45\textwidth}
\centering
\caption{Hayashi-Yoshida Results}
\begin{tabular}{|c|c|c|c|}
\hline
\textbf{Date} & \textbf{LLR} & \textbf{Lag (s)} & \textbf{Futures Leads?} \\ \hline
\textbf{March}     & 1.11  & 0.055  & Yes \\ \hline
\textbf{April}     & 1.15  & 0.066  & Yes \\ \hline
\textbf{May}       & 1.36  & 0.063  & Yes \\ \hline
\textbf{August}    & 0.97  & -0.571 & No   \\ \hline
\textbf{September} & 1.94  & 0.15   & Yes  \\ \hline
\end{tabular}
\end{subtable}

\label{table:summary_metrics}
\end{table}
We find that the futures market seems to lead on most dates according to both the Hasbrouck and Hayashi-Yoshida metrics, with the exception of August 5th. Interestingly, the Gonzalo-Granger methodology, unlike in the decentralized versus centralized case, proves to be less useful for BTC, as it does not provide definitive answers in most instances.

In summary, while the margin is not large, the futures market appears to generally lead the spot market for BTC.

We now compare our results to those discussed in the literature. \cite{HUIN} concludes that with the entrance of new information in the cryptocurrency market, it is first observed in Bitcoin futures, followed by Bitcoin spot prices. They analyzed the sample period of 2017-2020.
\cite{HUFU} also finds similar results with a 2020 sample size, using time-varying information share methodologies, concluding that futures prices Granger-cause spot prices and that futures prices dominate the price discovery process.

In general, previous literature indicates that the futures market leads the price discovery process.

We have obtained similar results, although the distinction between futures and spot prices is not as clear in most cases. While our findings confirm that futures seem to lead, the strength of this lead appears weaker than what was observed in analyses performed 4–5 years ago. Since that time, market inefficiencies have become more subtle, and although arbitrage opportunities still exist, as the metrics analyzed show, they are less pronounced than before. One possible reason is the increased access for both institutional and retail investors to new BTC spot ETFs and advances in electronic markets, which have improved market efficiency and made the differences between the two markets less pronounced.

In summary, our findings imply that both markets are integral to understanding the true value of Bitcoin. Although the futures market is likely more important to the process of price discovery, any analysis should consider the interaction between both markets.

\section{General conclusions of the empirical analysis}
We first compared the decentralized markets using three different methodologies. Our findings show that the centralized market incorporated the most information in most cases, both in the short run and the long run. One possible explanation is that centralized markets are more well-established compared to decentralized finance (DeFi), which is still relatively new.

It would be interesting to extend this analysis to other pools, such as Uniswap v3, which currently offers more liquidity than Uniswap v2 and incorporates several advantages over its predecessor \cite{Adams2021UniswapVC}.

Then, we analyzed the case of futures markets versus spot markets in the context of BTC. While there remains some uncertainty regarding which market—futures or spot—dominates in incorporating new information into Bitcoin prices, our analysis suggests that the futures market tends to incorporate more information overall. By comparing the price dynamics between CME futures and Binance spot Bitcoin markets, we observed distinct behaviors that arise from their structural differences. One key distinction is that the Binance spot market operates 24/7, while the CME futures market adheres to traditional trading hours, reflecting its institutional nature.

This raises the question of whether certain information is more efficiently processed or better utilized in one market over the other. Our findings indicate that, while both markets contribute valuable information to the overall price discovery process, the futures market appears to lead more often, although not as decisively as in the centralized versus decentralized case. The way each market absorbs and reflects new information seems to depend on market conditions, liquidity, and the participants involved.

For instance, the CME futures market often attracts institutional investors, who may react to macroeconomic events and regulatory developments, while the Binance spot market, with its 24/7 availability, tends to capture retail investor sentiment and reacts more quickly to events occurring outside traditional market hours. This suggests that while the futures market typically leads during institutional-driven events, the spot market may respond faster to global, retail-driven events or those happening outside traditional trading hours, making the two markets complementary in the price discovery process.

We now provide a table that presents the main results of our metrics for both centralized and decentralized markets, comparing ETH and BTC spot on Binance with BTC futures on CME.
\begin{sidewaystable}[h]
\centering
\caption{Summary of Results: Centralized vs Decentralized Markets and BTC Spot on Binance with BTC Futures on CME.}
\begin{tabular}{|c|c|c|c|c|c|c|c|}
\hline
\multirow{2}{*}{\textbf{Date}} & \multicolumn{2}{c|}{\textbf{Hasb. Info. Share (Fut.)}} & \multicolumn{1}{c|}{\textbf{Gonzalo-Granger (Fut.)}} & \multicolumn{3}{c|}{\textbf{Hayashi-Yoshida (Fut.)}} \\ \cline{2-7} 
                                & \textbf{Fut. Share} & \textbf{Fut. Leads?} & \textbf{Fut. Leads?} & \textbf{LLR} & \textbf{Lag (s)} & \textbf{Fut. Leads?} \\ \hline
\textbf{March}     & 0.563  & Yes & Neither  & 1.11  & 0.055 & Yes \\ \hline
\textbf{April}     & 0.552  & Yes & Neither  & 1.15  & 0.066 & Yes \\ \hline
\textbf{May}       & 0.521  & Yes & Neither  & 1.36  & 0.063 & Yes \\ \hline
\textbf{August}    & 0.158  & No  & Spot     & 0.97  & -0.571 & No   \\ \hline
\textbf{September} & 0.539  & Yes & Neither  & 1.94  & 0.15   & Yes  \\ \hline
\multirow{2}{*}{\textbf{Date}} & \multicolumn{2}{c|}{\textbf{Hasb. Info. Share (Centr.)}} & \multicolumn{1}{c|}{\textbf{Gonzalo-Granger (Centr.)}} & \multicolumn{3}{c|}{\textbf{Hayashi-Yoshida (Centr.)}} \\ \cline{2-7} 
                                & \textbf{Centr. Share} & \textbf{Centr. Leads?} & \textbf{Centr. Leads?} & \textbf{LLR} & & \textbf{Centr. Leads?} \\ \hline
\textbf{March}     & 0.959  & Yes & Yes  & 1.86  & & Yes \\ \hline
\textbf{April}     & 0.965  & Yes & Yes  & 1.47  & & Yes \\ \hline
\textbf{May}       & 0.628  & Yes & Yes  & 1.22  & & Yes \\ \hline
\textbf{August}    & 0.986  & Yes & Neither & 1.20  & & Yes \\ \hline
\textbf{September} & 0.995  & Yes & Yes  & 1.11  & & Yes \\ \hline
\end{tabular}
\label{table:summary_metrics}
\end{sidewaystable}

\clearpage  

We will now remind the reader of the data sources and specific dates used in our study. For the centralized market data, we obtained it from Binance \cite{binance}, one of the largest and most liquid cryptocurrency exchanges globally. The dataset includes detailed price and volume information, recorded at one-second intervals.
For the decentralized market data from Uniswap v2, we utilized Etherscan, a leading blockchain explorer for Ethereum. We retrieved all relevant transactions through the Etherscan API, covering the interactions that occurred in the blocks starting from May 19, 2020, when the Uniswap v2 pool was created. This dataset includes every interaction with the liquidity pool, such as swaps, liquidity additions, and removals, providing a comprehensive view of decentralized trading activity.

The specific dates selected for each month in our analysis are as follows: March 5th, April 30th, May 20th, August 5th, and September 6th, all in 2024. We also include the number of observations for both the centralized and decentralized datasets. For instance, in March, we have 43 observations from the Uniswap v2 pool and 841 observations from Binance. In May, there are 178 data points from the Uniswap v2 pool and 900 observations from Binance, all recorded at one-second intervals. For August, we recorded 49 observations from the Uniswap v2 pool and 901 from Binance. In September, there were 48 observations from the Uniswap v2 pool and 1,201 from Binance. We observe that in all cases we have more observations for the Binance time series.

In the second case of our analysis, we focused on Bitcoin. Here, the data collection followed the same structure as in the Ethereum analysis. The dates used are the same: March 5th, April 30th, May 20th, August 5th, and September 6th, all in 2024. The spot data was again sourced from Binance, in this case using tick-by-tick (t2t) data for these dates, ensuring a detailed and accurate depiction of spot market behavior. On the futures side, we used data from the CME Micro Bitcoin Futures, which represent one-tenth of the size of a standard Bitcoin futures contract. The futures data are also provided as tick-by-tick (t2t) data, ensuring consistency across both markets.

For both cases—centralized vs. decentralized markets and Bitcoin analysis—we used the same dates and data structure. The selected dates, March 5th, April 30th, May 20th, August 5th, and September 6th, 2024, allowed for a direct comparison across the different metrics we explored.

\chapter*{Annex: Database connection and usage}

All the data used in this research was retrieved from the data ecosystem created under the grant TED2021-131844B-I00. The database has been developed in accordance with the project’s scope and objectives. Documentation related to this dataset is available for replication purposes, as detailed in the annex of this document. Further information can be found at: \url{https://arfima.pages.arfima.com/arfima/crypto_project_uc3m/}.

In the analysis of cryptocurrency price discovery, accessing and working with large volumes of data is critical. This annex outlines the steps to establish a connection to the cryptocurrency database, retrieve data using SQL queries, and perform advanced queries using Python. 

\section*{Database Connection}
We can connect to the database using pgAdmin or programmatically through Python with PostgreSQL. Below is an example of how to retrieve data using SQL queries.

\subsection*{SQL Queries in pgAdmin}

The following are examples of queries used for retrieving data from the database:

\begin{lstlisting}[language=SQL]
SELECT DISTINCT(instrument)
FROM primarydata.okex_intrade;

SELECT *
FROM primarydata.erc20_transactions
ORDER BY time_stamp
LIMIT 20000000;

SELECT *
FROM primarydata.erc20_transactions
WHERE dtime >= '2024-08-06 00:00:00+00' AND dtime < '2024-08-07 00:00:00+00'
ORDER BY dtime;

SELECT *
FROM primarydata.binance_t2trade
WHERE dtime >= '2024-01-04'
AND instrument LIKE 'eth_%';
\end{lstlisting}

These queries retrieve distinct instruments, large transaction sets, transactions within a specific date range, and all Ethereum trade data from Binance since January 4th, 2024.

\section*{Querying in Python}
Below is an example of how to connect to the database and retrieve data programmatically using Python:

\begin{lstlisting}[language=Python]
import asyncio
from uglyData.ingestor.tslib import TSLib

conninfo = "postgresql://username:password@xx.xx.xx.xx:5432/cryptodbhist"

async def connect():
    tlib = TSLib()
    await tlib.connect(conninfo=conninfo)

async def fetch_data():
    ERC20_QUERY = """SELECT *
                     FROM primarydata.erc20_transactions
                     ORDER BY time_stamp
                     LIMIT 20000000;"""
    
    erc20_time_stamps = await tlib.conn.fetch(ERC20_QUERY, output="df")
    return erc20_time_stamps

# Execution
if __name__ == '__main__':
    asyncio.run(connect())
    df = asyncio.run(fetch_data())
\end{lstlisting}

This Python script demonstrates how to connect to the PostgreSQL database and retrieve the data in a pandas dataframe format for further analysis.

\printbibliography

\end{document}